\documentclass[12pt]{article}%
\pdfoutput=1
\usepackage[nosort]{cite}
\usepackage{graphicx}
\usepackage{multicol}
\usepackage{amsfonts}
\usepackage{amssymb}
\usepackage{amsmath}
\usepackage{heck}
\usepackage{afterpage}
\usepackage{setspace}
\usepackage{verbatim}
\usepackage{color}
\usepackage{longtable}
\usepackage{float}
\usepackage{subcaption}
\usepackage{epsfig}
\usepackage{epstopdf}
\usepackage{adjustbox}
\newsavebox{\mysavebox}
\newlength{\myrest}
\usepackage{tikz}
\usepackage[margin=1in]{geometry}
\usepackage{titletoc}%
\usepackage{hyperref}
\hypersetup{colorlinks,%
citecolor=black,%
filecolor=black,%
linkcolor=black,%
urlcolor=black,%
pdftex}
\providecommand{\U}[1]{\protect\rule{.1in}{.1in}}
\usetikzlibrary{decorations.markings}

\numberwithin{equation}{section}

\newcommand{\ba}{\begin{eqnarray}}
\newcommand{\ea}{\end{eqnarray}}

\newcommand{\mf}{\mathfrak}

\newcommand\im{\mathrm{Im}}

\newcommand{\be}{\begin{equation}}
\newcommand{\ee}{\end{equation}}

\tikzstyle{startstop} = [rectangle, rounded corners, minimum width=3cm, minimum height=1cm,text centered, draw=black, fill=blue!10]
\tikzstyle{startstop} = [rectangle, rounded corners, minimum width=3cm, minimum height=1cm,text centered, draw=black, fill=blue!10]

\tikzstyle{io} = [trapezium, trapezium left angle=70, trapezium right angle=110, minimum width=3cm, minimum height=1cm, text centered, draw=black, fill=blue!30]

\tikzstyle{process} = [rectangle, minimum width=3cm, minimum height=1cm, text centered, draw=black, fill=orange!30]
\tikzstyle{decision} = [diamond, minimum width=3cm, minimum height=1cm, text centered, draw=black, fill=green!30]

\tikzstyle{arrow} = [thick,->,>=stealth]

\begin{document}

\date{January 2016}

\title{6D RG Flows and Nilpotent Hierarchies}

\institution{UNC}{\centerline{${}^{1}$Department of Physics, University of North Carolina, Chapel Hill, NC 27599, USA}}

\institution{COLUMBIA}{\centerline{${}^{2}$Department of Physics, Columbia University, New York, NY 10027, USA}}

\institution{CUNY}{\centerline{${}^{3}$CUNY Graduate Center, Initiative for the Theoretical Sciences, New York, NY 10016, USA}}

\institution{HARVARD}{\centerline{${}^{4}$Jefferson Physical Laboratory, Harvard University, Cambridge, MA 02138, USA}}

\institution{BICOCCA}{\centerline{${}^{5}$Dipartimento di Fisica, Universit\`a di Milano-Bicocca, Milan, Italy}}

\institution{INFN}{\centerline{${}^6$INFN, sezione di Milano-Bicocca, Milan, Italy}}

\authors{Jonathan J. Heckman\worksat{\UNC, \COLUMBIA, \CUNY}\footnote{e-mail: {\tt jheckman@email.unc.edu}},
Tom Rudelius\worksat{\HARVARD}\footnote{e-mail: {\tt rudelius@physics.harvard.edu}},
and Alessandro Tomasiello\worksat{\BICOCCA,\INFN}\footnote{e-mail: {\tt alessandro.tomasiello@unimib.it}}}

\abstract{With the eventual aim of classifying
renormalization group flows between 6D superconformal field theories (SCFTs), we
study flows generated by the vevs of ``conformal matter,'' a generalization
of conventional hypermultiplets which naturally appear in the F-theory
classification of 6D SCFTs. We consider flows in which the parent UV theory
is (on its partial tensor branch) a linear chain of gauge groups connected by conformal matter,
with one flavor group $G$ at each end of the chain, and in which
the symmetry breaking of the conformal matter at each end is parameterized by
the orbit of a nilpotent element, i.e. T-brane data, of one of these flavor symmetries.
Such nilpotent orbits admit a partial ordering, which is reflected
in a hierarchy of IR fixed points. For each such nilpotent orbit, we determine
the corresponding tensor branch for the resulting SCFT. An important feature of
this algebraic approach is that it also allows us to systematically compute the
unbroken flavor symmetries inherited from the parent UV theory.}

\maketitle

\tableofcontents

\enlargethispage{\baselineskip}

\setcounter{tocdepth}{2}

\newpage

\section{Introduction \label{sec:INTRO}}

Renormalization group flows constitute a foundational element in the study of
quantum field theory. As fixed points of these flows, conformal field theories
are especially important in a range of physical phenomena. One of the surprises from
string theory is that suitable decoupling limits lead to the construction of
conformal fixed points in more than four spacetime dimensions \cite{Witten:1995zh}.

Though the full list of conformal field theories
is still unknown, there has recently been significant progress in
classifying six-dimensional superconformal field theories
(SCFTs). A top down classification of 6D\ SCFTs via
compactifications of F-theory has been completed in \cite{Heckman:2013pva,
DelZotto:2014hpa, Heckman:2014qba, DelZotto:2014fia,
Heckman:2015bfa} (see also \cite{Bhardwaj:2015xxa}
and \cite{Bhardwaj:2015oru} as well as the holographic
classification results of reference \cite{Apruzzi:2013yva}).\footnote{There are
still a few outlier theories which appear consistent with field theoretic
constructions, and also admit an embedding in perturbative IIA string theory (see e.g. \cite{Hanany:1997gh}).
As noted in \cite{Bhardwaj:2015oru}, these will likely yield to an embedding in a non-geometric
phase of F-theory since the elements of these constructions are so close to
those obtained in geometric phases of F-theory.} An important element
in this work is that in contrast to lower-dimensional systems, all of these
SCFTs have a simple universal structure given (on its partial tensor branch) by a
generalized quiver gauge theory consisting of a single spine of quiver nodes joined by links which in   \cite{DelZotto:2014hpa} were dubbed ``conformal matter.'' There can
also be a small amount of decoration by such links on the ends of this generalized quiver.

With such a list in place, the time is ripe to extract more detailed properties of these theories.
Though the absence of a Lagrangian construction is an obstruction,
it is nevertheless possible to extract some precision data such as the anomaly polynomial
\cite{Harvey:1998bx, Ohmori:2014pca, Ohmori:2014kda},
the scaling dimensions of certain protected operators
\cite{Heckman:2014qba} and the structure of the partition vector and its relation to the
spectrum of extended defects \cite{Tachikawa:2013hya, DelZotto:2015isa}.

It is also natural to expect that there is an overarching structure governing
possible RG\ flows between conformal fixed points. In recent work \cite{Heckman:2015ola}, the geometry of
possible deformations of the associated Calabi--Yau geometry of an F-theory
compactification has been used to characterize possible flows between theories, and has even been
used to give a \textquotedblleft proof by brute force\textquotedblright\
(i.e.~sweeping over a large list of possible flows) of $a$- and $c$-theorems in six
dimensions \cite{Heckman:2015axa} (see also \cite{Cordova:2015fha}
and \cite{Beccaria:2015ypa, Cremonesi:2015bld}).
In this geometric picture, there are two general classes of flows
parameterized by vevs for operators of the theory. On the tensor branch, we
consider vevs for the real scalars of 6D tensor multiplets, which in the
geometry translate to volumes of $\mathbb{P}^{1}$'s in the base of an F-theory
model. On the Higgs branch, we consider vevs for operators which break the
$SU(2)$\ R-symmetry of the SCFT. Geometrically, these correspond to complex
structure deformations. There are also mixed branches. Even so, a global
picture of how to understand the network of flows between theories
remains an outstanding open question.

Motivated by the fact that all 6D\ SCFTs are essentially just generalized
quivers, our aim in this note will be to study possible
RG\ flows for one such class of examples in which the decoration on the left
and right of a generalized quiver is \textquotedblleft
minimal.\textquotedblright\ These are theories which in M-theory are realized
by a stack of $k$ M5-branes probing the transverse geometry $\mathbb{R}_{\bot
}\times\mathbb{C}^{2}/\Gamma_{ADE}$, i.e.~the product of the real line with an
ADE\ singularity. In F-theory they are realized by a single linear chain of
$-2$ curves in the base which are wrapped by seven-branes with gauge group of
corresponding ADE type, in which there is a non-compact ADE\ seven-brane on
the very left and one on the very right as well. In M-theory, we reach the
SCFT point by making all the M5-branes coincident on the $\mathbb{R}_{\bot}$
factor (while still probing the orbifold singularity), while in F-theory this
is obtained by collapsing all of the $-2$ curves to zero volume.

One of the interesting features of these models is that on the partial tensor
branch, i.e.~where we separate all M5-branes along the transverse real line,
and in F-theory where we resolve all $-2$ curves, we can recognize that there are
additional degrees of freedom localized along defects of a higher-dimensional
bulk theory. Indeed, from the F-theory perspective, the degenerations of the
elliptic fibration at these points needs to be accompanied by additional
blowups in the base, leading to \textquotedblleft conformal
matter.\textquotedblright\ The reason for the suggestive terminology is
twofold. First, the actual structure of the geometries constructed
from M- and F-theory has the appearance of a generalized quiver. Second, and perhaps more
importantly, there is a precise notion in the F-theory description of
activating complex structure deformations at the places where conformal matter
is localized. For example, the breaking pattern for a conformal matter system
with $E_{8}\times E_{8}$ global symmetry to a system with only $E_{7}\times
E_{7}$ global symmetry is given by:%
\begin{equation}
y^{2}=x^{3}+\alpha u^{3}v^{3}x+u^{5}v^{5}.
\end{equation}
Such deformations trigger a decrease in the total number of tensor multiplets, and
also break the UV R-symmetry, with another emerging in the IR.

Since the structure of tensor branch flows is immediately captured by the
geometry of the F-theory model, i.e. K\"ahler resolutions of the base,
we shall primarily focus on Higgs branch flows.
Part of our aim will be to develop a general picture of how
vevs for conformal matter generate RG flows.

Along these lines, we provide supporting evidence
for this picture of conformal matter vevs, and use it as a way of
characterizing the induced flows for 6D\ SCFTs. In more detail, we
consider the class of theories called
\begin{equation}\label{eq:T}
	\mathcal{T}(G, \mu_{L},\mu_{R}, k)
\end{equation}
in reference \cite{DelZotto:2014hpa}. They are parameterized
by a choice of ADE group $G$; by a pair of nilpotent elements $\mu_{L}$ and $\mu_{R}$ in the complexification $\mathfrak{g}_{\mathbb{C}}$ of the Lie algebra of $G$;
and by a positive integer $k$. In the M-theory realization, $k$ is the number of M5-branes, and $\mu_L$, $\mu_R$ specify \textquotedblleft Nahm pole data\textquotedblright\ of a 7D super Yang-Mills theory. In the F-theory description, the theories (\ref{eq:T}) represent a chain of $-2$ curves with gauge group $G$ on each of them, with a \textquotedblleft T-brane\textquotedblright%
\cite{Donagi:2003hh, Cecotti:2010bp} (see also \cite{Donagi:2011jy, Anderson:2013rka, Heckman:2010qv, Collinucci:2014qfa, Collinucci:2014taa})
on each flavor curve at an end of the chain. The $\mu_L$ and $\mu_R$ appear as residues of a Higgs field for a Hitchin system on these flavor curves. These residues are in turn captured by operator vevs of the low energy effective field theory \cite{Anderson:2013rka, Beasley:2008dc}.
This provides the basic link between \textquotedblleft boundary
data\textquotedblright\ and the vevs of operators associated with conformal matter.

The first result of this paper is an explicit identification of the tensor branch for the theories $\mathcal{T}(G, \mu_{L},\mu_{R}, k)$
of line (\ref{eq:T}). To reach this, we shall find it convenient to take $k$ in (\ref{eq:T}) sufficiently large so that the effects of $\mu_{L}$ and $\mu_{R}$ decouple, so our aim will be to capture the effects of flows associated with just a single nilpotent element
of the flavor symmetry algebra. For $G=SU(N)$ or $SO(2N)$, nilpotent elements can be parameterized by partitions (i.e.~Young diagrams). For $G=E_n$, one cannot use partitions any more: their analogues are called \textit{Bala--Carter labels} (for a review of B--C labels, see for example \cite[Chap.~8]{NILPbook} or \cite{Chacaltana:2012zy}).\footnote{As a brief aside, let us note that the case of $k$ sufficiently large leads to a class of (singular) M-theory duals for these theories in which the T-brane data is localized near the orbifold fixed points of the classical gravity dual \cite{DelZotto:2014hpa}. For $G=SU(N)$, these theories also have a IIA realization \cite{Hanany:1997gh,Gaiotto:2014lca} and a non-singular holographic dual \cite{Apruzzi:2013yva,Apruzzi:2015wna}.}

Secondly, we will find that the well-known partial ordering on nilpotent elements also leads to a class of theories which can be connected by an RG\ flow:%
\begin{equation}\label{eq:RG}
\mu < \nu\Rightarrow\text{RG\ Flow: }\mathcal{T}(\mu)\rightarrow
\mathcal{T}(\nu).
\end{equation}

To provide further evidence in favor of our proposal, we also consider related theories where
the left flavor symmetry is replaced by a non-simply laced algebra. We get to such theories by first doing a blowdown of
some curves on the tensor branch for the $\mathcal{T}(G, \mu_{L},\mu_{R}, k)$ theories which are
then followed by a further vev for conformal matter. In this case, the flavor
symmetry does not need to be a simply laced ADE type algebra, but can also be
a non-simply laced BCFG algebra. All of this is quite transparent on the
F-theory side, and we again expect a parametrization of flows in terms of
nilpotent hierarchies. We find that this is indeed the case, again providing
highly non-trivial evidence for our proposal.

Another outcome from our analysis is that by phrasing everything
in terms of \textit{algebraic} data of the 6D\ SCFT flavor symmetry, we can
also read off the unbroken flavor symmetry, i.e.~those symmetry generators
which commute with our choice of nilpotent element. This provides a rather
direct way to determine the resulting IR flavor symmetry which is different
from working with the associated F-theory geometry. Indeed, there are
a few cases where we find that the geometric expectation from F-theory
predicts a flavor symmetry which is a proper subalgebra of the flavor symmetry
found through our field theoretic analysis. This is especially true in the case of
\textit{abelian} flavor symmetries. For more details on extracting
the geometric contribution to the flavor symmetry, see e.g. \cite{Bertolini:2015bwa}.

The rest of this paper is organized as follows. In section \ref{sec:CONFMATT}
we give a brief overview of some elements of conformal matter and how it
arises in both M- and F-theory constructions. After this, in section \ref{sec:CLASSICAL}
we give a first class of examples based on flows involving 6D\ SCFTs where the flavor symmetry
is a classical algebra. For the $SU$-type flavor symmetries, there is a
beautiful realization of nilpotent elements in terms of partitions of a brane
system. This is also largely true for the $SO$- and $Sp$-type algebras as well,
though there are a few cases where this correspondence breaks down. When this occurs,
we find that there is still a flow, but that some remnants of exceptional
algebras creep into the description of the 6D\ SCFT because of the presence of conformal matter in the system.
After this, we turn in section \ref{sec:EXCEPTIONAL} to flows for theories with an exceptional flavor symmetry. In some
cases there is a realization of these flows in terms of deformations of
$(p,q)$ seven-branes, though in general, we will find it more fruitful to work
in terms of the algebraic characterization of nilpotent orbits. Section \ref{sec:SHORT}
extends these examples to ``short'' generalized quivers where the breaking patterns of different
flavor symmetries are correlated, and in section \ref{sec:FLAVOR} we explain how this algebraic characterization
of flavor breaking sometimes leads to different predictions for the flavor symmetries of a 6D SCFT
compared with the geometric realization. In
section \ref{sec:CONC} we present our conclusions and potential directions for future research.
Some additional material on the correspondence between nilpotent orbits for
exceptional algebras and the corresponding
F-theory SCFTs is provided in an Appendix.

\section{Conformal Matter \label{sec:CONFMATT}}

In this section we discuss some of the salient features of 6D conformal matter introduced
in references \cite{DelZotto:2014hpa, Heckman:2014qba}, and the corresponding realization of these systems in
both M- and F-theory. We also extend these considerations, explaining the sense in which
conformal matter vevs provide a succinct way to describe
brane recombination in non-perturbatively realized configurations
of intersecting seven-branes.

Recall that to get a supersymmetric vacuum in 6D Minkowski space, we consider F-theory compactified on an elliptically fibered
Calabi--Yau threefold. Since we are interested in a field theory limit, we always take the base of the elliptic model
to be non-compact so that gravity is decoupled. Singularities of the elliptic fibration lead to divisors in the base, i.e.
these are the loci where seven-branes are wrapped. When the curve is compact,
this leads to a gauge symmetry in the low energy theory, and when the curve is
non-compact, we instead have a flavor symmetry.

In F-theory, we parameterize the profile of the axio-dilaton using the
Minimal Weierstrass model:%
\begin{equation}
y^{2}=x^{3}+fx+g,
\end{equation}
where here, $f$ and $g$ are sections of bundles defined over the base. As
explained in reference \cite{Heckman:2013pva}, the \textquotedblleft non-Higgsable
clusters\textquotedblright\ of reference \cite{Morrison:2012np} can be used to
construct the base for the tensor branch of all 6D\ SCFTs. The basic idea is
that a collapsed $-1$ curve in isolation defines the \textquotedblleft E-string
theory,\textquotedblright  that is, a theory with an $E_{8}$ flavor symmetry.
By gauging an appropriate subalgebra of this flavor symmetry, we can start to
produce larger bases, provided these additional compact curves are part of a
small list of irreducible building blocks known as \textquotedblleft
non-Higgsable clusters.\textquotedblright

For the present work, we will not need to know much about the structure of
these non-Higgsable clusters, so we refer the interested reader for example to \cite{Heckman:2013pva, Morrison:2012np}
for further details. The essential feature we require is that the
self-intersection of a curve --- or a configuration of curves --- dictates the
minimal gauge symmetry algebra supported over the curve. In some limited
situations, additional seven-branes can be wrapped over some of these curves.
Let us briefly recall the minimal gauge symmetry for the various building
blocks of an F-theory base:%
\begin{align}
\text{single curve}  &  \text{: }\overset{\mathfrak{su}_{3}}{3}\text{,}%
\overset{\mathfrak{so}_{8}}{4}\text{,}\overset{\mathfrak{f}_{4}}{5}%
\text{,}\overset{\mathfrak{e}_{6}}{6}\text{,}\overset{\mathfrak{e}_{7}%
}{7}\text{,}\overset{\mathfrak{e}_{7}}{8}\text{,}\overset{\mathfrak{e}_{8}%
}{9}\text{,}\overset{\mathfrak{e}_{8}}{10}\text{,}\overset{\mathfrak{e}%
_{8}}{11}\text{,}\overset{\mathfrak{e}_{8}}{12}\\
\text{two curves}  &  \text{: }\overset{\mathfrak{su}_{2}}{2}\text{
}\overset{\mathfrak{g}_{2}}{3}\\
\text{three curves}  &  \text{: }2\text{ }\overset{\mathfrak{sp}_{1}}{2}\text{
}\overset{\mathfrak{g}_{2}}{3}\text{,}\overset{\mathfrak{su}_{2}}{2}\text{
}\overset{\mathfrak{so}_{7}}{3}\text{ }\overset{\mathfrak{su}_{2}}{2}.
\end{align}
In some cases, there are also matter fields localized at various points of
these curves. This
occurs, for example, for a half hypermultiplet in the $\mathbf{56}$ of an $\mathfrak{e}_7$ gauge algebra supported
on a $-7$ curve, and also occurs for a half hypermultiplet in the $\mathbf{2}$ of an $\mathfrak{su}_2$ gauge
algebra supported on the $-2$ curve of the non-Higgsable cluster $2,3$. When the fiber type is minimal, we shall leave
these matter fields implicit.

For non-minimal fiber enhancements, we indicate the corresponding matter fields which arise from a collision of the
compact curve supporting a gauge algebra, and a non-compact component of the discriminant locus. We use the notation $[N_f = n]$ and
$[N_s = n]$ to indicate $n$ hypermultiplets respectively in the fundamental representation or spinor representation (as can happen for the
$\mathfrak{so}$-type gauge algebras). Note that when the representation is pseudo-real, $n$ can be a half-integer. We shall
also use the notation $[G]$ to indicate a corresponding non-abelian flavor symmetry which is localized in the geometry.\footnote{Here we do not distinguish between the algebra and the global structure of the flavor symmetry group.}

One of the hallmarks of 6D SCFTs is the generalization of the conventional notion of hypermultiplets to ``conformal matter.''
An example of conformal matter comes from the geometry:
\begin{equation}
y^{2}=x^{3}+u^{5}v^{5}.
\end{equation}
At the intersection point, the order of vanishing for $f$ and $g$ becomes too
singular, and blowups in the base are required.
Let us list the minimal conformal matter for
the collision of two ADE singularities which are the same
\cite{Heckman:2013pva, DelZotto:2014hpa, Heckman:2014qba, Morrison:2012np, Bershadsky:1996nu}:
\begin{align}
&  \lbrack E_{8}]1,2,2,3,1,5,1,3,2,2,1[E_{8}]\\
&  \lbrack E_{7}]1,2,3,2,1[E_{7}]\\
&  \lbrack E_{6}]1,3,1[E_{6}]\\
&  \lbrack SO_{2n}]\overset{\mathfrak{sp}_{n}}{1}[SO_{2n}]\\
&  \lbrack SU_{n}]\cdot\lbrack SU_{n}].
\end{align}
In the case of the collision of D-type symmetry algebras, there are also half
hypermultiplets localized at the $\mathfrak{so}/\mathfrak{sp}$ intersections,
and in the case of the A-type symmetry algebras, we have a conventional
hypermultiplet in the bifundamental representation.

Given this conformal matter, we can then proceed to gauge these flavor
symmetries to produce longer generalized quivers. Assuming that the flavor
symmetries are identical, we can then label these theories according to the
number of gauge groups $(k-1)$:%
\begin{equation}
[G_{0}]-G_{1}-....-G_{k-1}-[G_{k}],
\end{equation}
in the obvious notation. Implicit in the above description is the charge of the tensor multiplets
paired with each such gauge group factor. In F-theory, we write the partial tensor branch for this theory as:
\begin{equation}
\lbrack G]\overset{\mathfrak{g}}{2}...\overset{\mathfrak{g}}{2}[ G ],
\end{equation}
i.e. there are $(k-1)$ compact $-2$ curves, each with a singular fiber type giving a corresponding
gauge group of ADE type, and on the left and the right we have a flavor symmetry supported on a non-compact curve.
This is a partial tensor branch because at the collision of two components of the discriminant locus, the elliptic fiber ceases to be
in Kodaira-Tate form. Indeed, such a collision point is where the conformal matter of the system is localized.
Performing the minimal required number of blowups in the base to reach a model where all fibers
remain in Kodaira-Tate form, we get the tensor branch of the associated conformal matter.
Let us note that this class of theories also has a
straightforward realization in M-theory via $k$ spacetime filling M5-branes
probing the transverse geometry $\mathbb{R}_{\bot}\times\mathbb{C}^{2}%
/\Gamma_{G}$, where $\Gamma_{G}$ is a discrete ADE\ subgroup of $SU(2)$. In
that context, the conformal matter is associated with localized
\textquotedblleft edge modes\textquotedblright\ which are trapped on the M5-brane.

Starting from such a configuration, we can also consider various boundary
conditions for our configuration. In the context of intersecting seven-branes,
these vacua are dictated by the Hitchin system associated with the $G_{0}$ and
$G_{k}$ flavor branes. In particular, the collision point between $G_{0}$ and
$G_{1}$ and that between $G_{k-1}$ and $G_{k}$ allows us to add an additional
source term for the Higgs field at these punctures:%
\begin{equation}\label{HitchinSource}
\overline{\partial}\Phi_{0}=\mu_{0,1}^{(L)}\text{ }\delta_{G_{0}\cap G_{1}%
}\text{ \ \ and \ \ }\overline{\partial}\Phi_{k}=\mu_{k-1,k}^{(R)}\text{
}\delta_{G_{k-1}\cap G_{k}},
\end{equation}
where the $\delta$'s denote $(1,1)$-form delta functions with support at the
collision of the two seven-branes. Here, $\mu_{0,1}^{(L)}$ and $\mu
_{k-1,k}^{(R)}$ are elements in the complexifications of the Lie algebras
$\mathfrak{g}_{0}$ and $\mathfrak{g}_{k}$. The additional subscripts indicate
that these elements are localized at the intersection point of two
seven-branes. Moreover, the superscript serves to remind us that the source is
really an element of the left or right symmetry algebra. In what follows, we
shall often refer to these nilpotent elements as $\mu_{L}$ and $\mu_{R}$ in
the obvious notation.

When the collision corresponds to ordinary localized matter,
there is an interpretation in terms of the vevs of these matter fields
\cite{Anderson:2013rka, Beasley:2008dc}. More generalized source terms localized at a point
correspond to vevs for conformal matter \cite{DelZotto:2014hpa, Heckman:2014qba}.
Let us also note that similar considerations apply for the boundary conditions of 7D super Yang
Mills-theory, and so can also be phrased in M-theory as well.

In principle, there can also be more singular source terms on the righthand sides of
line (\ref{HitchinSource}). Such higher order singularities translate in turn into
higher (i.e. degree 2 or more) order poles for the Higgs field at a given puncture. Far
from the marked point, these singularities are subleading contributions to the boundary data
of the intersecting seven-brane configuration so we expect that the effects of possible breaking patterns
(as captured by the residue of the Higgs field simple pole) will suffice to parameterize possible RG flows.

Proceeding in this way, we see that for each collision point, we get two such
source terms, which we can denote by $\left(  \mu_{i,i+1}^{(L)},\mu
_{i,i+1}^{(R)}\right)  $. On general grounds, we expect that the possible
flows generated by conformal matter vevs are specified by a sequence of such
pairs. Even so, these sequences are rather rigid, and in many cases simply
stating $\mu_{0,1}^{(L)}$ and $\mu_{k-1,k}^{(R)}$ is typically enough to specify
the flow.\footnote{Note that when we initiate a larger breaking pattern from say
$[E_8]-E_8-....-E_8-[E_8]$ to $[E_7]-E_7-....-E_7-[E_7]$, the effects of the breaking pattern are not localized
and propagate from one end of the generalized quiver to the other. We leave a detailed
analysis of such flows for future work.} In this case, the invariant data is really given by the conjugacy
class of the element in the flavor symmetry algebra, i.e.~the orbit of the
element inside the complexified Lie algebra \cite{DelZotto:2014hpa, Gaiotto:2014lca}.

In the context of theories with weakly coupled
hypermultiplets, the fact that neighboring hypermultiplet vevs are coupled
together through D-term constraints means that specifying one set of vevs will
typically propagate out to additional vev constraints for matter on
neighboring quiver nodes. Part of our aim in this note will be to determine
what sorts of constraints are imposed by just the leftmost element of such a
sequence. Given a sufficiently long generalized quiver gauge theory, the particular
elements $\mu_{L}$ and $\mu_{R}$ can be chosen independently from one another
\cite{DelZotto:2014hpa}.
For this reason, we shall often reference the flow for a theory by only
listing the leftmost quiver nodes:%
\begin{equation}
\lbrack G_{0}]-G_{1}-G_{2}-....
\end{equation}

Now, although the M-theory realization is simplest in the case where the left
flavor symmetry is of ADE-type, there is no issue in the F-theory realization
with performing a partial tensor branch flow to reach more general flavor
symmetries of BCFG-type. Indeed, to reach such configurations we can
simply consider the corresponding non-compact seven-brane with this symmetry.
From the perspective of the conformal field theory, we can reach these cases
by starting with a theory with ADE\ flavor symmetry and flowing through a
combination of Higgs and tensor branch flows. We shall therefore view these
flavor symmetries on an equal footing with their simply laced cousins.

What then are the available choices for our boundary data $\mu\in
\mathfrak{g}_{\mathbb{C}}$? It is helpful at this point to recall that any
element of a simple Lie algebra can be decomposed into a semi-simple and
nilpotent part:%
\begin{equation}
\mu=\mu_{s}+\mu_{n},
\end{equation}
so that for any representation of $\mathfrak{g}_{\mathbb{C}}$, the image of
$\mu_{s}$ is a diagonalizable matrix, and $\mu_{n}$ is nilpotent.
Geometrically, the contribution from the semi-simple elements is
described by an unfolding which is directly visible in the complex geometry.

Less straightforward is the contribution from the nilpotent elements. Indeed,
such \textquotedblleft T-brane\textquotedblright\ contributions (so-named
because they often look like upper triangular matrices) have a degenerate
spectral equation, and as such do not appear directly in the deformations of the
complex geometry. Rather, they appear in the limiting behavior of
deformations associated with the Weil intermediate Jacobian of the
Calabi--Yau threefold and its fibration over the complex structure moduli of the threefold
\cite{Anderson:2013rka}. For flows between SCFTs, however, the
key point is that all we really need to keep track of is the relevant hierarchies of scales induced by such flows.
This is where the hyperkahler nature of the Higgs branch moduli space, and in particular its geometric avatar becomes
quite helpful. We recall from \cite{Anderson:2013rka} that there is a direct match between the geometric
realization of the Higgs branch moduli space of the seven-brane gauge theory in terms of the fibration of the Weil intermediate Jacobian of the Calabi-Yau threefold over the complex structure moduli. In this picture, the base of the Hitchin moduli space
is captured by complex structure deformations. Provided we start at a smooth point of the geometric moduli space, we can interpret this in the associated Hitchin system as a diagonalizable Higgs field vev. As we approach singular points in the geometric moduli space, we can thus reach
T-brane configurations. From the geometric perspective, however, this leads to the \textit{same} endpoint for an RG flow, so we can
either label the resulting endpoint of the flow by a nilpotent orbit of the flavor symmetry group or by an explicit F-theory geometry. Said
differently, T-brane vacua do not lead to non-geometric phases for 6D SCFTs \cite{Heckman:2015bfa}.
One of our aims will be to determine the \textit{explicit}
Calabi--Yau geometry for the F-theory SCFT associated with a given nilpotent orbit.\footnote{A
related class of explicit F-theory models classified by group theoretic data
was studied in references \cite{DelZotto:2014hpa, Heckman:2015bfa}. Though this data is purely geometric on the F-theory side,
in the dual heterotic description, we have
small instantons of heterotic string theory on an ADE singularity in which the boundary data of the small instantons leads
to different classes of 6D SCFTs. This boundary data is classified by homomorphisms
from discrete ADE subgroups of $SU(2)$ to $E_8$.}

In general, given a nilpotent element $\mu \in \mathfrak{g}_{\mathbb{C}}$ a semisimple Lie algebra,
the Jacobson--Morozov theorem tells us that there is a corresponding homomorphism
\begin{equation} \label{JacMor}
\rho:\mathfrak{sl}(2,\mathbb{C})\rightarrow\mathfrak{g}%
_{\mathbb{C}}%
\end{equation}
where the nilpotent element $\mu$ defines a raising operator in the image. The commutant subalgebra of $\im(\rho)$ in $\mathfrak{g}_{\mathbb{C}}$ then tells us the unbroken flavor symmetry for this conformal matter vev.
Though a microscopic characterization of conformal matter is still an outstanding open question, we can therefore expect that
an analysis of symmetry breaking patterns can be deduced using this purely algebraic characterization.
Indeed, more ambitiously, one might expect that once the analogue of F- and D-term constraints have been determined for conformal matter,
we can use such conformal matter vevs as a pragmatic way to extend the characterization of bound states of perturbative branes in terms of
such breaking patterns. From this perspective one can view the analysis of the present paper as determining these constraints for a
particular class of operator vevs.

One of the things we would like to determine are properties of
the IR fixed point associated with a given nilpotent orbit. For example, we would like to know both the characterization
on the tensor branch, as well as possible flavor symmetries of the system.
As explained in reference \cite{Heckman:2015ola}, a flow from a UV SCFT to an IR SCFT in F-theory is given by some combination of K\"ahler and complex structure deformations. In all the flows, we will indeed be able to track the rank of the gauge groups, as well as the total number of tensor multiplets for each proposed IR theory. The decrease in the rank of gauge groups (on the tensor branch) translates to a less singular elliptic fiber, and is a strong indication of a complex structure deformation. So, to verify that we have indeed realized a flow, it will suffice to provide an explicit match between a given nilpotent orbit and a corresponding F-theory geometry where the tensor branch of the SCFT is given by
a smaller number of tensor multiplets and a smaller gauge group.

With this in mind, our plan in much of this note will be to focus on the flows induced
by nilpotent elements, i.e.~T-branes, and to determine the endpoints of these
flows. An added benefit of this analysis will be that by tracking the
commutant subalgebra of the parent flavor symmetry, we will arrive at a
proposal for the unbroken flavor symmetry for these theories.

\subsection{E-String Flows} \label{ssec:estring}

As we have already mentioned, one of the important structural features of
6D\ SCFTs is that on their tensor branch, they are built up via a gluing
construction using the E-string theory. As one might expect, the RG\ flows
associated with this building block will therefore be important in our more
general discussion of flows induced by conformal matter vevs.

With this in mind, let us recall a few additional features of this theory.
Recall that in M-theory, the rank $k$ E-string theory is given by $k$
M5-branes probing an $E_{8}$ Ho\v{r}ava--Witten nine-brane \cite{Horava:1995qa, Horava:1996ma}.
In F-theory, it is realized on the tensor branch by a collection of curves in the base:%
\begin{equation}
\text{E-string theory base: }[E_{8}]\underset{k}{\text{ }%
\underbrace{1,2,...,2}}.
\end{equation}
We reach the 6D\ SCFT by collapsing all of these curves to zero size. Now,
provided $k<12$, we also get an SCFT by gauging this $E_{8}$ flavor symmetry.
This gauge group is supported on a $-12$ curve:%
\begin{equation}
\overset{\mathfrak{e}_{8}}{(12)}\underset{k}{\text{ }\underbrace{1,2,...,2}}.
\end{equation}
Starting from the UV SCFT, we reach various IR fixed points by moving onto a partial Higgs branch.
These have the interpretation of moving onto the Higgs branch of the 6D\ SCFT. In the
heterotic picture, we can picture this as moving onto various branches of the
multi-instanton moduli space. For example, we can consider moving some of the
small instantons to a different point of the $-12$ curve. This complex
structure deformation amounts to partitioning the small instantons into
separate chains (after moving onto the tensor branch for the corresponding fixed point):
\begin{equation}
\overset{\mathfrak{e}_{8}}{(12)}\underset{k}{\text{ }\underbrace{1,2,...,2}%
}\rightarrow\text{ }\underset{l}{\underbrace{2,...,2,1}}\overset{\mathfrak{e}%
_{8}}{(12)}\underset{k-l}{\text{ }\underbrace{1,2,...,2}},\label{manure}%
\end{equation}
and as can be verified by an analysis of the corresponding anomaly
polynomials, this does indeed define an RG\ flow \cite{Heckman:2015ola}.
In equations, the deformation of the singular Weierstrass model for the UV theory to the less
singular IR theory is given by:%
\begin{equation}
y^{2}=x^{3}+u^{5}v^{k}\rightarrow x^{3}+u^{5}(v-v_{1})^{l}(v-v_{2})^{k-l},
\end{equation}
where $u=0$ denotes the $\mathfrak{e}_{8}$ locus, and $v=v_{1}$ and $v=v_{2}$
indicate the two marked points on $u=0$ where the small instantons touch this seven-brane.

We can also consider dissolving the instantons back into flux in the
$\mathfrak{e}_{8}$ gauge theory. Geometrically, this is described by a
sequence of blowdowns involving the $-1$ curve, which in turn increases the
self-intersection of its neighboring curves by $+1$. Moving to a generic point
of complex structure moduli then Higgses the $\mathfrak{e}_{8}$ down to a
lower gauge symmetry (on the tensor branch). For example, after combining four
small instantons we reach a $-8$ curve with an $\mathfrak{e}_{7}$ gauge
symmetry:%
\begin{equation}
\overset{\mathfrak{e}_{8}}{(12)}\underset{k}{\text{ }\underbrace{1,2,...,2}%
}\rightarrow\text{ }\overset{\mathfrak{e}_{7}}{(8)}\underset{k-4}{\text{
}\underbrace{1,2,...,2}}.
\end{equation}

An important feature of this class of deformations is that they are localized.
What this means is that when we encounter larger SCFT structures, the same set
of local deformations will naturally embed into more elaborate RG\ flows, and
can be naturally extended to small instanton tails attached to other curves of
self-intersection $-x$.

For example, in all cases other than the A-type symmetry algebras, we will
encounter examples of a blowdown of a $-1$ curve, and a corresponding complex
structure deformation. Additionally, in the case of the exceptional flavor
symmetries, we will sometimes have to consider \textquotedblleft small
instanton maneuvers\textquotedblright\ of the type given in line
(\ref{manure}):
\begin{equation}
...(x)\text{ }1,2,...\rightarrow...\underset{1}{(x)}\text{ }1,...,
\end{equation}
that is, we move one of the small instantons to a new location on the $-x$
curve. Doing this may in turn require further deformations, since now the
curve touching the $-1$ curve on the right is now closer to the $-x$ curve.

\section{Flows for Classical Flavor Symmetries \label{sec:CLASSICAL}}

As a warmup for our general analysis, in this section we consider the case of
RG\ flows parameterized by nilpotent orbits of the classical algebras of
$SU$-, $SO$- and $Sp$-type. Several aspects of nilpotent elements of the
classical algebras can be found in \cite{NILPbook}, and we shall also follow the
discussion found in \cite{Chacaltana:2012zy}.

There is a simple algebraic characterization of all nilpotent orbits of
$\mathfrak{sl}(N,\mathbb{C})$. First, note that given an $N\times N$
nilpotent matrix we can then decompose it (in a suitable basis) as a collection of nilpotent Jordan blocks
of size $\mu_{i}\times\mu_{i}$. Without loss of generality, we can organize
these from largest to smallest, i.e.~$\mu_{1}\geq...\geq\mu_{N}\geq0$, so we
also define a partition, i.e.~a choice of Young diagram. Note that we allow
for the possibility that some $\mu_{i}$ are zero. When this occurs, it simply
means that the partition has terminated earlier for some $l\leq N$. Similar
considerations also hold for the other classical algebras with a few
restrictions \cite{NILPbook}:%
\begin{align}
\mathfrak{so}  &  :\text{even multiplicity of each even }\mu_{i}\\
\mathfrak{sp}  &  :\text{even multiplicity of each odd }\mu_{i},
\end{align}
where we note that if all $\mu_{i}$ are even for $\mathfrak{so}(2N,\mathbb{C})$,
we get two nilpotent elements which are related to each other by a $\mathbb{Z}_2$ outer automorphism
of the algebra.

\begin{figure}[ptb]%
\centering
\includegraphics[
trim=2.485037in 3.193803in 2.501044in 2.267488in,
height=0.9003in,
width=3.6167in
]%
{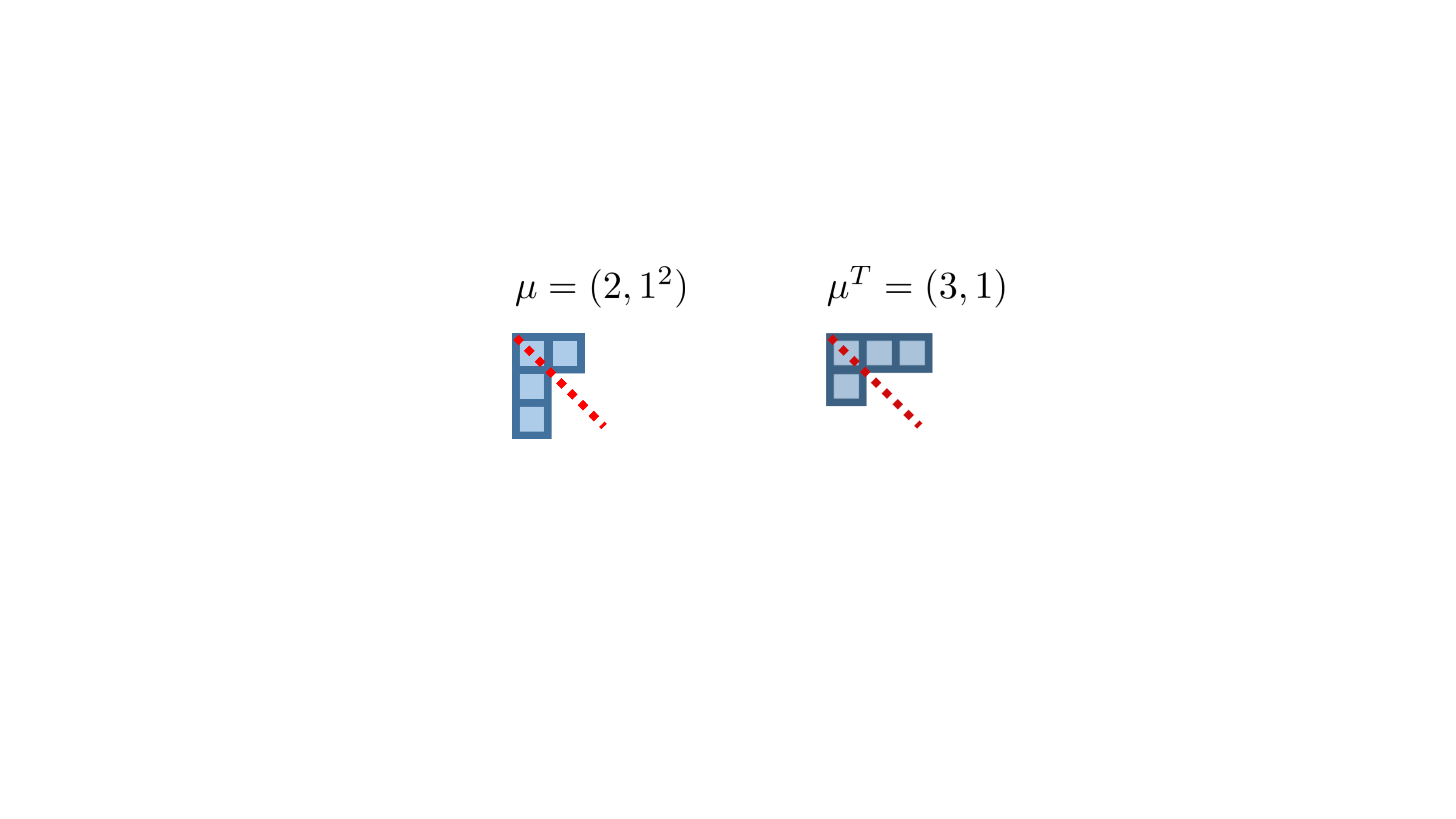}%
\caption{Example of a transposition of a partition.}%
\label{fig:transpose}%
\end{figure}

There is also a natural ordering of these partitions. Given partitions
$\mu=(\mu_{1},...,\mu_{N})$ and $\nu=(\nu_{1},...,\nu_{N})$, we say that:%
\begin{equation}\label{eq:order}
\mu\geq\nu\text{ \ \ if and only if \ \ }\underset{i=1}{\overset{k}{\sum}}%
\mu_{i}\geq\underset{i=1}{\overset{k}{\sum}}\nu_{i}\text{ for all \ \ }1\leq
k\leq N.
\end{equation}
There is a related ordering specified by taking the transpose of a given partition, i.e.~by reflecting a
Young diagram along a $45$ degree angle (see figure \ref{fig:transpose} for an example). The ordering
for the transposed partitions reverses the ordering of the original partitions, i.e.~we have $\mu > \nu $ if and only if
$\mu^T < \nu^T$. Finally, as a point of notation we shall often write a partition in the
shorthand $(\mu_{1}^{d_{1}},...,\mu_{l}^{d_{l}})$ to indicate that $\mu_i$ has multiplicity $d_i$.

As an example, see the first column of figure \ref{fig:su4} for an example of the ordering of partitions of $N=4$ according to (\ref{eq:order}). The diagrams are reverse ordered so that for $\mu < \nu$ (or equivalently for $\mu^T > \nu^T$), the partition $\mu$ appears higher up than $\nu$. Intuitively, if one takes a Young diagram and moves a box at the end of a row to a lower row, one obtains a ``smaller'' Young diagram. In the example of figure \ref{fig:su4}, the ordering is total (i.e.~any two diagrams can be compared). This ceases to be the case for larger $N$.

Given such a partition, we can also readily read off the \textquotedblleft
unbroken\textquotedblright\ symmetry, i.e.~the generators which will commute
with this choice of partition. As reviewed for example in \cite{Chacaltana:2012zy}, for a
partition $\mu$ where the entry $\mu_{i}$ has multiplicity $d_{i}$, these are:%
\begin{align}
\mathfrak{su}  &  :\mathfrak{g}_{\text{unbroken}}=\mathfrak{s}\left(
\underset{i}{\oplus}\mathfrak{u}(d_{i})\right) \label{suflavor}\\
\mathfrak{so}  &  :\mathfrak{g}_{\text{unbroken}}=\underset{i\text{
odd}}{\oplus}\mathfrak{so}\left(  d_{i}\right)  \oplus\underset{i\text{
even}}{\oplus}\mathfrak{sp}\left(  d_{i}/2\right) \label{eq:soflavor} \\
\mathfrak{sp}  &  :\mathfrak{g}_{\text{unbroken}}=\underset{i\text{
even}}{\oplus}\mathfrak{so}\left(  d_{i}\right)  \oplus\underset{i\text{
odd}}{\oplus}\mathfrak{sp}\left(  d_{i}/2\right) , \label{eq:spflavor}
\end{align}
where in the above ``$i$ odd'' or ``$i$ even'' is shorthand
for indicating that $\mu_i$ is odd or even, respectively.

Observe that in the case of the $\mathfrak{su}$-type flavor symmetries, there is an overall trace condition
on a collection of unitary algebras. This leads to a general expectation that such theories will have many $\mathfrak{u}(1)$
symmetry algebra factors. Similarly, for the $\mathfrak{so}$ and $\mathfrak{sp}$ algebras, we get
$\mathfrak{so}(2) \simeq \mathfrak{u}(1)$ factors when $d_i = 2$. Such symmetry factors can sometimes be
subtle to determine directly from the associated F-theory geometry, a point we return to later on in section \ref{sec:FLAVOR}.

For $\mathfrak{su}$ gauge groups, there is also a physical realization in terms of
IIA suspended brane configurations \cite{Hanany:1997gh, Gaiotto:2014lca}; we will return
to this picture in subsection \ref{sub:su}. For the $\mathfrak{so/sp}$-type
gauge algebras, which we will discuss in subsection \ref{sub:so}, a similar story involves
the use of O6 orientifold planes. In these cases,
the best we should in general hope for is that the nilpotent elements which
embed in a maximal $\mathfrak{su}(N)$ subalgebra can also be characterized in
terms of partitions of branes (and their images under the orientifold
projection). Indeed, we will see some striking examples where the
``na\"ive'' semi-classical intuition fails in a
rather spectacular way: Starting from a perturbative IIA\ configuration, we
will generate SCFT\ flows which land us on non-perturbatively realized SCFTs
i.e.~those in which the string coupling is order one!

The rest of this section is organized as follows. Mainly focusing on a broad
class of examples, we first explain for the $\mathfrak{su}$-type flavor
symmetries how hierarchies for nilpotent elements translate to corresponding
hierarchies for RG\ flows. We then turn to a similar analysis for the
$\mathfrak{so}_{\text{even}}$ flavor symmetries where we encounter our first
examples of flows involving conformal matter vevs. These cases are a strongly
coupled analogue of weakly coupled Higgsing, and we shall indeed see that
including these flows is necessary to maintain the expected correspondence
between nilpotent elements and RG\ flows. Finally, we turn to the cases of
$\mathfrak{so}_{\text{odd}}$ and $\mathfrak{sp}$-type flavor symmetry algebras.

\subsection{Flows from $\mathfrak{su}_{N}$}\label{sub:su}

As a first class of examples, we consider flows starting from the 6D\ SCFT
with tensor branch:
\begin{equation}\label{eq:m0}
\lbrack SU(N)]\overset{\mathfrak{su}_{N}}{2}...\overset{\mathfrak{su}_{N}%
}{2}[SU(N)],
\end{equation}
that is, we have colliding seven-branes with a hypermultiplet localized at each
point of intersection. One can Higgs each of the two $SU(N)$ flavor symmetries in a way parameterized by two partitions $\mu_L$, $\mu_R$ of $N$; this results in the SCFT ${\cal T}(SU(N),\mu_L, \mu_R,k)$, where
$(k-1)$ is the number of gauge groups in (\ref{eq:m0}).

These theories can be realized in terms of D6-branes suspended in between NS5-branes \cite{Hanany:1997gh, Gaiotto:2014lca}. At the very left and right, these D6-branes attach to D8-branes,
and the choice of boundary condition on each D8-brane is controlled by
\textquotedblleft Nahm pole data,\textquotedblright which in turn dictates the flavor symmetry for the
resulting 6D\ SCFT. \ These Nahm poles are boundary conditions for the Nahm equations living on the D6-brane worldvolume; they describe a ``fuzzy funnel," namely a fuzzy sphere configuration on the D6s which expands into a D8. This description is T-dual to the Hitchin pole description of section \ref{sec:CONFMATT}.

\begin{figure}[ptb]
\centering
\includegraphics[scale = 0.7]{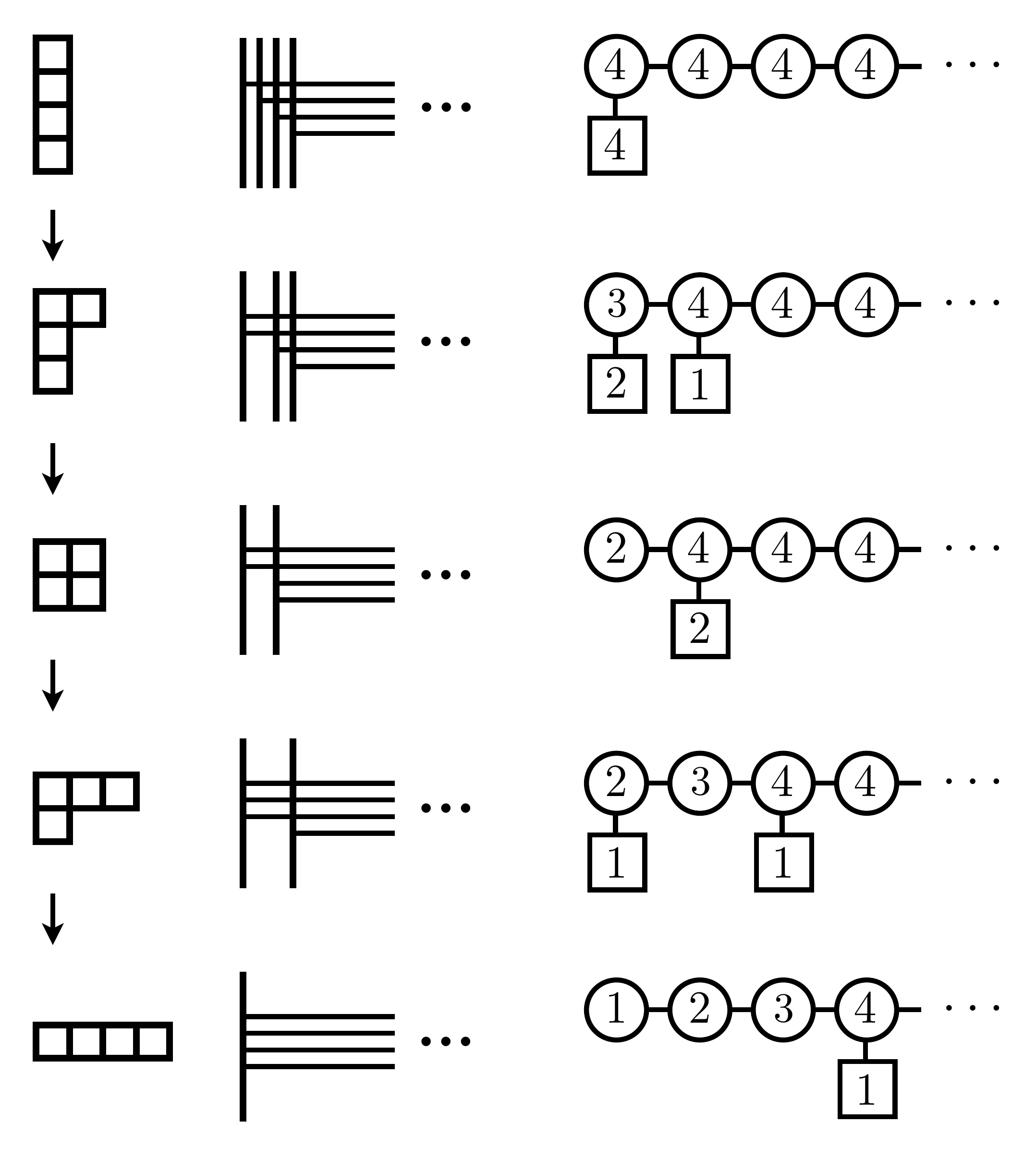}
\caption{Depiction of the IIA suspended brane configuration for a 6D SCFT with $\mathfrak{su}_4$  flavor symmetry. The partitioning of the branes
is specified by taking the transpose of the corresponding partition. In the figure, the vertical lines indicate D8-branes, and the horizontal lines denote D6-branes which attach on the left to D8-branes and on the right to NS5-branes.}%
\label{fig:su4}%
\end{figure}

As described in the introduction, we will at first consider theories where the number of gauge groups $(k-1)$ is sufficiently large enough so that the effect of Higgsing the left and right flavor groups are decoupled. (We will comment on the situation where that does not happen in section \ref{sec:SHORT}.) Given partitions $\mu_L$ and $\mu_R$ for the theory ${\cal T}(SU(N), \mu_L, \mu_R, k)$, there is a straightforward algorithm
for determining the associated suspended brane configuration \cite{DelZotto:2014hpa, Hanany:1997gh, Gaiotto:2014lca} (for a longer review, see also section 2 of \cite{Cremonesi:2015bld}). To illustrate, let us focus on the left partition $\mu_L=\mu$. Consider now the transposed Young diagram $\mu^{T}=(\mu_{1}^{T},...,\mu_{N}^{T})$. The gauge groups are now given by $SU(N_i(\mu))$, with
\begin{equation}\label{eq:Ni}
	\mu_{i}^{T} = N_i - N_{i-1},
\end{equation}
where $N_0 = 0$. The gauge group $SU(N_i)$ also has $f_i$ hypermultiplets in the fundamental representation.
Anomaly cancellation requires $2 N_i=N_{i-1}+N_{i+1}+f_i$. So in
fact the function $i\mapsto N_i$ is convex; moreover, the $f_i$ are equal to the jump in the slope of this function.
This accounts for the presence of the product flavor symmetry factors
in (\ref{suflavor}). See figure \ref{fig:su4} for a depiction of the suspended brane
configurations, associated partitions and quivers for the $N=4$ case.

Let us now verify that if we have a two partitions $\mu$ and $\nu$ such that
$\mu < \nu$, that there is then a corresponding RG flow between the theories,
i.e.~$\mathcal{T(\mu)}\rightarrow\mathcal{T(\nu)}$. For each choice of partition, we get a
sequence of gauge groups:
\begin{equation}
\left\{  N_{i}(\mu)\right\}_{i}\text{ \ \ and \ \ }\left\{  N_{i}(\nu)\right\}_{i}.
\end{equation}
From (\ref{eq:Ni}) we see that $N_i(\mu)= \sum_{j=1}^i \mu_j^T$. So the condition that $\mu < \nu$, or $\mu^T > \nu^T$, translates to a related condition on
the values of each of these ranks:%
\begin{equation}
N_{i}(\mu) \geq N_{i}(\nu).
\end{equation}
In some cases this condition is vacuously true since $N_{i}(\nu)$ may
be zero after initiating some breaking pattern. The resulting nilpotent
hierarchy therefore directly translates back to allowed RG\ flows for our system.  We also note that this correspondence between hierarchies and RG flows applies even for partitions of different sizes.  More precisely, for theories with a different number of
boxes in the respective Young diagrams, we first consider the transposed partition, and then use the partial ordering for these partitions. In other words, $\mu^{T} > \nu^{T}$ implies the existence of an RG flow between the corresponding theories even if $|\mu| \neq |\nu|$.

\subsection{Flows from $\mathfrak{so}_{\text{even}}$}\label{sub:so}

One of the significant simplifications in studying RG\ flows for the theories
with $\mathfrak{su}$-type flavor symmetries is that there is a direct match between
nilpotent orbits of the flavor symmetry and geometric maneuvers for the configuration of suspended branes.
This is mainly due to the fact that the resulting theories on the tensor branch have conventional matter fields.
In all other cases, we will inevitably need to include the effects of vevs for conformal matter.

As a first example of this type, we now turn to
examples where the flavor symmetry on one side of our generalized quiver
theory is an $\mathfrak{so}_{\text{even}}$-type flavor symmetry. One way to
engineer these examples is to consider the case of a stack of M5-branes
probing a D-type orbifold singularity. In the F-theory realization, we then
get our UV theory on the tensor branch:%
\begin{equation}
\lbrack SO(2N)]\overset{\mathfrak{sp}_{N}}{1}\text{ }\overset{\mathfrak{so}%
_{2N}}{4}\text{ }\overset{\mathfrak{sp}_{N}}{1}...\overset{\mathfrak{so}%
_{2N}}{4}\text{ }\overset{\mathfrak{sp}_{N}}{1}[SO(2N)].
\end{equation}
We shall primarily focus on the effects of nilpotent flows associated with
just one flavor symmetry factor, so we will typically assume a sufficiently
large number of tensor multiplets are present to make such genericity assumptions.

\begin{figure}
\centering
\begin{subfigure}[b]{1.0\textwidth}
       \includegraphics[scale = 0.75]{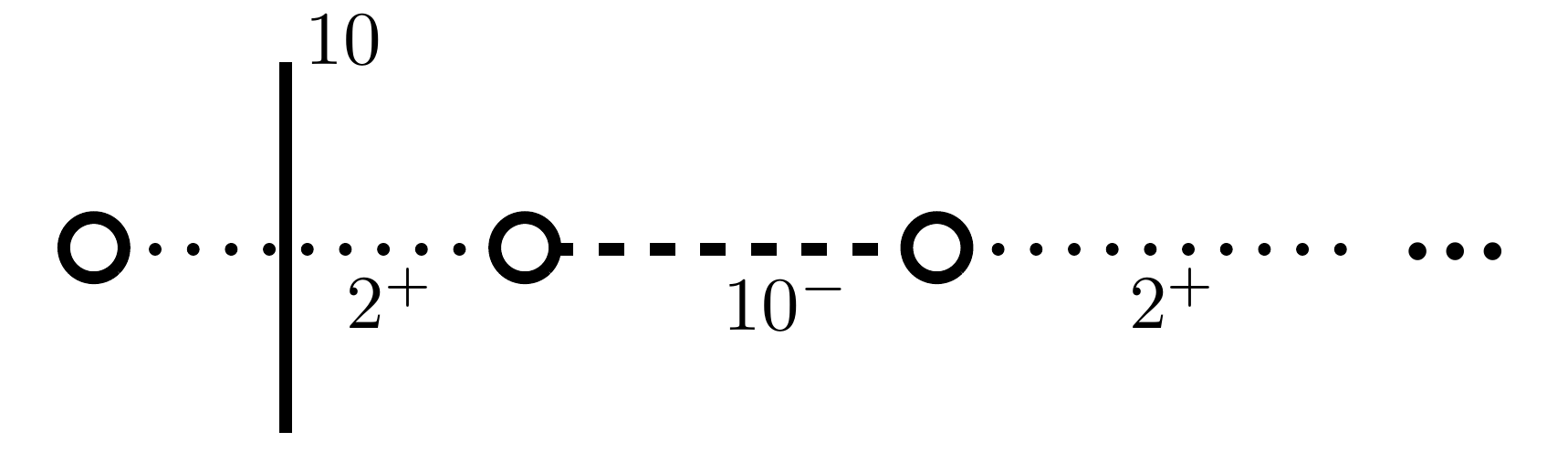}
        \label{fig:so10a}
    \end{subfigure}
 \begin{subfigure}[b]{1.0\textwidth}
      \includegraphics[scale = 0.75]{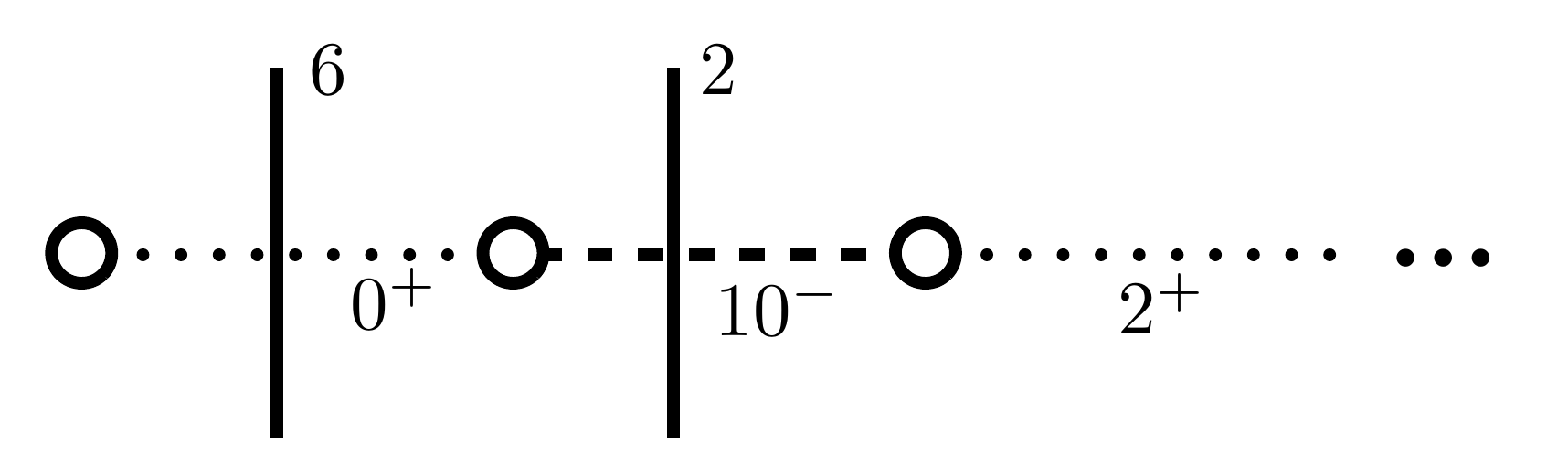}
        \label{fig:so10b}
    \end{subfigure}
 \begin{subfigure}[b]{1.0\textwidth}
     \includegraphics[scale = 0.75]{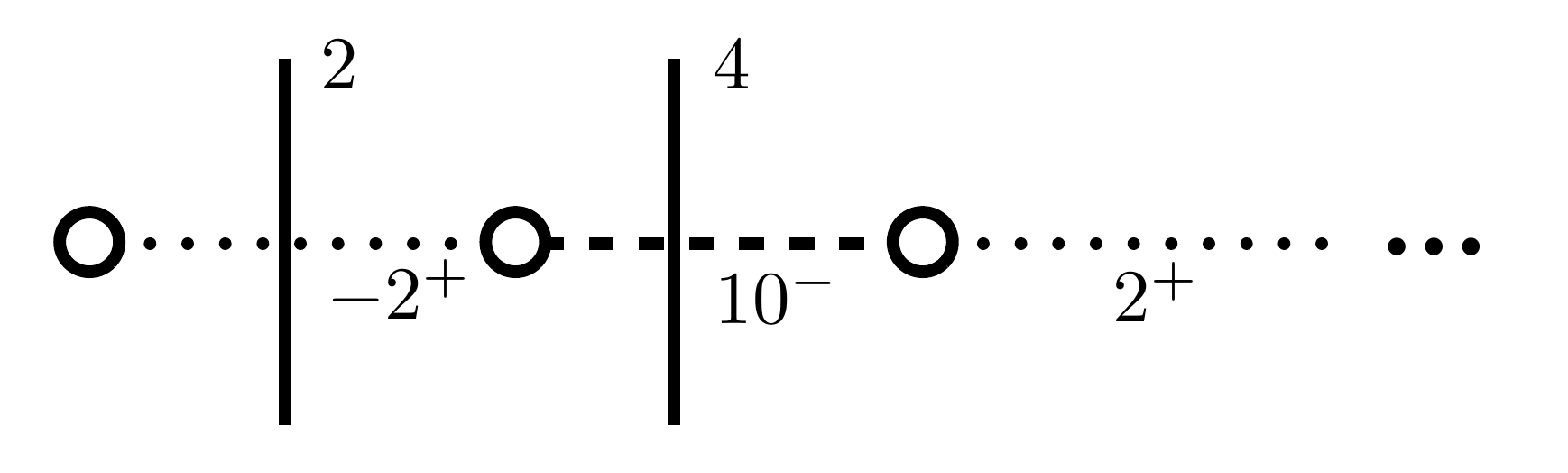}
        \label{fig:so10c}
    \end{subfigure}
\caption{IIA realizations of $SO(10)$ nilpotent orbits. TOP: $(1^{10})$, MIDDLE: $(2^2,1^6)$, BOTTOM: $(2^4,1^2)$.  Vertical lines indicate the presence of D8-branes, horizontal lines indicate D6-branes, and $\times$'s indicate NS5-branes.  The numbers of D6's and D8's are displayed.  A $^+$ superscript indicates an $O6^+$ while a $^-$ indicates an $O6^-$.    The $(2^4, 1^2)$ theory formally requires a negative number of D6-branes, indicating a breakdown of the IIA description and the presence of $Spin(10)$ spinor representations.}
\label{SO10brane}
\end{figure}

Because this is still a classical algebra, all of the nilpotent orbits are
labeled by a suitable partition of $2N$, but where each even entry occurs
with even multiplicity. Additionally, there is clearly a partial
ordering of these partitions. However, in this case we can expect the breaking
patterns to be more involved in part because now, we can also give vevs to
conformal matter. Indeed, we shall present examples where matter in a spinor
representation inevitably makes an appearance. In the IIA setup, these \textquotedblleft oddities\textquotedblright\
formally require the presence of a negative number of branes in a suspended
brane configuration, as shown in Figure \ref{SO10brane}. In such cases, we must instead pass to the F-theory realization of these models.

We shall primarily focus on some illustrative examples. Figure \ref{fig:so8flows} summarizes RG flows among theories ${\cal T}(SO(8),\mu_L,\mu_R,k)$ where we vary $\mu_L$ and for simplicity hold fixed $\mu_R=(1^{8})$. In figures \ref{fig:SO10} and \ref{fig:SO12} we show similar diagrams for ${\cal T}(SO(10),\mu_L,1^{10},k)$ and ${\cal T}(SO(12),\mu_L,1^{12},k)$, respectively. As already noted in section \ref{sec:CONFMATT}, we omit the flavors which are implicit for theories with minimal fiber types, i.e. those which arise on non-Higgsable clusters (e.g. $2,3$ and $7$).

All of the flows we consider are associated with motion on the Higgs branch. So, even though these flows
are parameterized by nilpotent orbits (i.e. T-branes), the hyperkahler structure of the Higgs branch
ensures that we can also understand these flows in terms of a complex structure deformation \cite{Heckman:2015bfa}.
It is easiest to exhibit them after shrinking some $-1$ curves present on the tensor branch.
For example, in the first flow of figure \ref{fig:SO10}, one can shrink the leftmost $-1$ curve on both the $(1^{10})$ and $(2^2,1^6)$ configuration. The complex deformation from $(1^{10})$ to  $(2^2, 1^6)$ is actually
realized by a two-parameter family of deformations which in Tate form (see e.g. \cite{Bershadsky:1996nh})
is given by the Weierstrass model:
\begin{equation}
y^2 + (u + \epsilon_1) v xy + (uv)^2 y = x^3 + (uv) x^2 + (u + \epsilon_2) (u^2 v^3) x + (uv)^4,
\end{equation}
so that when $\epsilon_1 , \epsilon_2 = 0$, we realize the original $(1^{10})$ configuration, while
for $\epsilon_{1}, \epsilon_2 \neq 0$, we have an $\mathfrak{su}_4 \simeq \mathfrak{so}_6$ flavor symmetry
localized along $u = 0$, with an $\mathfrak{so}_{10}$ localized along $v = 0$. The
appearance of the two unfolding parameters is instructive and
illustrates that specifying a T-brane configuration imposes further restrictions on
the allowed deformations. Indeed, although there is no
nilpotent generator which breaks $\mathfrak{so}_{10}$
to either $\mathfrak{so}_8$ or $\mathfrak{su}_5$, there are
of course semisimple generators which do.

In principle, one could proceed in this way for all the flows in figures \ref{fig:so8flows}, \ref{fig:SO10}, and \ref{fig:SO12}.
In practice, however, one can speed up the computation by using information coming from field theory, from anomaly cancellation, and from the known properties of the E-string theory. The possible gauge algebras on a curve, depending on its self-intersection number, are listed for example in \cite[Pages 45--46]{Heckman:2015bfa}. The expected representations of the matter fields, and the corresponding flavor group symmetries acting on them, are listed in \cite[Table 5.1]{Bertolini:2015bwa}. For example, we
sometimes encounter an $\mathfrak{so}$-type gauge theory on a
$-4$, $-3$, $-2$ and $-1$ curve. Anomaly cancellation uniquely fixes the
spectrum of hypermultiplets transforming in a non-trivial representation of
the gauge symmetry algebra. For a $-4$ curve with $\mathfrak{so}$-type gauge algebra,
all matter transforms in the fundamental representation. For the last three cases, there are always spinor representations, and the number of spinors is $16/d_{s}$, $32/d_{s}$ and
$48/d_{s}$ respectively, where $d_{s}$ is the dimension of the irreducible
spinor representation of this algebra. Note that this also places an upper
bound on the rank of the gauge groups, i.e.~the maximal rank $\mathfrak{so}%
$-type algebra for a system with spinors is in these cases respectively
$\mathfrak{so}(12)$, $\mathfrak{so}(13)$ and $\mathfrak{so}(12)$.

Finally, one should keep in mind that the E-string living on an empty $-1$ curve has an $E_8$ flavor symmetry; thus, when we gauge a product
subalgebra, we necessarily have $\mathfrak{g}_1 \times \mathfrak{g}_2 \subset \mathfrak{e}_8$ (see \cite{Heckman:2013pva}).
If this subalgebra is not maximal, we also expect there to be a residual flavor symmetry given by the commutant subalgebra.

Let us now turn to some examples. The first flow of figure \ref{fig:SO10} simply corresponds to giving a vev to a fundamental hypermultiplet
for the leftmost $\mathfrak{sp}_1$ gauge algebra, breaking it completely. One ends up with an E-string, and $\mathfrak{so}_6 \oplus \mathfrak{so}_{10}\subset \mathfrak{so}_{16}$ is indeed a subalgebra of $\mathfrak{e}_8$. The gauge algebra $\mathfrak{so}_{10}$  on the leftmost $-4$ curve should still have 2 fundamental hypermultiplets; given the presence of the $\mathfrak{sp}_1$ on the right, we deduce the presence of a ``side link'' of conformal matter with
flavor symmetry $SU(2)$. In the next step, we can give a vev to this side link, which breaks $\mathfrak{so}_{10}\to \mathfrak{so}_9$ and leads to $(3,1^7)$; or alternatively we can shrink the empty $-1$ curve. In this second case, the $-4$ curve becomes a $-3$ curve, and now the $\mathfrak{so}_{10}$ should support three fundamental hypers; again, given the presence of the $\mathfrak{sp}_1$ on the right, we deduce an $\mathfrak{sp}_2$ flavor symmetry. We can now iterate the process until no further Higgsing is possible; this leads to figure \ref{fig:SO10}.

The diagram precisely corresponds with the ordering of partitions,
in agreement with line (\ref{eq:RG}). That we achieve a perfect match between the hierarchies of nilpotent
elements and a corresponding hierarchy of RG flows again provides strong
evidence for our proposed picture of RG\ flows induced by conformal matter vevs.

As an examples of a Higgsing operation, consider the SCFT with tensor branch:
\begin{equation}
[SO(12)] \,\, \overset{\mathfrak{sp}_{2}}{1} \,\, \overset{\mathfrak{so}_{12}}{4}\,\,\overset{\mathfrak{sp}_{2}}{1} \,\, ....
\end{equation}
We can flow to another SCFT in the IR by activating a vev for a
fundamental hypermultiplet of the leftmost $\mathfrak{sp}_2$ gauge algebra. The
resulting tensor branch for this IR SCFT is then:
\begin{equation}
[SO(8)] \,\, \overset{\mathfrak{sp}_{1}}{1} \,\, \underset{[Sp(1)]}{\overset{\mathfrak{so}_{12}}{4}}\,\,\overset{\mathfrak{sp}_{2}}{1} \,\, ....
\end{equation}
The hypermultiplet in the bifundamental representation, i.e. the $\frac{1}{2}\bf(4,12)$ decomposes as $\frac{1}{2}\textbf{(2,12)}\oplus\textbf{(1,12)}$, yielding the single fundamental on the leftmost $\mathfrak{so}_{12}$ of the IR theory, which transforms under a global $Sp(1)$ symmetry.

We can also see that vevs of conformal matter can sometimes drive us away from a perturbative IIA realization of the tensor branch.
For example, by starting on the tensor branch, we can collapse the leftmost $-1$ curve of the configuration:
\begin{equation}
[SO(7)] \,\, {1}\,\,\overset{\mathfrak{so}_{9}}{4}\,\,\overset{\mathfrak{sp}_{1}}{1} \,\, ...
\end{equation}
so a vev for conformal matter can trigger a flow to the configuration with tensor branch:
\begin{equation}
[SU(2) \times SU(2)] \,\,\overset{\mathfrak{so}_{9}}{3}\,\,\overset{\mathfrak{sp}_{1}}{1} \,\, ....
\end{equation}
That is, collapsing the $-1$ curve converts the $-4$ curve to a $-3$ curve and the remnants of conformal matter not eaten by the Higgs mechanism
show up as matter in possibly ``exotic'' representations.  In this case, a spinor and a fundamental of $\mathfrak{so}_9$ appear on the -3 curve after blowdown. Note that at the SCFT point, we are always dealing with collapsed curves anyway, so we should properly view this as a complex
structure deformation.  Such deformations may also involve collapsing $-1$ curves located in the interior of the tensor branch quiver.  For instance, the bottom flow in figure \ref{fig:so8flows} corresponds to a blowdown of the leftmost $-1$ curve of the theory
\begin{equation}
{\overset{\mathfrak{su}_2}2} \, {\overset{\mathfrak{g_{2}}}3}  \,\, 1\,\, \overset{\mathfrak{so_{8}}}4  \,\,  1\,\, ...[SO(8)]
\end{equation}
and produces a theory with quiver
\begin{equation}
2 \,\, \overset{\mathfrak{su}_2}2 \,\, {\overset{\mathfrak{g_{2}}}3}  \,\, 1\,\, \overset{\mathfrak{so_{8}}}4  \,\,  1 \,\, ...[SO(8)]
\end{equation}
Note that the $-3$ curve of the UV theory has become the second $-2$ curve of the IR theory, and the leftmost $-4$ curve of the UV theory has become the $-3$ curve of the IR theory.

Additionally, recall that partitions of $2N$ with only even entries give rise to two distinct nilpotent orbits of $\mathfrak{so}(2N)$, which are related to each other by outer automorphism.  However, matching with the hierarchy of RG flows reveals that these distinct nilpotent elements do \emph{not} give rise to distinct 6D SCFTs.  Thus, we conclude that RG flows parametrized by nilpotent orbits related by an outer automorphism lead to physically equivalent IR fixed points.  This is illustrated most poignantly in the $\mathfrak{so}(8)$ case shown in figure \ref{fig:so8flows}: here, not only the two $(2^4)$ orbits but also the  $(3,1^5)$ partition are related by the triality outer automorphism (likewise for $(4^2)_I$, $(4^2)_{II}$ and $(5,1^3)$).\footnote{One way to see this triality is to note that the weighted Dynkin diagrams associated with these nilpotent orbits are related by permutation of the three external nodes \cite[Page 84]{NILPbook}.}  We see that in both cases, all three of these nilpotent orbits correspond to the same 6D SCFT.  In the $\mathfrak{so}(10)$ and $\mathfrak{so}(12)$ figures, we therefore display only a single theory for each partition.

We also observe that just as in the case of theories with an $\mathfrak{su}$-type
flavor symmetry, we can extend the nilpotent hierarchy to partitions with a
different number of boxes, i.e. by working in terms of the
transposed Young diagrams:
\begin{equation}
\mu^T > \nu^T \Rightarrow T(\mu) \rightarrow T(\nu).
\end{equation}
For instance, comparing the list of $SO(10)$ theories with the list of $SO(12)$ theories, we see that there is clearly a flow from the $(2^2,1^8) $ theory of $SO(12)$ to the $(1^{10})$ theory of $SO(10)$, as expected since $(2^2, 1^8)^T > (1^{10})^T$.  However, there is no flow that will take us from the $(2^4, 1^4)$ theory of $SO(12)$ to the $(1^{10})$ theory of $SO(10)$, and indeed $(2^4, 1^4)^T \ngtr (1^{10})^T$.

\begin{center}

\begin{figure}

\begin{tikzpicture}[node distance=2cm]

\node (1) [startstop] {
$
1^{8}: [SO(8)]\,\, 1 \,\, {\overset{\mathfrak{so_{8}}}4}  \,\, 1\,\, \overset{\mathfrak{so_{8}}}4  \,\,  1 \,\, ... [SO(8)]
$};

\node (2) [startstop, below of=1] {
$
2^2,1^4: [SU(2)\times SU(2) \times SU(2)] \,\, {\overset{\mathfrak{so_{8}}}3}  \,\, 1\,\, \overset{\mathfrak{so_{8}}}4  \,\,  1 \,\, ... [SO(8)]
$};

\node (3) [startstop, below of=2] {
$
3,1^5:  [Sp(2)] \,\, {\overset{\mathfrak{so_{7}}}3}  \,\, 1\,\, \overset{\mathfrak{so_{8}}}4  \,\,  1 \,\, ... [SO(8)]
$};

\node (4) [startstop, right of=3,xshift=4.3cm] {
$
2^4_{II}:  [Sp(2)] \,\, {\overset{\mathfrak{so_{7}}}3}  \,\, 1\,\, \overset{\mathfrak{so_{8}}}4  \,\,  1 \,\, ... [SO(8)]
$};

\node (4b) [startstop, right of=3,xshift=-8.3cm] {
$
2^4_{I}:  [Sp(2)] \,\, {\overset{\mathfrak{so_{7}}}3}  \,\, 1\,\, \overset{\mathfrak{so_{8}}}4  \,\,  1 \,\, ... [SO(8)]
$};

\node (5) [startstop, below of=3] {
$
3,2^2,1: [SU(2)] \,\, {\overset{\mathfrak{g_{2}}}3}  \,\, 1\,\, \overset{\mathfrak{so_{8}}}4  \,\,  1 \,\, ... [SO(8)]
$};

\node (6) [startstop, below of=5] {
$
3^2, 1^2:  {\overset{\mathfrak{su_{3}}}3}  \,\, 1\,\, \overset{\mathfrak{so_{8}}}4  \,\,  1 \,\, ... [SO(8)]
$};

\node (7) [startstop, below of=6] {
$
5, 1^3: {\overset{\mathfrak{su}_2}2} \,\, \underset{[SU(2)]}{\overset{\mathfrak{so_{7}}}3}  \,\, 1\,\, \overset{\mathfrak{so_{8}}}4  \,\,  1\,\,  ... [SO(8)]
$};

\node (7b) [startstop, right of=7, xshift=-8.3cm] {
$
4^2_{I}: {\overset{\mathfrak{su}_2}2} \, \underset{[SU(2)]}{\overset{\mathfrak{so_{7}}}3}  \,\, 1\,\, \overset{\mathfrak{so_{8}}}4  \,\,  1\,\,  ... [SO(8)]
$};

\node (8) [startstop, right of=7, xshift=4.3cm] {
$
4^2_{II}: {\overset{\mathfrak{su}_2}2} \, \underset{[SU(2)]}{\overset{\mathfrak{so_{7}}}3}  \,\, 1\,\, \overset{\mathfrak{so_{8}}}4  \,\,  1\,\,  ... [SO(8)]
$};

\node (9) [startstop, below of=7] {
$
5,3: {\overset{\mathfrak{su}_2}2} \, {\overset{\mathfrak{g_{2}}}3}  \,\, 1\,\, \overset{\mathfrak{so_{8}}}4  \,\,  1\,\, ...[SO(8)]
$};

\node (10) [startstop, below of=9] {
$
7,1: 2 \,\, \overset{\mathfrak{su}_2}2 \,\, {\overset{\mathfrak{g_{2}}}3}  \,\, 1\,\, \overset{\mathfrak{so_{8}}}4  \,\,  1 \,\, ...[SO(8)]
$};

\draw [arrow, color=blue] (1) -- (2);
\draw [arrow] (2) -- (3);
\draw [arrow] (2) -- (4);
\draw [arrow] (3) -- (5);
\draw [arrow] (2) -- (4b);
\draw [arrow] (4) -- (5);
\draw [arrow] (5) -- (6);
\draw [arrow] (4b) -- (5);
\draw [arrow, color=blue] (6) -- (7);
\draw [arrow, color=blue] (6) -- (7b);
\draw [arrow] (7) -- (9);
\draw [arrow] (7b) -- (9);
\draw [arrow, color=blue] (6) -- (8);
\draw [arrow] (8) -- (9);
\draw [arrow, color=blue] (9) -- (10);
\end{tikzpicture}

\caption{Flows for $SO(8)$ nilpotent orbits.  Blue arrows indicate flows where one or more free tensors appears in the IR.}
\label{fig:so8flows}
\end{figure}
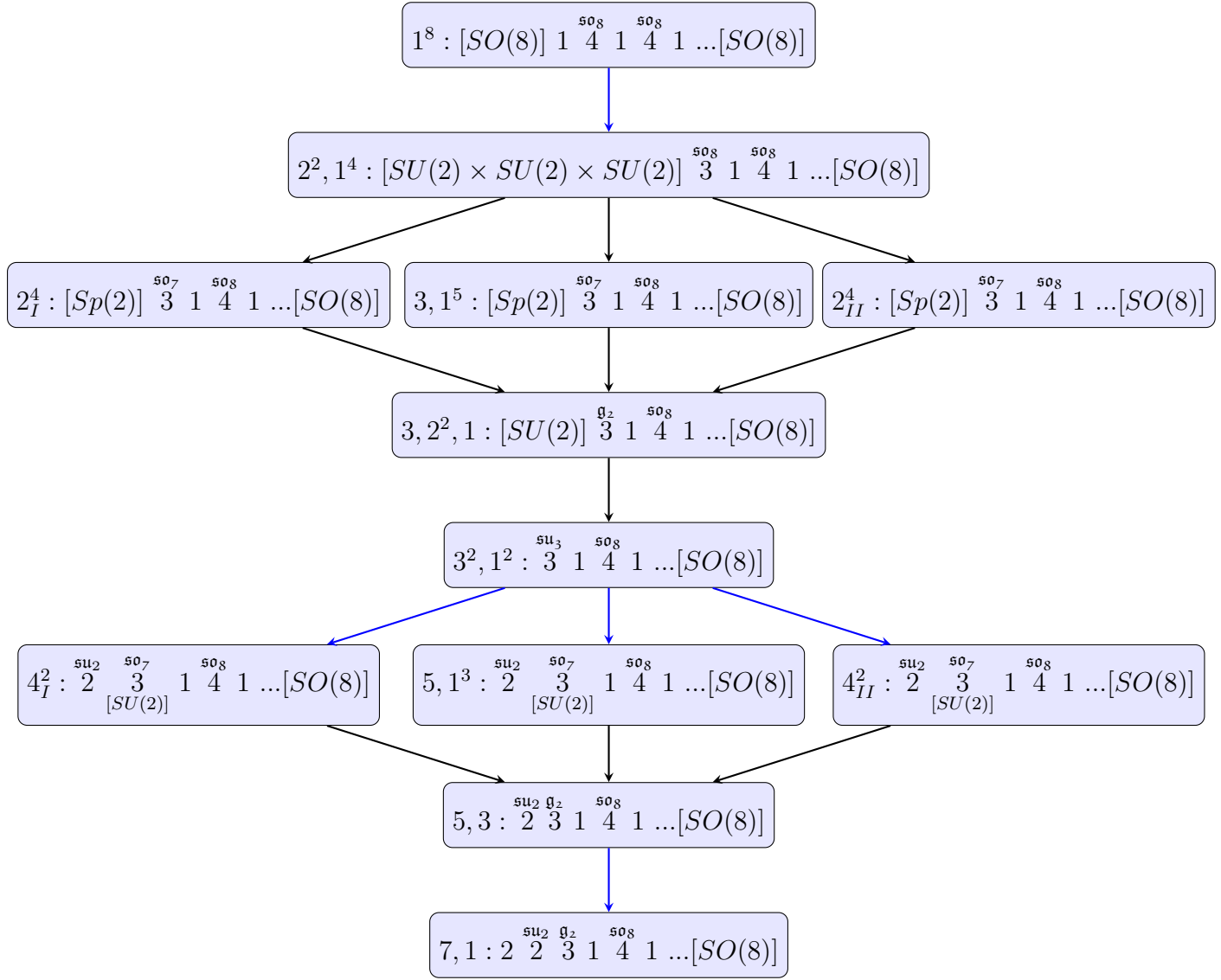

\begin{figure}
\begin{tikzpicture}[node distance=1.6cm]

\node (1) [startstop] {
$
1^{10}: [SO(10)]\,\,\overset{\mathfrak{sp_1}}1 \,\, {\overset{\mathfrak{so_{10}}}4}  \,\, \overset{\mathfrak{sp_1}}1 \,\, \overset{\mathfrak{so_{10}}}4  \,\,  \overset{\mathfrak{sp_1}}1  \,\, ...[SO(10)]
$};

\node (2) [startstop, below of=1] {
$
2^2,1^6: [SO(6)]\,\, 1 \,\,  \underset{[SU(2)]}{\overset{\mathfrak{so_{10}}}4} \,\, \overset{\mathfrak{sp_1}}1 \,\, \overset{\mathfrak{so_{10}}}4  \,\,  \overset{\mathfrak{sp_1}}1  \,\, ...[SO(10)]
$};

\node (3) [startstop, below of=2] {
$
3,1^7: [SO(7)]\,\, 1 \,\, \overset{\mathfrak{so_{9}}}4  \,\, \underset{[N_f=\frac{1}{2}]}{\overset{\mathfrak{sp_1}}1} \,\, \overset{\mathfrak{so_{10}}}4  \,\,  \overset{\mathfrak{sp_1}}1  \,\, ...[SO(10)]
$};

\node (4) [startstop, right of=3,xshift=7cm] {
$
2^4,1^2:[Sp(2)]\,\, \underset{[N_s=1]}{\overset{\mathfrak{so_{10}}}3}  \,\, \overset{\mathfrak{sp_1}}1 \,\, \overset{\mathfrak{so_{10}}}4  \,\,  \overset{\mathfrak{sp_1}}1  \,\, ...[SO(10)]
$};

\node (5) [startstop, below of=3] {
$
3,2^2,1^3: [SU(2) \times SU(2)]\,\, \overset{\mathfrak{so_{9}}}3  \,\, \underset{[N_f=\frac{1}{2}]}{\overset{\mathfrak{sp_1}}1} \,\, \overset{\mathfrak{so_{10}}}4  \,\,  \overset{\mathfrak{sp_1}}1  \,\, ...[SO(10)]
$};

\node (6) [startstop, below of=5] {
$
3^2,1^4:[SU(2) \times SU(2)]\,\,\overset{\mathfrak{so_{8}}}3  \,\,  \underset{[N_f=1]}{\overset{\mathfrak{sp_1}}1} \,\, \overset{\mathfrak{so_{10}}}4  \,\,  \overset{\mathfrak{sp_1}}1  \,\, ...[SO(10)]
$};

\node (7) [startstop, below of=6] {
$
3^2,2^2:[SU(2) ]\,\, \overset{\mathfrak{so_{7}}}3  \,\,  \underset{[N_f=1]}{\overset{\mathfrak{sp_1}}1} \,\, \overset{\mathfrak{so_{10}}}4  \,\,  \overset{\mathfrak{sp_1}}1  \,\, ...[SO(10)]
$};

\node (8) [startstop, below of=7] {
$
 3^3,1: \overset{\mathfrak{g_2}}3  \,\, \underset{[SU(2)]}{\overset{\mathfrak{sp_1}}1} \,\, \overset{\mathfrak{so_{10}}}4  \,\,  \overset{\mathfrak{sp_1}}1  \,\, ...[SO(10)]
$};

\node (9) [startstop, right of=7, xshift=6cm] {
$
5,1^5: [Sp(2) ]\,\, \overset{\mathfrak{so_7}}3  \,\, 1 \,\, \overset{\mathfrak{so_{9}}}4  \,\,   \underset{[N_f=\frac{1}{2}]}{\overset{\mathfrak{sp_1}}1}  \,\, ...[SO(10)]
$};

\node (10) [startstop, below of=8] {
$
4^2,1^2: \overset{\mathfrak{su_3}}3  \,\, 1 \,\, \underset{[SU(2)]}{\overset{\mathfrak{so_{10}}}4}  \,\,  \overset{\mathfrak{sp_1}}1  \,\, ...[SO(10)]
$};

\node (11) [startstop, right of=10, xshift=6cm] {
$
5,2^2,1: [SU(2) ] \,\, \overset{\mathfrak{g_2}}3  \,\, 1 \,\, \overset{\mathfrak{so_{9}}}4  \,\,   \underset{[N_f=\frac{1}{2}]}{\overset{\mathfrak{sp_1}}1}  \,\, ...[SO(10)]
$};

\node (12) [startstop, below of=10] {
$
5,3,1^2: \overset{\mathfrak{su_3}}3  \,\, 1 \,\, \overset{\mathfrak{so_{9}}}4  \,\,   \underset{[N_f=\frac{1}{2}]}{\overset{\mathfrak{sp_1}}1}  \,\, ...[SO(10)]
$};

\node (13) [startstop, below of=12] {
$
5^2: \overset{\mathfrak{su_2}}2  \,\, \overset{\mathfrak{so_{7}}}3  \,\,   \underset{[N_f=1]}{\overset{\mathfrak{sp_1}}1} \,\, \overset{\mathfrak{so_{10}}}4  \,\, ...[SO(10)]
$};

\node (14) [startstop, right of=13, xshift=5cm] {
$
7,1^3: \overset{\mathfrak{su_2}}2  \,\, \underset{[SU(2)]}{\overset{\mathfrak{so_{7}}}3}  \,\,  1 \,\, \overset{\mathfrak{so_{9}}}4  \,\, ...[SO(10)]
$};

\node (15) [startstop, below of=13] {
$
7,3: {\overset{\mathfrak{su_2}}2}  \,\, \overset{\mathfrak{g_2}}3  \,\,  1 \,\, \overset{\mathfrak{so_{9}}}4  \,\, ...[SO(10)]
$};

\node (16) [startstop, below of=15] {
$
9,1: 2 \,\, \overset{\mathfrak{su_2}}2  \,\, \overset{\mathfrak{g_2}}3  \,\,  1 \,\, \overset{\mathfrak{so_{9}}}4  \,\, ...[SO(10)]
$};

\draw [arrow] (1) -- (2);
\draw [arrow] (2) -- (3);
\draw [arrow, color=blue] (2) -- (4);
\draw [arrow, color=blue] (3) -- (5);
\draw [arrow] (4) -- (5);
\draw [arrow] (5) -- (6);
\draw [arrow] (6) -- (7);
\draw [arrow] (7) -- (8);
\draw [arrow] (6) -- (9);
\draw [arrow] (8) -- (10);
\draw [arrow] (8) -- (11);
\draw [arrow] (9) -- (11);
\draw [arrow] (10) -- (12);
\draw [arrow] (11) -- (12);
\draw [arrow, color=blue] (12) -- (13);
\draw [arrow, color=blue] (12) -- (14);
\draw [arrow] (13) -- (15);
\draw [arrow] (14) -- (15);
\draw [arrow, color=blue] (15) -- (16);

\end{tikzpicture}
\caption{\label{fig:SO10} Flows for $SO(10)$ nilpotent orbits.  Blue arrows indicate flows where one or more free tensors appears in the IR.}
\end{figure}

\end{center}

\begin{center}

\begin{lrbox}{\mysavebox}%

\begin{tikzpicture}[node distance=1.5cm]

\node (1) [startstop, yshift=-2cm] {
$
1^{12}: [SO(12)]\,\,\overset{\mathfrak{sp_2}}1 \,\, {\overset{\mathfrak{so_{12}}}4}  \,\, \overset{\mathfrak{sp_2}}1 \,\, \overset{\mathfrak{so_{12}}}4  \,\,  \overset{\mathfrak{sp_2}}1  \,\, ... [SO(12)]
$};

\node (2) [startstop, below of=1] {
$
2^2, 1^{8}: [SO(8)]\,\,\overset{\mathfrak{sp_1}}1 \,\, \underset{[SU(2)]}{\overset{\mathfrak{so_{12}}}4}  \,\, \overset{\mathfrak{sp_2}}1 \,\, \overset{\mathfrak{so_{12}}}4  \,\,  \overset{\mathfrak{sp_2}}1  \,\, ... [SO(12)]
$};

\node (3) [startstop, below of=2] {
$
2^4,1^4: [SO(4)]\,\, 1 \,\, \underset{[Sp(2)]}{\overset{\mathfrak{so_{12}}}4}  \,\, \overset{\mathfrak{sp_2}}1 \,\, \overset{\mathfrak{so_{12}}}4  \,\,  \overset{\mathfrak{sp_2}}1  \,\, ... [SO(12)]
$};

\node (4) [startstop, right of=3,xshift=8cm] {
$
3,1^9: [SO(9)]\,\, \overset{\mathfrak{sp}_1}1 \,\, {\overset{\mathfrak{so_{11}}}4}  \,\, \overset{\mathfrak{sp_2}}1 \,\, \overset{\mathfrak{so_{12}}}4  \,\,  \overset{\mathfrak{sp_2}}1  \,\, ... [SO(12)]
$};

\node (5) [startstop, below of=3] {
$
2^6: [Sp(3)]\,\,  {\overset{\mathfrak{so_{12}}}3}  \,\, \overset{\mathfrak{sp_2}}1 \,\, \overset{\mathfrak{so_{12}}}4  \,\,  \overset{\mathfrak{sp_2}}1  \,\, ... [SO(12)]
$};

\node (6) [startstop, right of=5, xshift=8cm] {
$
3,2^2,1^5: [SO(5)] \,\, 1\,\,  \underset{[Sp(1)]}{\overset{\mathfrak{so_{11}}}4}  \,\, \overset{\mathfrak{sp_2}}1 \,\, \overset{\mathfrak{so_{12}}}4  \,\,  \overset{\mathfrak{sp_2}}1  \,\, ... [SO(12)]
$};

\node (7) [startstop, below of=6] {
$
3^2,1^6: [SO(6)] \,\, 1\,\,  {\overset{\mathfrak{so_{10}}}4}  \,\, \overset{\mathfrak{sp_2}}1 \,\, \overset{\mathfrak{so_{12}}}4  \,\,  \overset{\mathfrak{sp_2}}1  \,\, ... [SO(12)]
$};

\node (8) [startstop, below of=5] {
$
3,2^4,1: [Sp(2)] \,\,   {\overset{\mathfrak{so_{11}}}3}  \,\, \overset{\mathfrak{sp_2}}1 \,\, \overset{\mathfrak{so_{12}}}4  \,\,  \overset{\mathfrak{sp_2}}1  \,\, ... [SO(12)]
$};

\node (9) [startstop, below of=8] {
$
3^2,2^2,1^2: [Sp(1)] \,\,   {\overset{\mathfrak{so_{10}}}3}  \,\, \overset{\mathfrak{sp_2}}1 \,\, \overset{\mathfrak{so_{12}}}4  \,\,  \overset{\mathfrak{sp_2}}1  \,\, ... [SO(12)]
$};

\node (10) [startstop, below of=9] {
$
3^3,1^3: [Sp(1)] \,\,   {\overset{\mathfrak{so_{9}}}3}  \,\, \underset{[SO(3)]}{\overset{\mathfrak{sp_2}}1} \,\, \overset{\mathfrak{so_{12}}}4  \,\,  \overset{\mathfrak{sp_2}}1  \,\, ... [SO(12)]
$};

\node (12) [startstop, below of=10] {
$
4^2,1^4:
[SU(2) \times SU(2)] \,\,   \overset{\mathfrak{so}_{8}}3  \,\, \overset{\mathfrak{sp_1}}1 \,\, \underset{[Sp(1)]}{\overset{\mathfrak{so_{12}}}4}  \,\,  \overset{\mathfrak{sp_2}}1  \,\, ... [SO(12)]
$};

\node (11) [startstop, right of=12, xshift=7cm] {
$
3^4:
\overset{\mathfrak{so}_{7}}3  \,\, \underset{[SO(4)]}{\overset{\mathfrak{sp_2}}1} \,\, {\overset{\mathfrak{so_{12}}}4}  \,\,  \overset{\mathfrak{sp_2}}1  \,\, ... [SO(12)]
$};

\node (13) [startstop, below of=12] {
$
4^2,2^2:
[SU(2)] \,\, \overset{\mathfrak{so}_{7}}3  \,\, {\overset{\mathfrak{sp_1}}1} \,\, \underset{[SU(2)]}{\overset{\mathfrak{so_{12}}}4}  \,\,  \overset{\mathfrak{sp_2}}1  \,\, ... [SO(12)]
$};

\node (14) [startstop, below of=13] {
$
4^2,3,1:
[SU(2)] \,\,   \overset{\mathfrak{g}_{2}}3  \,\, \overset{\mathfrak{sp_1}}1 \,\, \underset{[SU(2)]}{\overset{\mathfrak{so_{12}}}4}  \,\,  \overset{\mathfrak{sp_2}}1  \,\, ... [SO(12)]
$};

\node (15) [startstop, right of=14, xshift=7cm] {
$
5,1^7:
[SO(7)] \,\,  1 \,\, \overset{\mathfrak{so}_{9}}4  \,\, \overset{\mathfrak{sp_1}}1 \,\, {\overset{\mathfrak{so_{11}}}4}  \,\,  \overset{\mathfrak{sp_2}}1  \,\, ... [SO(12)]
$};

\node (16) [startstop, below of=15, xshift=-1cm] {
$
5,2^2,1^3:
[SU(2) \times SU(2)] \,\,   \overset{\mathfrak{so}_{9}}3  \,\, \overset{\mathfrak{sp_1}}1 \,\, \overset{\mathfrak{so_{11}}}4  \,\,  \overset{\mathfrak{sp_2}}1  \,\, ... [SO(12)]
$};

\node (17) [startstop, below of=16, xshift=-1cm] {
$
5,3,1^4:
[SU(2) \times SU(2)] \,\,   \overset{\mathfrak{so}_{8}}3  \,\, \overset{\mathfrak{sp_1}}1 \,\, {\overset{\mathfrak{so_{11}}}4}  \,\,  \overset{\mathfrak{sp_2}}1  \,\, ... [SO(12)]
$};

\node (18) [startstop, below of=17, xshift=-1cm] {
$
5,3,2^2:
[SU(2)] \,\, \overset{\mathfrak{so}_{7}}3  \,\, {\overset{\mathfrak{sp_1}}1} \,\, {\overset{\mathfrak{so_{11}}}4}  \,\,  \overset{\mathfrak{sp_2}}1  \,\, ... [SO(12)]
$};

\node (19) [startstop, below of=18,xshift=-2cm] {
$
5,3^2,1:
\overset{\mathfrak{g}_{2}}3  \,\, {\overset{\mathfrak{sp_1}}1} \,\, {\overset{\mathfrak{so_{11}}}4}  \,\,  \overset{\mathfrak{sp_2}}1  \,\, ... [SO(12)]
$};

\node (20) [startstop, below of=19] {
$
5^2,1^2:
\overset{\mathfrak{su}_{3}}3  \,\, 1 \,\, {\overset{\mathfrak{so_{10}}}4}  \,\,  \overset{\mathfrak{sp_2}}1  \,\, ... [SO(12)]
$};

\node (21) [startstop, below of=20, xshift=-3cm] {
$
6^2:
 \overset{\mathfrak{su}_{2}}2 \,\, \overset{\mathfrak{so}_{7}}3  \,\, \overset{\mathfrak{sp_1}}1 \,\, \underset{[SU(2)]}{\overset{\mathfrak{so_{12}}}4}  \,\,  \overset{\mathfrak{sp_2}}1  \,\, ... [SO(12)]
$};

\node (22) [startstop, right of=21, xshift=6cm] {
$
7,1^5:
[Sp(2)] \,\, \overset{\mathfrak{so}_{7}}3  \,\, 1 \,\, {\overset{\mathfrak{so_{9}}}4} \,\, {\overset{\mathfrak{sp_1}}1} \,\, {\overset{\mathfrak{so_{11}}}4}  \,\,  \overset{\mathfrak{sp_2}}1  \,\, ... [SO(12)]
$};

\node (23) [startstop, below of=22, xshift=-1cm] {
$
7,2^2,1:
[Sp(1)] \,\, \overset{\mathfrak{g}_{2}}3  \,\, 1 \,\, {\overset{\mathfrak{so_{9}}}4} \,\, {\overset{\mathfrak{sp_1}}1} \,\, {\overset{\mathfrak{so_{11}}}4}  \,\,  \overset{\mathfrak{sp_2}}1  \,\, ... [SO(12)]
$};

\node (24) [startstop, below of=23, xshift=-1cm] {
$
7,3,1^2:
\overset{\mathfrak{su_3}}3  \,\, 1 \,\, {\overset{\mathfrak{so_{9}}}4} \,\, {\overset{\mathfrak{sp_1}}1} \,\, {\overset{\mathfrak{so_{11}}}4}  \,\,  \overset{\mathfrak{sp_2}}1  \,\, ... [SO(12)]
$};

\node (25) [startstop, below of=24, xshift=-4cm] {
$
7,5:
\overset{\mathfrak{su_2}}2  \,\, \overset{\mathfrak{so_7}}3  \,\, {\overset{\mathfrak{sp_1}}1} \,\, {\overset{\mathfrak{so_{11}}}4}  \,\,  \overset{\mathfrak{sp_2}}1  \,\, ... [SO(12)]
$};

\node (26) [startstop, right of=25, xshift=6cm] {
$
9,1^3:
\overset{\mathfrak{su_2}}2  \,\, \underset{[SU(2)]}{\overset{\mathfrak{so_7}}3}  \,\, 1 \,\, {\overset{\mathfrak{so_{9}}}4} \,\, {\overset{\mathfrak{sp_1}}1} \,\, {\overset{\mathfrak{so_{11}}}4}  \,\,  \overset{\mathfrak{sp_2}}1  \,\, ... [SO(12)]
$};

\node (27) [startstop, below of=25,xshift=3cm] {
$
9,3:
\overset{\mathfrak{su_2}}2  \,\, \overset{\mathfrak{g_2}}3  \,\, 1 \,\, {\overset{\mathfrak{so_{9}}}4} \,\, {\overset{\mathfrak{sp_1}}1} \,\, {\overset{\mathfrak{so_{11}}}4}  \,\,  \overset{\mathfrak{sp_2}}1  \,\, ... [SO(12)]
$};

\node (28) [startstop, below of=27] {
$
11,1:
2\,\, \overset{\mathfrak{su_2}}2  \,\, \overset{\mathfrak{g_2}}3  \,\, 1 \,\, {\overset{\mathfrak{so_{9}}}4} \,\, {\overset{\mathfrak{sp_1}}1} \,\, {\overset{\mathfrak{so_{11}}}4}  \,\,  \overset{\mathfrak{sp_2}}1  \,\, ... [SO(12)]
$};

\draw [arrow] (1) -- (2);
\draw [arrow] (2) -- (3);
\draw [arrow] (2) -- (4);
\draw [arrow, color=blue] (3) -- (5);
\draw [arrow] (3) -- (6);
\draw [arrow] (4) -- (6);
\draw [arrow] (5) -- (8);
\draw [arrow] (6) -- (7);
\draw [arrow, color=blue] (6) -- (8);
\draw [arrow, color=blue] (7) -- (9);
\draw [arrow] (8) -- (9);
\draw [arrow] (9) -- (10);
\draw [arrow] (10) -- (11);
\draw [arrow] (10) -- (12);
\draw [arrow] (11) -- (13);
\draw [arrow] (12) -- (13);
\draw [arrow] (13) -- (14);
\draw [arrow] (12) -- (17);
\draw [arrow] (7) -- (15);
\draw [arrow, color=blue] (15) -- (16);
\draw [arrow] (10) -- (16);
\draw [arrow] (16) -- (17);
\draw [arrow] (13) -- (18);
\draw [arrow] (14) -- (19);
\draw [arrow] (17) -- (18);
\draw [arrow] (18) -- (19);
\draw [arrow] (19) -- (20);
\draw [arrow, color=blue] (20) -- (21);
\draw [arrow] (17) -- (22);
\draw [arrow] (22) -- (23);
\draw [arrow] (19) -- (23);
\draw [arrow] (23) -- (24);
\draw [arrow] (20) -- (24);
\draw [arrow, color=blue] (24) -- (25);
\draw [arrow] (21) -- (25);
\draw [arrow, color=blue] (24) -- (26);
\draw [arrow] (25) -- (27);
\draw [arrow] (26) -- (27);
\draw [arrow, color=blue] (27) -- (28);

\end{tikzpicture}
\end{lrbox}%
\ifdim\ht\mysavebox>\textheight
    \setlength{\myrest}{\ht\mysavebox}%
    \loop\ifdim\myrest>\textheight
        \newpage\par\noindent
        \clipbox{0 {\myrest-\textheight+.5cm} 0 {\ht\mysavebox-\myrest}}{\usebox{\mysavebox}}%
        \addtolength{\myrest}{-\textheight}%
    \repeat
    \newpage\par\noindent
\begin{figure}
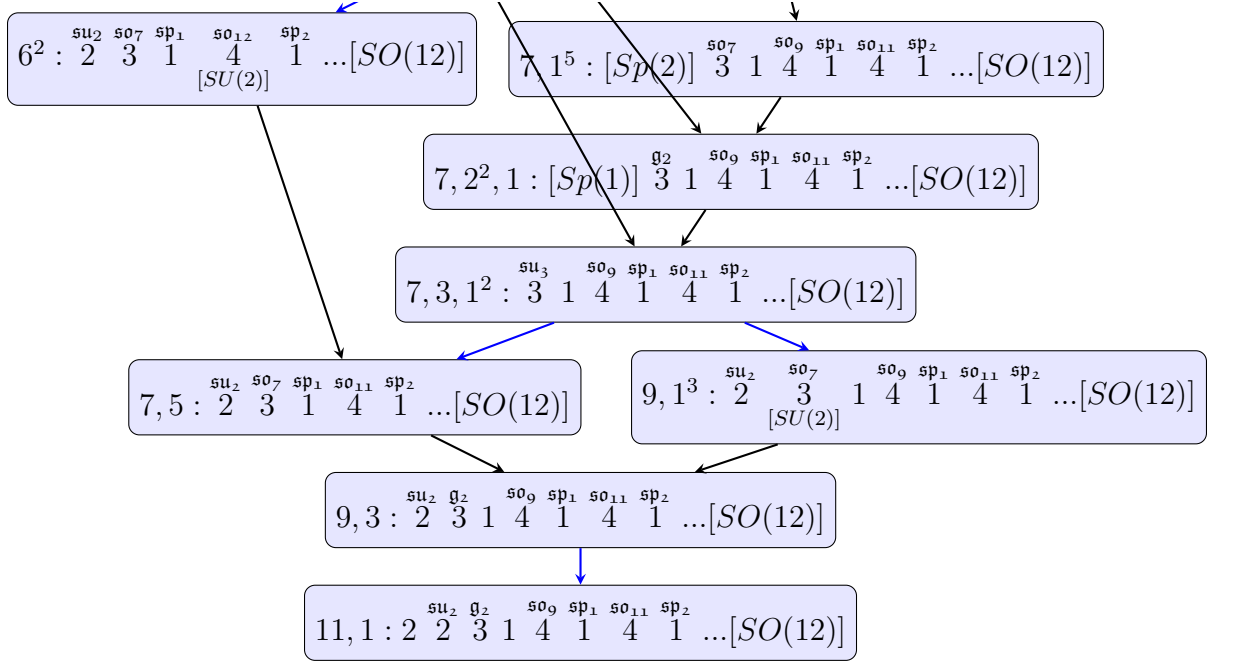

    \clipbox{0 0 0 {\ht\mysavebox-\myrest-.6cm}}{\usebox{\mysavebox}}%
\caption{\label{fig:SO12} Flows for $SO(12)$ nilpotent orbits.  Blue arrows indicate flows where one or more free tensors appears in the IR.}
\end{figure}
\else
    \usebox{\mysavebox}%
\fi

\end{center}

\subsection{Flows from $\mathfrak{so}_{\text{odd}}$ and $\mathfrak{sp}_{N}$}\label{ssec:SOodd}

Finally, we come to the analysis of flows involving the non-simply laced
classical algebras $\mathfrak{so}(2N+1)$ and $\mathfrak{sp}(N)$. In these
cases, we do not directly reach the desired flavor symmetry from M5-branes
probing an ADE\ singularity. Rather, we must first consider the case of a
partial tensor branch flow and / or some contribution from conformal matter vevs.
For example, to reach the $\mathfrak{sp}$-type flavor symmetries, we can
start from:%
\begin{equation}
\lbrack SO(2N)]\overset{\mathfrak{sp}_{N}}{1}\text{ }\overset{\mathfrak{so}%
_{2N}}{4}\text{ }\overset{\mathfrak{sp}_{N}}{1}...\overset{\mathfrak{so}%
_{2N}}{4}\text{ }\overset{\mathfrak{sp}_{N}}{1}[SO(2N)],
\end{equation}
and by decompactifying the leftmost and rightmost $-1$ curves, we reach the system:%
\begin{equation}
\lbrack Sp(N)]\overset{\mathfrak{so}_{2N}}{4}\text{ }\overset{\mathfrak{sp}%
_{N}}{1}...\overset{\mathfrak{so}_{2N}}{4}[Sp(N)].
\end{equation}
In the case of an $SO(2N+1)$ flavor symmetry we can also start from a theory
with $SO(2N+2p)$ flavor symmetry. For sufficiently large $p$, we can then
reach the desired $SO(2N+1)$ flavor symmetry by activating a conformal matter
vev associated with the partition $(2p-1,1^{2N+1})$. See figures \ref{fig:sp3flows} and \ref{fig:so9flows}
for examples of the flow diagrams and associated F-theory models for these systems.

\begin{center}

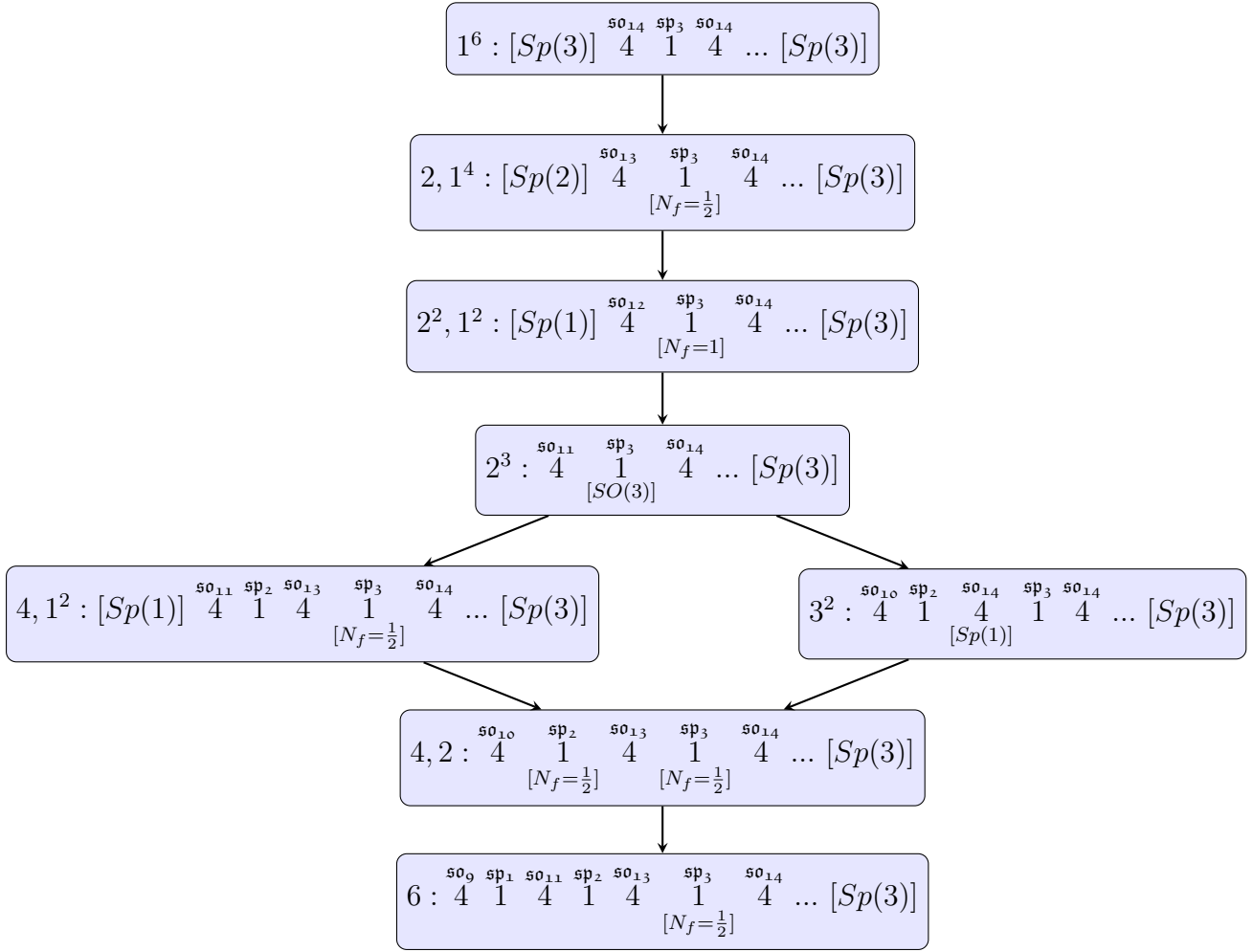
\begin{figure}
\begin{tikzpicture}[node distance=2cm]

\node (1) [startstop, xshift=-1cm] {
$
1^6: [Sp(3)] \,\,   \overset{\mathfrak{so_{14}}}4 \,\, \overset{\mathfrak{sp_{3}}}1 \,\, \overset{\mathfrak{so_{14}}}4 \,\, ... \,\, [Sp(3)]
$};

\node (2) [startstop, below of=1] {
$
2,1^4: [Sp(2)] \,\,   \overset{\mathfrak{so_{13}}}4 \,\, \underset{[N_f=\frac{1}{2}]}{\overset{\mathfrak{sp_{3}}}1} \,\, \overset{\mathfrak{so_{14}}}4 \,\, ... \,\, [Sp(3)]
$};

\node (3) [startstop, below of=2] {
$
2^2,1^2: [Sp(1)] \,\,   \overset{\mathfrak{so_{12}}}4 \,\, \underset{[N_f=1]}{\overset{\mathfrak{sp_{3}}}1} \,\, \overset{\mathfrak{so_{14}}}4 \,\, ... \,\, [Sp(3)]
$};

\node (4) [startstop, below of=3] {
$
2^3:    \overset{\mathfrak{so_{11}}}4 \,\, \underset{[SO(3)]}{\overset{\mathfrak{sp_{3}}}1} \,\, \overset{\mathfrak{so_{14}}}4 \,\, ... \,\, [Sp(3)]
$};

\node (5) [startstop, below of=4, xshift=5cm] {
$
3^2:    \overset{\mathfrak{so_{10}}}4 \,\, {\overset{\mathfrak{sp_{2}}}1} \,\,  \underset{[Sp(1)]}{\overset{\mathfrak{so_{14}}}4} \,\, \overset{\mathfrak{sp_{3}}}1 \,\,  {\overset{\mathfrak{so_{14}}}4}\,\, ... \,\, [Sp(3)]
$};

\node (6) [startstop, below of=4, xshift=-5cm] {
$
4,1^2:   [Sp(1)] \,\,  \overset{\mathfrak{so_{11}}}4 \,\, {\overset{\mathfrak{sp_{2}}}1} \,\,  {\overset{\mathfrak{so_{13}}}4} \,\, \underset{[N_f=\frac12]}{\overset{\mathfrak{sp_{3}}}1} \,\,  {\overset{\mathfrak{so_{14}}}4}\,\, ... \,\, [Sp(3)]
$};

\node (7) [startstop, below of=6, xshift=5cm] {
$
4,2:    \overset{\mathfrak{so_{10}}}4 \,\, \underset{[N_f = \frac{1}{2}]}{\overset{\mathfrak{sp_{2}}}1} \,\,  {\overset{\mathfrak{so_{13}}}4} \,\, \underset{[N_f = \frac{1}{2}]}{\overset{\mathfrak{sp_{3}}}1} \,\,  {\overset{\mathfrak{so_{14}}}4}\,\, ... \,\, [Sp(3)]
$};

\node (8) [startstop, below of=7] {
$
6:    \overset{\mathfrak{so_{9}}}4 \,\, {\overset{\mathfrak{sp_{1}}}1} \,\,  {\overset{\mathfrak{so_{11}}}4} \,\, \overset{\mathfrak{sp_{2}}}1 \,\,  {\overset{\mathfrak{so_{13}}}4}\,\, \underset{[N_f=\frac{1}{2}]}{\overset{\mathfrak{sp_{3}}}1}  \,\,  {\overset{\mathfrak{so_{14}}}4}\,\, ... \,\, [Sp(3)]
$};

\draw [arrow] (1) -- (2);
\draw [arrow] (2) -- (3);
\draw [arrow] (3) -- (4);
\draw [arrow] (4) -- (5);
\draw [arrow] (4) -- (6);
\draw [arrow] (5) -- (7);
\draw [arrow] (6) -- (7);
\draw [arrow] (7) -- (8);

\end{tikzpicture}
\caption{Flows for $Sp(3)$ nilpotent orbits.}
\label{fig:sp3flows}
\end{figure}

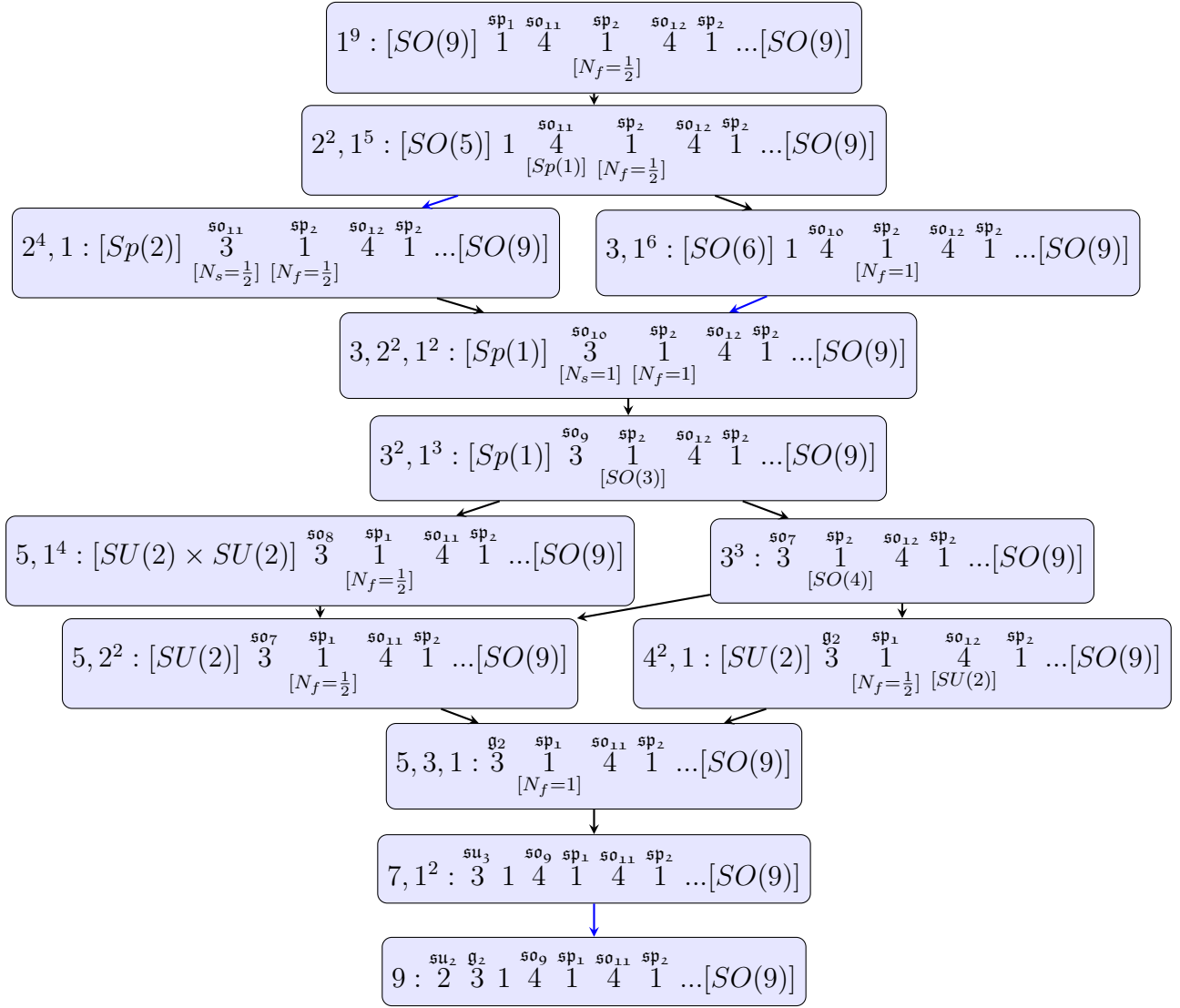
\begin{figure}

\begin{tikzpicture}[node distance=1.5cm]

\node (1) [startstop, xshift=-2cm]{
$
1^9: [SO(9)]\,\, \overset{\mathfrak{sp}_1}1 \,\, {\overset{\mathfrak{so_{11}}}4}  \,\, \underset{[N_f=\frac12]}{\overset{\mathfrak{sp_2}}1} \,\, \overset{\mathfrak{so_{12}}}4  \,\,  \overset{\mathfrak{sp_2}}1  \,\, ... [SO(9)]
$};

\node (2) [startstop, below of=1, xshift=0cm] {
$
2^2,1^5: [SO(5)] \,\, 1\,\,  \underset{[Sp(1)]}{\overset{\mathfrak{so_{11}}}4}  \,\, \underset{[N_f=\frac12]}{\overset{\mathfrak{sp_2}}1} \,\, \overset{\mathfrak{so_{12}}}4  \,\,  \overset{\mathfrak{sp_2}}1  \,\, ... [SO(9)]
$};

\node (3) [startstop, below of=2, xshift=4cm] {
$
3,1^6: [SO(6)] \,\, 1\,\,  {\overset{\mathfrak{so_{10}}}4}  \,\, \underset{[N_f=1]}{\overset{\mathfrak{sp_2}}1} \,\, \overset{\mathfrak{so_{12}}}4  \,\,  \overset{\mathfrak{sp_2}}1  \,\, ... [SO(9)]
$};

\node (4) [startstop, right of=3, xshift=-10cm] {
$
2^4,1: [Sp(2)] \,\,   \underset{[N_s=\frac12]}{\overset{\mathfrak{so_{11}}}3}  \,\, \underset{[N_f=\frac12]}{\overset{\mathfrak{sp_2}}1} \,\, \overset{\mathfrak{so_{12}}}4  \,\,  \overset{\mathfrak{sp_2}}1  \,\, ... [SO(9)]
$};

\node (5) [startstop, below of=4, xshift=5cm] {
$
3,2^2,1^2: [Sp(1)] \,\,   \underset{[N_s=1]}{\overset{\mathfrak{so_{10}}}3}  \,\, \underset{[N_f=1]}{\overset{\mathfrak{sp_2}}1} \,\, \overset{\mathfrak{so_{12}}}4  \,\,  \overset{\mathfrak{sp_2}}1  \,\, ... [SO(9)]
$};

\node (6) [startstop, below of=5] {
$
3^2,1^3: [Sp(1)] \,\,   {\overset{\mathfrak{so_{9}}}3}  \,\, \underset{[SO(3)]}{\overset{\mathfrak{sp_2}}1} \,\, \overset{\mathfrak{so_{12}}}4  \,\,  \overset{\mathfrak{sp_2}}1  \,\, ... [SO(9)]
$};

\node (7) [startstop, below of=6, xshift=4cm] {
$
3^3:
\overset{\mathfrak{so}_{7}}3  \,\, \underset{[SO(4)]}{\overset{\mathfrak{sp_2}}1} \,\, {\overset{\mathfrak{so_{12}}}4}  \,\,  \overset{\mathfrak{sp_2}}1  \,\, ... [SO(9)]
$};

\node (8) [startstop, below of=7, xshift=0cm] {
$
4^2,1:
[SU(2)] \,\,   \overset{\mathfrak{g}_{2}}3  \,\, \underset{[N_f=\frac12]}{\overset{\mathfrak{sp_1}}1} \,\, \underset{[SU(2)]}{\overset{\mathfrak{so_{12}}}4}  \,\,  \overset{\mathfrak{sp_2}}1  \,\, ... [SO(9)]
$};

\node (9) [startstop,right of=7, xshift=-10cm] {
$
5,1^4:
[SU(2) \times SU(2)] \,\,   \overset{\mathfrak{so}_{8}}3  \,\, \underset{[N_f=\frac12]}{\overset{\mathfrak{sp_1}}1} \,\, {\overset{\mathfrak{so_{11}}}4}  \,\,  \overset{\mathfrak{sp_2}}1  \,\, ... [SO(9)]
$};

\node (10) [startstop, below of=9, xshift=0cm] {
$
5,2^2:
[SU(2)] \,\, \overset{\mathfrak{so}_{7}}3  \,\, \underset{[N_f=\frac12]}{\overset{\mathfrak{sp_1}}1} \,\, {\overset{\mathfrak{so_{11}}}4}  \,\,  \overset{\mathfrak{sp_2}}1  \,\, ... [SO(9)]
$};

\node (11) [startstop, below of=10, xshift=4cm] {
$
5,3,1:
\overset{\mathfrak{g}_{2}}3  \,\, \underset{[N_f=1]}{\overset{\mathfrak{sp_1}}1} \,\, {\overset{\mathfrak{so_{11}}}4}  \,\,  \overset{\mathfrak{sp_2}}1  \,\, ... [SO(9)]
$};

\node (12) [startstop, below of=11, xshift=0cm] {
$
7,1^2:
\overset{\mathfrak{su_3}}3  \,\, 1 \,\, {\overset{\mathfrak{so_{9}}}4} \,\, {\overset{\mathfrak{sp_1}}1} \,\, {\overset{\mathfrak{so_{11}}}4}  \,\,  \overset{\mathfrak{sp_2}}1  \,\, ... [SO(9)]
$};

\node (13) [startstop, below of=12] {
$
9:
\overset{\mathfrak{su_2}}2  \,\, \overset{\mathfrak{g_2}}3  \,\, 1 \,\, {\overset{\mathfrak{so_{9}}}4} \,\, {\overset{\mathfrak{sp_1}}1} \,\, {\overset{\mathfrak{so_{11}}}4}  \,\,  \overset{\mathfrak{sp_2}}1  \,\, ... [SO(9)]
$};

\draw [arrow] (1) -- (2);
\draw [arrow] (2) -- (3);
\draw [arrow,color=blue] (2) -- (4);
\draw [arrow] (4) -- (5);
\draw [arrow,color=blue] (3) -- (5);
\draw [arrow] (5) -- (6);
\draw [arrow] (6) -- (7);
\draw [arrow] (7) -- (8);
\draw [arrow] (7) -- (10);
\draw [arrow] (6) -- (9);
\draw [arrow] (9) -- (10);
\draw [arrow] (8) -- (11);
\draw [arrow] (10) -- (11);
\draw [arrow] (11) -- (12);
\draw [arrow,color=blue] (12) -- (13);

\end{tikzpicture}
\caption{Flows for $SO(9)$ nilpotent orbits.  Blue arrows indicate flows where one or more free tensors appears in the IR.}
\label{fig:so9flows}
\end{figure}
\end{center}

\section{Exceptional Flavor Symmetries \label{sec:EXCEPTIONAL}}

In the previous section we focused on examples with classical flavor symmetry algebras
where there is a combinatorial construction of all nilpotent
orbits in terms of partitions of positive integers (with suitable restrictions).

But we have also seen that for all cases other than the A-type flavor
symmetry, conformal matter vevs can sometimes drive us to a conformal fixed
point where spinor representations are present, indicating that the
construction really requires non-perturbative elements (i.e., an embedding in F-theory).

Now, in the case of flows from a theory with exceptional flavor symmetries, we
must resort to the F-theory realization from the start. Nevertheless, we still
expect that some (but not all!) of the RG\ flows induced by nilpotent orbits
can be understood in terms of partitions of perturbative D7-branes. For
example, in the terminology of \cite{Gaberdiel:1997ud}, a seven-brane with $E_{8}$ gauge
symmetry is given by a non-perturbative bound state of seven-branes of
different $(p,q)$ type, i.e.~$A^{7}BC^{2}$. In a suitable duality frame, the
$A$-type seven-branes are just the perturbative D7-branes, and so we can
expect some of the nilpotent orbits to be described by partitions of these
seven seven-branes. By a similar token, there are six such seven-branes for
$E_{7}$ and five for $E_{6}$. Nevertheless, there are also more general
nilpotent orbits which do not appear to admit such a simple characterization
in terms of partitions.

To deal with this more general class of nilpotent orbits, and to verify that
we indeed get a corresponding match with hierarchies expected from RG flows, we will instead
need to rely on some results from the Bala--Carter (B--C) theory of nilpotent orbits
for exceptional algebras. The main point is that for each nilpotent element
$\mu\in\mathfrak{g}_{\mathbb{C}}$, we get a corresponding homomorphism via the
Jacobson--Morozov theorem (see line (\ref{JacMor})). So, to
characterize possible homomorphisms, we simply need to specify the embedding
in a subalgebra of $\mathfrak{g}_{\mathbb{C}}$. Indeed, there is also a notion
of partial ordering for these nilpotent orbits, which is reviewed in great
detail in reference \cite{Chacaltana:2012zy}. For this reason, we should expect there to be
a similar correspondence between nilpotent orbits and RG\ flows.

Since there is a finite list of nilpotent orbits for each exceptional flavor symmetry, we can
explicitly determine the induced flow for each case.
For the simply laced algebras $E_{6}$, $E_{7}$ and $E_{8}$, our starting
point will be a long generalized quiver of the form:%
\begin{align}
&  \lbrack E_{6}]-E_{6}-E_{6}-...,\\
&  \lbrack E_{7}]-E_{7}-E_{7}-...,\\
&  \lbrack E_{8}]-E_{8}-E_{8}-...,
\end{align}
i.e.~we take a stack of M5-branes probing an E-type singularity. The links
here \textquotedblleft$-$\textquotedblright\ denote the corresponding
conformal matter for these systems. In F-theory terms, the resolved theory on
the tensor branch for each of these cases is:%
\begin{align}
&  [E_{6}]\,1\overset{\mathfrak{su}_{3}}{3}1\overset{\mathfrak{e}_{6}%
}{6}...,\label{esixconfmatt}\\
&  \lbrack E_{7}]\,1\text{ }\overset{\mathfrak{su}_{2}}{2}\text{ }%
\overset{\mathfrak{s0}_{7}}{3}\text{ }\overset{\mathfrak{su}_{2}%
}{2}1\overset{\mathfrak{e}_{7}}{8}...,\\
&  [E_{8}]\,1\text{ }2\text{ }\overset{\mathfrak{sp}_{1}}{2}\text{
}\overset{\mathfrak{g}_{2}}{3}\text{ }1\text{ }\overset{\mathfrak{f}_{4}%
}{5}\text{ }1\text{ }\overset{\mathfrak{g}_{2}}{3}\text{ }%
\overset{\mathfrak{sp}_{1}}{2}\text{ }2\text{ }1\text{ }(\overset{\mathfrak{e}%
_{8}}{12})..., \label{eeightconfmatt}%
\end{align}
We can also reach SCFTs with non-simply laced flavor symmetry algebras
$\mathfrak{g}_{2}$ and $\mathfrak{f}_{4}$ by decompactifying the $-3$ and $-5$
curves of the $(E_{8},E_{8})$ conformal matter system:%
\begin{align}
&  \lbrack G_{2}]\overset{\mathfrak{sp}_{1}}{2}\text{ }2\text{ }1\text{
}(\overset{\mathfrak{e}_{8}}{12})...,\\
&  \lbrack F_{4}]\text{ }1\text{ }\overset{\mathfrak{g}_{2}}{3}\text{
}\overset{\mathfrak{sp}_{1}}{2}\text{ }2\text{ }1\text{ }%
(\overset{\mathfrak{e}_{8}}{12})....
\end{align}
In these cases, the \textquotedblleft$...$\textquotedblright\ indicates that
we continue beyond this point with a sequence of $E_{8}$ gauge groups with
conformal matter between each such factor.

The rest of this section is organized as follows. We begin by giving an
analysis of the nilpotent orbits of the simply laced exceptional algebras and
the corresponding F-theory models associated with each such element. Using
Bala--Carter theory, we also determine the flavor symmetries expected from
the commutant of the nilpotent orbit in the parent flavor symmetry algebra
and compare it with those flavor symmetries
visible on the tensor branch of an F-theory model. We then turn to a similar
analysis for the non-simply laced exceptional algebras.

\subsection{Flows from $\mathfrak{e}_{6}$, $\mathfrak{e}_{7}$, $\mathfrak{e}%
_{8}$}

Let us begin with an analysis of the flows for the exceptional algebras
$\mathfrak{e}_{6}$, $\mathfrak{e}_{7}$ and $\mathfrak{e}_{8}$.
Proceeding as in the previous examples, we start from the theories  (\ref{esixconfmatt})--(\ref{eeightconfmatt}) and break the flavor symmetry on the left in various ways while holding fixed the flavor symmetry on the right. That is, we consider the theories
${\cal T}(E_n,\mu_L, \mu_R,k)$ obtained by varying $\mu_L$ whilst holding $\mu_R$ fixed and trivial. We now show how
the hierarchy on nilpotent orbits determines hierarchies of RG fixed points.

For $\mathfrak{e}_6$ we show the results in a diagram similar to the ones given so far, in figure \ref{fig:E6}. In the cases with $\mathfrak{e}_{7}$ and $\mathfrak{e}_{8}$ flavor symmetry, the full list of nilpotent hierarchies does not easily fit on a few pages,
but is presented for example in \cite[App.~C]{Chacaltana:2012zy}. Thus in Appendix \ref{app:nilp} we give the full list of Bala--Carter labels, the corresponding global flavor symmetries (expected from the commutants of $\im(\rho)$ in $\mathfrak{g}_{\mathbb{C}}$; see (\ref{JacMor})) and the corresponding realization in an F-theory model, with the understanding that there is an RG flow whenever there is an ordering relation between the corresponding label as in \cite[App.~C]{Chacaltana:2012zy}, in agreement with line (\ref{eq:RG}).

The methods we used to produce these results are the same as the ones for the previous tables, as described in section \ref{sub:so}. Once again, each flow corresponds to a complex deformation, which can be exhibited most easily by shrinking some $-1$ curve; for example, the very first flow corresponds to the deformation $y^2 = x^3 + (u^2+ \epsilon x)^2 v^2$. At $\epsilon=0$ this describes a collision between an $\mathfrak{e}_6$ at
$u = 0$ and an $\mathfrak{su}_3$ curve at $v = 0$; for $\epsilon\neq 0$ the $u = 0$ curve instead supports an $\mathfrak{su}_6$ gauge algebra. Once again, however, it is quicker to use a combination of field theory techniques and F-theory intuition. There is a new type of Higgs flow that did not appear earlier: see for example the flows $3A_1\to A_2$ or $D_4(a_1)\to D_4$ in figure \ref{fig:E6}. This type of flow was discussed around \cite[Eq.(4.24)]{Heckman:2015ola}. In $3A_1$, we can shrink the leftmost $-1$ curve, we reveal another $-1$ curve; if we also shrink that one as well, we have a special point on the leftmost $\mathfrak{e}_6$ curve of multiplicity 2. The flow consists of going to a more generic situation where there are two special points of multiplicity 1; blowing them up produces two separate $-1$ curves touching the $\mathfrak{e}_6$ curve, which we see in the $A_2$ theory. These are examples of ``small instanton maneuvers'' of the type encountered in
subsection \ref{ssec:estring}.

Another new point is that in some examples the flavor symmetry expected from the B-C
labels refines the ``na\"ive'' expectation one would have from
just treating subsectors of a field theory on its tensor branch
in isolation. In some cases, this also conforms with
restrictions on non-abelian flavor symmetries expected
from F-theory considerations. In other cases, however, we find
that ---especially for abelian symmetry factors--- the B-C label analysis provides a
systematic way to extract such flavor symmetries which are difficult to deduce
using other techniques. We develop this point further
in section \ref{sec:FLAVOR}.

An important aspect of the tight match found here
is that in general, we find several gauge groups
of $\mathfrak{e}_{n}$ type will generically be Higgsed
in a given flow by conformal matter vevs. This is
not altogether surprising since related phenomena are already present for
models with weakly coupled hypermultiplets. Indeed in
the quivers in the third column of figure \ref{fig:su4}
we see that the ranks of the gauge groups decrease in an
RG flow not only in the rightmost position.  There, it is
a consequence of the fact that there will typically be a
propagating sequence of D-term constraints.

\begin{center}

\begin{lrbox}{\mysavebox}%

\begin{tikzpicture}[node distance=1.8cm]

\node (1) [startstop, xshift=-1cm] {
$
0: [E_6] \,\, 1 \,\, \overset{\mathfrak{su_{3}}}3  \,\, 1 \,\, \overset{\mathfrak{e_{6}}}6  \,\, 1 \,\,  \overset{\mathfrak{su_{3}}}3 \,\, 1  \,\, ...[E_6]
$};

\node (2) [startstop, below of=1] {
$
A_1: [SU(6)] \,\,  \overset{\mathfrak{su_{3}}}2  \,\, 1 \,\, \overset{\mathfrak{e_{6}}}6  \,\, 1 \,\,  \overset{\mathfrak{su_{3}}}3 \,\, 1  \,\, ...[E_6]
$};

\node (3) [startstop, below of=2] {
$
2 A_1: [SO(7)] \,\,  \overset{\mathfrak{su_{2}}}2  \,\, 1 \,\, \overset{\mathfrak{e_{6}}}6  \,\, 1 \,\,  \overset{\mathfrak{su_{3}}}3 \,\, 1  \,\, ...[E_6]
$};

\node (4) [startstop,  below of=3] {
$
3 A_1:[SU(2)] \,\,  2  \,\, \underset{[SU(3)]}1 \,\, \overset{\mathfrak{e_{6}}}6  \,\, 1 \,\,  \overset{\mathfrak{su_{3}}}3 \,\, 1  \,\, ...[E_6]
$};

\node (5) [startstop, xshift=0cm, below of=4] {
$
A_2: [SU(3)] \,\, 1 \,\, \underset{[SU(3)]}{\underset{1}{\overset{\mathfrak{e_{6}}}6}}  \,\, 1 \,\,  \overset{\mathfrak{su_{3}}}3 \,\, 1  \,\, ...[E_6]
$};

\node (6) [startstop, xshift=0cm, below of=5] {
$
A_2 + A_1: [SU(3)] \,\, 1 \,\, {\underset{[N_f=1]}{\overset{\mathfrak{e_{6}}}5}}  \,\, 1 \,\,  \overset{\mathfrak{su_{3}}}3 \,\, 1  \,\, ...[E_6]
$};

\node (7) [startstop, xshift=-4cm, below of=6] {
$
2 A_2: [G_2] \,\, 1 \,\, {\overset{\mathfrak{f_{4}}}5}  \,\, 1 \,\,  \overset{\mathfrak{su_{3}}}3 \,\, 1  \,\,\overset{\mathfrak{e_{6}}}6  \,\, 1 \,\,  ...[E_6]
$};

\node (8) [startstop, xshift=6cm, below of=6] {
$
A_2 +2 A_1:  [SU(2)]\,\,\overset{\mathfrak{e_{6}}}4  \,\, 1 \,\,  \overset{\mathfrak{su_{3}}}3 \,\, 1  \,\, \overset{\mathfrak{e_{6}}}6  \,\, 1 \,\, ...[E_6]
$};

\node (9) [startstop, xshift=-10cm, below of=8] {
$
2 A_2 + A_1: [SU(2)]\,\,\overset{\mathfrak{f_{4}}}4  \,\, 1 \,\,  \overset{\mathfrak{su_{3}}}3 \, \, 1  \,\,\overset{\mathfrak{e_{6}}}6  \,\, 1 \,\, ...[E_6]
$};

\node (10) [startstop, xshift=0cm, below of=8] {
$
A_3: [Sp(2)] \,\, {\overset{\mathfrak{so_{10}}}4}  \,\, 1 \,\,  \overset{\mathfrak{su_{3}}}3 \,\, 1  \,\, \overset{\mathfrak{e_{6}}}6  \,\, 1 \,\,...[E_6]
$};

\node (11) [startstop, xshift=-6cm, below of=10] {
$
A_3+A_1: [SU(2)] \,\, {\overset{\mathfrak{so_{9}}}4}  \,\, 1 \,\,  \overset{\mathfrak{su_{3}}}3 \,\, 1  \,\, \overset{\mathfrak{e_{6}}}6  \,\, 1 \,\, ...[E_6]
$};

\node (12) [startstop, xshift=0cm, below of=11] {
$
D_4(a_1):  {\overset{\mathfrak{so_{8}}}4}  \,\, 1 \,\,  \overset{\mathfrak{su_{3}}}3 \,\, 1  \,\,\overset{\mathfrak{e_{6}}}6  \,\, 1 \,\,  ...[E_6]
$};

\node (13) [startstop, xshift=-4cm, below of=12] {
$
A_4:  [SU(2)] \,\, {\overset{\mathfrak{so_{7}}}3}  \,\,   \overset{\mathfrak{su_{2}}}2 \,\, 1  \,\,\overset{\mathfrak{e_{6}}}6  \,\, 1 \,\,  ...[E_6]
$};

\node (14) [startstop, xshift=0cm, below of=13] {
$
A_4+A_1:   {\overset{\mathfrak{g_{2}}}3}  \,\,   {\overset{\mathfrak{su_{2}}}2} \,\, 1  \,\,\overset{\mathfrak{e_{6}}}6  \,\, 1 \,\,  ...[E_6]
$};

\node (15) [startstop, xshift=8cm, right of=14] {
$
D_4:   {\overset{\mathfrak{su_{3}}}3}  \,\,   1  \,\, \underset{[SU(3)]}{\underset{1}{\overset{\mathfrak{e_{6}}}6}}  \,\, 1 \,\,   \overset{\mathfrak{su_{3}}}3 \,\, 1  \,\,\overset{\mathfrak{e_{6}}}6  \,\, 1 \,\,...[E_6]
$};

\node (16) [startstop, xshift=0cm, below of=14] {
$
A_5:    [SU(2)] \,\, {\overset{\mathfrak{g_{2}}}3}  \,\,    1  \,\,\overset{\mathfrak{f_{4}}}5  \,\, 1 \,\,   \overset{\mathfrak{su_{3}}}3 \,\, 1  \,\, \overset{\mathfrak{e_{6}}}6  \,\, 1 \,\,...[E_6]
$};

\node (17) [startstop, xshift=0cm, below of=15] {
$
D_5 (a_1):  {\overset{\mathfrak{su_{3}}}3}  \,\,   1  \,\, \underset{[N_f=1]}{\overset{\mathfrak{e_{6}}}5}  \,\, 1 \,\,   \overset{\mathfrak{su_{3}}}3 \,\, 1  \,\,\overset{\mathfrak{e_{6}}}6  \,\, 1 \,\,...[E_6]
$};

\node (18) [startstop, xshift=-5cm, below of=17] {
$
E_6 (a_3):  {\overset{\mathfrak{su_{3}}}3}  \,\,   1  \,\, {\overset{\mathfrak{f_{4}}}5}  \,\, 1 \,\,   \overset{\mathfrak{su_{3}}}3 \,\, 1  \,\,\overset{\mathfrak{e_{6}}}6  \,\, 1 \,\,...[E_6]
$};

\node (19) [startstop, xshift=0cm, below of=18] {
$
D_5:   {\overset{\mathfrak{su_{2}}}2}  \,\,   {\overset{\mathfrak{so_{7}}}3}  \,\,  \overset{\mathfrak{su_{2}}}2 \,\, 1  \,\,\overset{\mathfrak{e_{6}}}6  \,\, 1 \,\,...[E_6]
$};

\node (20) [startstop, xshift=0cm, below of=19] {
$
E_6 (a_1):   {\overset{\mathfrak{su_{2}}}2}  \,\,   {\overset{\mathfrak{g_2}}3}  \,\, 1 \,\, \overset{\mathfrak{f_{4}}}5 \,\, 1 \,\,  {\overset{\mathfrak{su_{3}}}3}  \,\,   1 \,\,\overset{\mathfrak{e_{6}}}6  \,\, 1 \,\,...[E_6]
$};

\node (21) [startstop, xshift=0cm, below of=20] {
$
E_6:    2  \,\,   {\overset{\mathfrak{su_{2}}}2}  \,\,   {\overset{\mathfrak{g_2}}3}  \,\, 1 \,\, \overset{\mathfrak{f_{4}}}5 \,\, 1 \,\,  {\overset{\mathfrak{su_{3}}}3}  \,\,   1 \,\,\overset{\mathfrak{e_{6}}}6  \,\, 1 \,\,...[E_6]
$};

\draw [arrow, color=blue] (1) -- (2);
\draw [arrow] (2) -- (3);
\draw [arrow] (3) -- (4);
\draw [arrow] (4) -- (5);
\draw [arrow, color=blue] (5) -- (6);
\draw [arrow] (6) -- (7);
\draw [arrow, color=blue] (6) -- (8);
\draw [arrow, color=blue] (7) -- (9);
\draw [arrow] (8) -- (9);
\draw [arrow] (8) -- (10);
\draw [arrow] (10) -- (11);
\draw [arrow] (9) -- (11);
\draw [arrow] (11) -- (12);
\draw [arrow, color=blue] (12) -- (13);
\draw [arrow] (13) -- (14);
\draw [arrow, color=blue] (12) -- (15);
\draw [arrow, color=blue] (14) -- (16);
\draw [arrow, color=blue] (14) -- (17);
\draw [arrow, color=blue] (15) -- (17);
\draw [arrow] (16) -- (18);
\draw [arrow] (17) -- (18);
\draw [arrow, color=blue] (18) -- (19);
\draw [arrow, color=blue] (19) -- (20);
\draw [arrow, color=blue] (20) -- (21);

\end{tikzpicture}
\end{lrbox}%
\ifdim\ht\mysavebox>\textheight
    \setlength{\myrest}{\ht\mysavebox}%
    \loop\ifdim\myrest>\textheight
        \newpage\par\noindent
        \clipbox{0 {\myrest-\textheight+0cm} 0 {\ht\mysavebox-\myrest}}{\usebox{\mysavebox}}%
        \addtolength{\myrest}{-\textheight}%
    \repeat
    \newpage\par\noindent
\begin{figure}
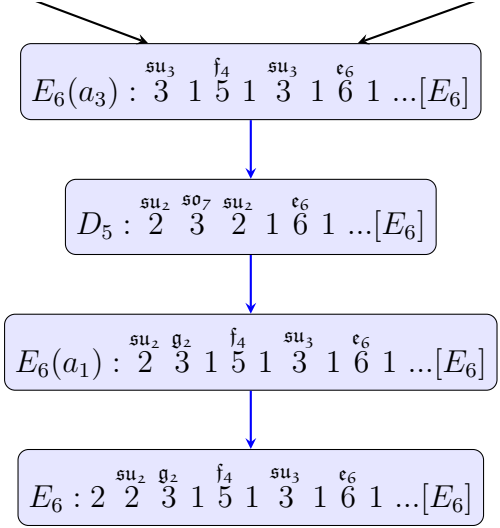

    \clipbox{0 0 0 {\ht\mysavebox-\myrest-0cm}}{\usebox{\mysavebox}}%
\caption{\label{fig:E6} Flows for $E_6$ nilpotent orbits. Blue arrows indicate flows where one or more free tensors appears in the IR. In the above, we always take a trivial nilpotent orbit on the right; on the left we present the B-C label for the nilpotent orbit. The same information is also presented in the table of appendix \ref{app:E6}, where both abelian and non-abelian flavor symmetries are also shown.}
\end{figure}
\else
    \usebox{\mysavebox}%
\fi

\end{center}

\begin{figure}[h!]
\begin{center}
\begin{tikzpicture}[node distance=1.5cm]

\node (1) [startstop, xshift=-1cm] {
$
 1:   [G_2] \,\, \overset{\mathfrak{su}_2}2 \,\, 2 \,\, 1\,\, \overset{\mathfrak{e_{8}}}{12}  \,\, 1 \,\,  2 \,\, {\overset{\mathfrak{su_{2}}}2}  \,\,   {\overset{\mathfrak{g_2}}3}  \,\, 1 \,\, \overset{\mathfrak{f_{4}}}5 \,\, 1\,\, ...[G_2]
$};

\node (2) [startstop, below of=1] {
$
 A_1:   [SU(2)] \,\, 2 \,\, 2 \,\, 1\,\, \overset{\mathfrak{e_{8}}}{12}  \,\, 1 \,\,  2 \,\, {\overset{\mathfrak{su_{2}}}2}  \,\,   {\overset{\mathfrak{g_2}}3}  \,\, 1 \,\, \overset{\mathfrak{f_{4}}}5 \,\, 1\,\, ... [G_2]
$};

\node (3) [startstop, below of=2] {
$
\widetilde A_1:   [SU(2)] \,\, 2 \,\, 1\,\, \overset{\mathfrak{e_{8}}}{11}  \,\, 1 \,\,  2 \,\, {\overset{\mathfrak{su_{2}}}2}  \,\,   {\overset{\mathfrak{g_2}}3}  \,\, 1 \,\, \overset{\mathfrak{f_{4}}}5 \,\, 1\,\, ...[G_2]
$};

\node (4) [startstop, below of=3] {
$
 G_2(a_1):   \overset{\mathfrak{e_{8}}}{9}  \,\, 1 \,\,  2 \,\, {\overset{\mathfrak{su_{2}}}2}  \,\,   {\overset{\mathfrak{g_2}}3}  \,\, 1 \,\, \overset{\mathfrak{f_{4}}}5 \,\, 1\,\, ...[G_2]
$};

\node (5) [startstop, below of=4] {
$
 G_2:   \overset{\mathfrak{e_{7}}}{8}  \,\, 1 \,\, {\overset{\mathfrak{su_{2}}}2}  \,\,   {\overset{\mathfrak{g_2}}3}  \,\, 1 \,\, \overset{\mathfrak{f_{4}}}5 \,\, 1\,\, ...[G_2]
$};

\draw [arrow] (1) -- (2);
\draw [arrow] (2) -- (3);
\draw [arrow] (3) -- (4);
\draw [arrow] (4) -- (5);

\end{tikzpicture}

\end{center}
\caption{Flows for $G_2$ nilpotent orbits.}
\label{fig:g2flows}
\end{figure}
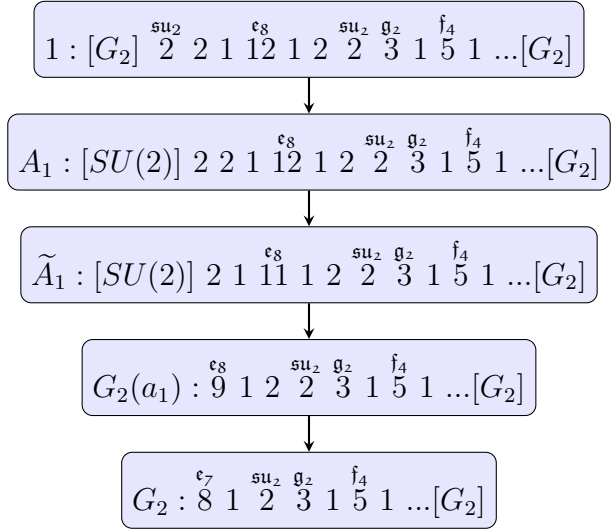

\newpage

\subsection{Flows from $\mathfrak{f}_{4}$, $\mathfrak{g}_{2}$}

\begin{figure}
\begin{center}

\begin{tikzpicture}[node distance=1.75cm]

\node (1) [startstop, xshift=-1cm] {
$
1: [F_4] \,\,  1 \,\, \overset{\mathfrak{g_{2}}}3 \,\, \overset{\mathfrak{su_{2}}}2 \,\, 2 \,\, 1\,\, \overset{\mathfrak{e_{8}}}{12}  \,\, 1 \,\,  ...[F_4]
$};

\node (2) [startstop, below of=1] {
$
A_1: [Sp(3)] \,\,   \overset{\mathfrak{g_{2}}}2 \,\, \overset{\mathfrak{su_{2}}}2 \,\, 2 \,\, 1\,\, \overset{\mathfrak{e_{8}}}{12}  \,\, 1 \,\,  ...[F_4]
$};

\node (3) [startstop, below of=2] {
$
\widetilde A_1: [SU(4)] \,\,   \overset{\mathfrak{su_{3}}}2 \,\, \overset{\mathfrak{su_{2}}}2 \,\, \overset{\mathfrak{su_{1}}}2 \,\, 1\,\, \overset{\mathfrak{e_{8}}}{12}  \,\, 1 \,\,  ...[F_4]
$};

\node (4) [startstop, below of=3] {
$
A_1 + \widetilde A_1: [SO(4)] \,\,   \overset{\mathfrak{su_{2}}}2 \,\, \underset{[N_f=1]}{\overset{\mathfrak{su_{2}}}2} \,\, \overset{\mathfrak{su_{1}}}2 \,\, 1\,\, \overset{\mathfrak{e_{8}}}{12}  \,\, 1 \,\,  ...[F_4]
$};

\node (5) [startstop, below of=4, xshift=-4cm] {
$
A_2:  \overset{\mathfrak{su_{1}}}2 \,\, \underset{[SU(3)]}{\overset{\mathfrak{su_{2}}}2} \,\, \overset{\mathfrak{su_{1}}}2 \,\, 1\,\, \overset{\mathfrak{e_{8}}}{12}  \,\, 1 \,\,  ...[F_4]
$};

\node (6) [startstop, below of=4, xshift=4cm] {
$
\widetilde A_2:   [G_2] \,\, \overset{\mathfrak{su_{2}}}2 \,\, \overset{\mathfrak{su_{1}}}2 \,\, 1\,\, \overset{\mathfrak{e_{8}}}{11}  \,\, 1 \,\,  ...[F_4]
$};

\node (7) [startstop, below of=6, yshift=-2cm] {
$
\widetilde A_2 +A_1:   [SU(2)] \,\, 2 \,\, 2 \,\, 1\,\, \overset{\mathfrak{e_{8}}}{11}  \,\, 1 \,\,  ...[F_4]
$};

\node (8) [startstop, below of=5] {
$
 A_2 + \widetilde A_1:   [SU(2)] \,\,2\,\, 2 \,\, 2 \,\, 1\,\, \overset{\mathfrak{e_{8}}}{12}  \,\, 1 \,\,  ...[F_4]
$};

\node (9) [startstop, below of=8] {
$
 B_2:   [SU(2)]  \,\, 2 \,\, 1\,\, \underset{[SU(2)]}{\underset{2}{\underset{1}{\overset{\mathfrak{e_{8}}}{12}}}}  \,\, 1 \,\,  ...[F_4]
$};

\node (10) [startstop, below of=9, xshift=5cm] {
$
 C_3(a_1):   [SU(2)]  \,\, 2 \,\, 1\,\, \overset{\mathfrak{e_{8}}}{10}  \,\, 1 \,\,  ...[F_4]
$};

\node (11) [startstop, below of=10, xshift=0cm] {
$
 F_4(a_3):   \overset{\mathfrak{e_{8}}}{8}  \,\, 1 \,\,  2 \,\, {\overset{\mathfrak{su_{2}}}2}  \,\,   {\overset{\mathfrak{g_2}}3}  \,\, 1 \,\, \overset{\mathfrak{f_{4}}}5 \,\, 1\,\, ...[F_4]
$};

\node (12) [startstop, below of=11, xshift=-4cm] {
$
 B_3:   \overset{\mathfrak{e_{7}}}{8}  \,\, \underset{[SU(2)]}1 \,\,  2 \,\, {\overset{\mathfrak{su_{2}}}2}  \,\,   {\overset{\mathfrak{g_2}}3}  \,\, 1 \,\, \overset{\mathfrak{f_{4}}}5 \,\, 1\,\, ...[F_4]
$};

\node (13) [startstop, below of=11, xshift=4cm] {
$
C_3:  [SU(2)] \,\, 1 \,\, \overset{\mathfrak{e_{7}}}{8}  \,\, 1 \,\,  {\overset{\mathfrak{su_{2}}}2}  \,\,   {\overset{\mathfrak{g_2}}3}  \,\, 1 \,\, \overset{\mathfrak{f_{4}}}5 \,\, 1\,\, ...[F_4]
$};

\node (14) [startstop, below of=13, xshift=-4cm, yshift=.1cm] {
$
F_4(a_2):  \overset{\mathfrak{e_{7}}}{7}  \,\, 1 \,\,  {\overset{\mathfrak{su_{2}}}2}  \,\,   {\overset{\mathfrak{g_2}}3}  \,\, 1 \,\, \overset{\mathfrak{f_{4}}}5 \,\, 1\,\, ...[F_4]
$};

\node (15) [startstop, below of=14, xshift=0cm, yshift=.2cm] {
$
F_4(a_1):  {\overset{\mathfrak{e_{6}}}{6}}  \,\, 1 \,\,   {\overset{\mathfrak{su_3}}3}  \,\, 1 \,\, \overset{\mathfrak{f_{4}}}5 \,\, 1\,\, ...[F_4]
$};

\node (16) [startstop, below of=15, xshift=0cm, yshift=.3cm] {
$
F_4:  \overset{\mathfrak{f_{4}}}5   \,\, 1 \,\,   {\overset{\mathfrak{g_2}}3}  \,\,  {\overset{\mathfrak{su_{2}}}2} \,\, 2 \,\, 1 \,\,  \overset{\mathfrak{e_{8}}}{11} \,\, 1\,\, ...[F_4]
$};

\draw [arrow, color=blue] (1) -- (2);
\draw [arrow] (2) -- (3);
\draw [arrow] (3) -- (4);
\draw [arrow] (4) -- (5);
\draw [arrow] (4) -- (6);
\draw [arrow] (5) -- (8);
\draw [arrow] (6) -- (7);
\draw [arrow] (8) -- (7);
\draw [arrow] (8) -- (9);
\draw [arrow] (7) -- (10);
\draw [arrow] (9) -- (10);
\draw [arrow] (10) -- (11);
\draw [arrow, color=blue] (11) -- (12);
\draw [arrow, color=blue] (11) -- (13);
\draw [arrow, color=blue] (12) -- (14);
\draw [arrow, color=blue] (13) -- (14);
\draw [arrow, color=blue] (14) -- (15);
\draw [arrow, color=blue] (15) -- (16);

\end{tikzpicture}

\end{center}
\caption{Flows for $F_4$ nilpotent orbits. Blue arrows indicate flows where one or more free tensors appears in the IR.}
\label{fig:f4flows}

\end{figure}
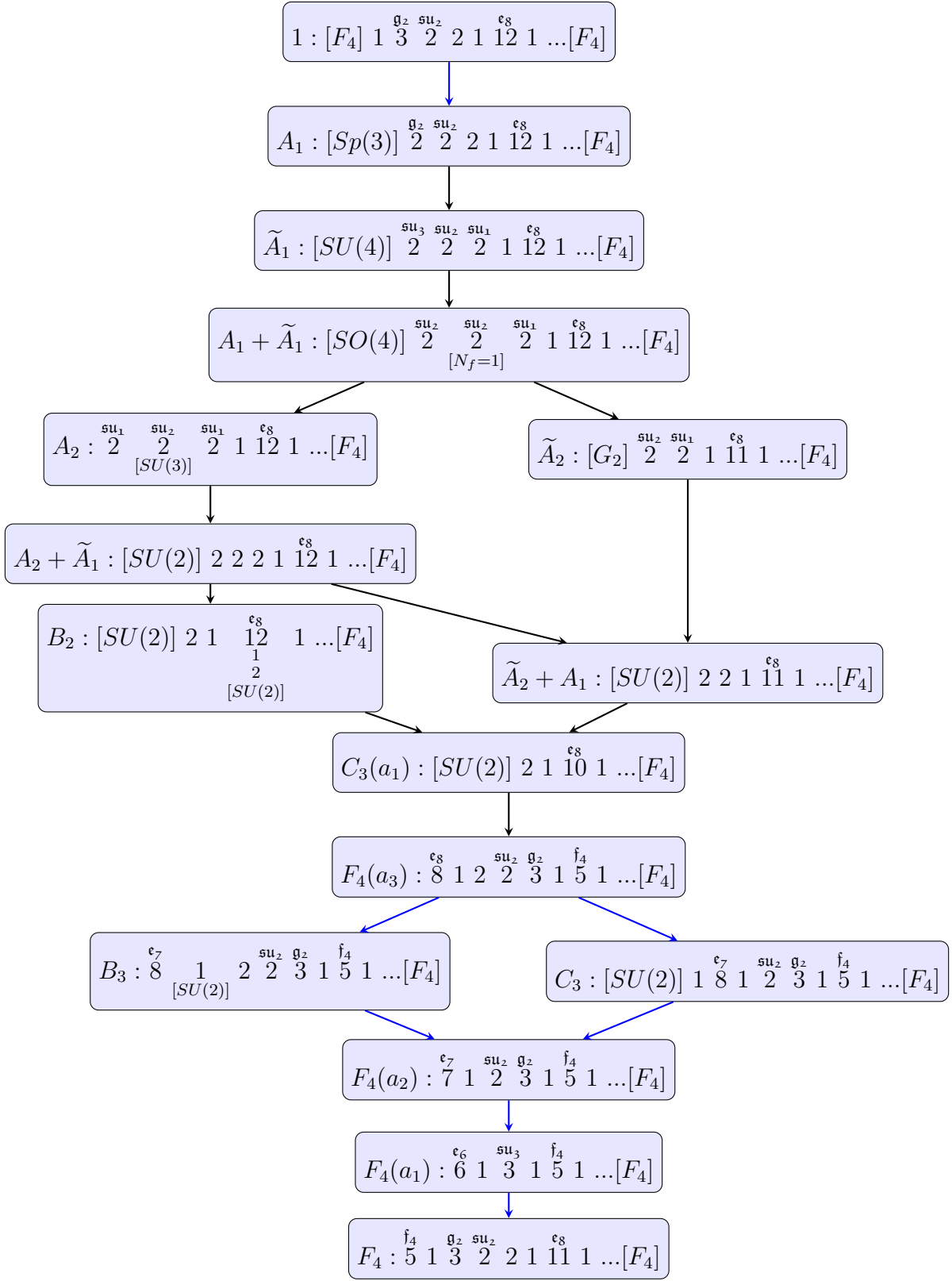

Finally, as a last class of examples, we also consider flows
induced by nilpotent orbits for the non-simply laced algebras $\mathfrak{f}%
_{4}$ and $\mathfrak{g}_{2}$. Actually, we can reach all of these flows by
first considering a nilpotent orbit which has commutant subalgebra
$\mathfrak{f}_{4}$ and $\mathfrak{g}_{2}$, and then adding an additional
nilpotent element which embeds in this subalgebra. This is quite similar to
our analysis of flavor symmetries of $\mathfrak{so}_{\text{odd}}$ type. Alternatively, we
can work out the F-theory geometries obtained from such nilpotent
orbits. The results of this final set of analyses, along with the partially
ordered set of RG flows / nilpotent elements is displayed in figures \ref{fig:g2flows} and \ref{fig:f4flows}.

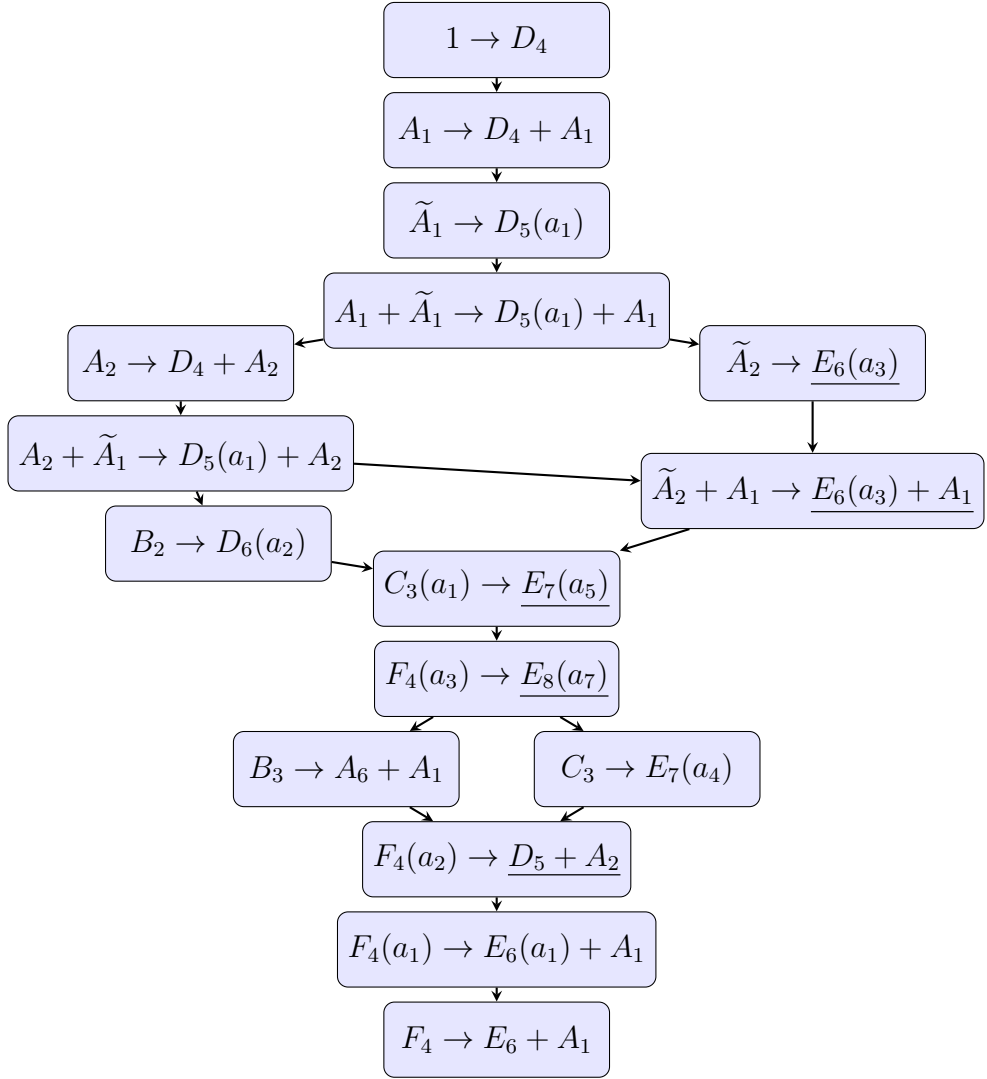
\begin{figure}[h!]
\begin{center}

\begin{tikzpicture}[node distance=1.2cm]

\node (1) [startstop, xshift=-1cm] {
$
1 \to D_4 $};

\node (2) [startstop, below of=1] {
$
A_1 \to D_4 + A_1
$};

\node (3) [startstop, below of=2] {
$
\widetilde A_1 \to D_5(a_1)
$};

\node (4) [startstop, below of=3] {
$
A_1 + \widetilde A_1 \to D_5(a_1)+A_1
$};

\node (5) [startstop, below of=4, xshift=-4.2cm, yshift=.5cm] {
$
A_2\to D_4+ A_2
$};

\node (6) [startstop, below of=4, xshift=4.2cm, yshift=.5cm] {
$
\widetilde A_2\to \underline{E_6(a_3)}
$};

\node (7) [startstop, below of=6, yshift=-.5cm] {
$
\widetilde A_2 +A_1\to \underline{E_6(a_3)+A_1}
$};

\node (8) [startstop, below of=5] {
$
 A_2 + \widetilde A_1\to D_5(a_1)+A_2
$};

\node (9) [startstop, below of=8,xshift=.5cm] {
$
 B_2 \to D_6(a_2)
$};

\node (10) [startstop, below of=4, yshift=-2.5cm] {
$
 C_3(a_1)\to \underline{E_7(a_5)}
$};

\node (11) [startstop, below of=10, xshift=0cm] {
$
 F_4(a_3)\to \underline{E_8(a_7)}
$};

\node (12) [startstop, below of=11, xshift=-2cm] {
$
 B_3 \to A_6+A_1
$};

\node (13) [startstop, below of=11, xshift=2cm] {
$
C_3\to E_7(a_4)
$};

\node (14) [startstop, below of=13, xshift=-2cm] {
$
F_4(a_2)\to \underline{D_5+A_2}
$};

\node (15) [startstop, below of=14, xshift=0cm] {
$
F_4(a_1)\to E_6(a_1)+A_1
$};

\node (16) [startstop, below of=15, xshift=0cm] {
$
F_4 \to E_6+A_1
$};

\draw [arrow] (1) -- (2);
\draw [arrow] (2) -- (3);
\draw [arrow] (3) -- (4);
\draw [arrow] (4) -- (5);
\draw [arrow] (4) -- (6);
\draw [arrow] (5) -- (8);
\draw [arrow] (6) -- (7);
\draw [arrow] (8) -- (7);
\draw [arrow] (8) -- (9);
\draw [arrow] (7) -- (10);
\draw [arrow] (9) -- (10);
\draw [arrow] (10) -- (11);
\draw [arrow] (11) -- (12);
\draw [arrow] (11) -- (13);
\draw [arrow] (12) -- (14);
\draw [arrow] (13) -- (14);
\draw [arrow] (14) -- (15);
\draw [arrow] (15) -- (16);

\end{tikzpicture}

\end{center}
\caption{\small For each of the theories of figure \ref{fig:f4flows}, we show here the corresponding theory in the $\mathfrak{e}_8$ nilpotent hierarchy of the table in appendix \ref{app:E8}. This realizes the $\mathfrak{f}_4$ nilpotent hierarchy as a sub-hierarchy of the $\mathfrak{e}_8$ one. The underlined labels realize in a similar way the $\mathfrak{g}_2$ hierarchy of figure \ref{fig:g2flows}.}
\label{fig:subF4G2}
\end{figure}

As a curiosity, we also notice that the diagrams for  $\mathfrak{f}_{4}$ and $\mathfrak{g}_{2}$ can be embedded into the one for $\mathfrak{e}_8$. The reason is that both  $\mathfrak{f}_{4}$ and $\mathfrak{g}_{2}$  appear in the $E_8-E_8$ conformal matter theory. In the $\mathfrak{e}_8$ nilpotent hierarchy, the theory labeled $D_4$ (see the table of appendix \ref{app:E8}), for example, is almost identical to the theory labeled 1 in figure \ref{fig:f4flows}; the only difference is that the leftmost $\mathfrak{e}_8$ is on a $-11$ curve rather than on a $-12$ curve. Starting from this $D_4$ theory, then, we can reproduce all the flows that appear in the $\mathfrak{f}_4$ diagram of figure \ref{fig:f4flows}; the theories of that figure have almost identical avatars in the $\mathfrak{e}_8$ nilpotent hierarchy. We show the correspondence in figure \ref{fig:subF4G2}. One can check in \cite[Table 19]{Chacaltana:2012zy} that the theories shown in that diagram are indeed in the correct inclusion relation for $\mathfrak{e}_8$. Thus, the $\mathfrak{f}_4$ nilpotent hierarchy is isomorphic to a sub-hierarchy of the $\mathfrak{e}_8$ nilpotent hierarchy. Similarly, one can check that the $\mathfrak{g}_2$ is also a sub-hierarchy of the $\mathfrak{f}_4$ hierarchy, as also summarized in figure \ref{fig:subF4G2}.

\newpage

\section{Short Quivers} \label{sec:SHORT}

Up to this point, we have assumed that the generalized quivers of our 6D SCFTs were sufficiently long to Higgs the left and right of the quiver independently. Strictly speaking, even when this is not the case we can continue to parameterize all flows according to two independent nilpotent orbits. However, the resulting flow will then contain various redundancies since the data associated with this pair will inevitably become correlated. Our plan in this section will be to extend our analysis of flows to theories where this happens, which we will call ``short quivers.''

The picture is clearest in the case of flows from $\mathfrak{su}_N$.  Here, the allowed Higgsings are characterized by a partition on the left of the quiver and a partition on the right.  The non-redundant data of such flows is captured by a pair of partitions of equal size. Moreover, each column of a partition corresponds to the change in gauge group rank between neighboring nodes.  If there are $(k-1)$ tensor multiplets in the theory, then there can be up to $k$ changes in the rank of the associated symmetry algebra (including the leftmost and rightmost flavor symmetries). So, there are at most total $k$ columns in the two partitions.  For a large quiver $k \gg N$, and the restriction on the number of columns of the partition simply comes from the size of each partition, $N$.  For small quivers, on the other hand, the requirement that the total number or columns should be at most $k$ places important constraints.

As an example, we list the theories with three tensor multiplets and partitions of size three:
\begin{gather}
(1^3): [SU(3)] \,\, \overset{\mathfrak{su3}}2 \,\,\overset{\mathfrak{su3}}2 \,\, \overset{\mathfrak{su3}}2 \,\, [SU(3)] : (1^3) \\
(1^3): [SU(3)] \,\, \overset{\mathfrak{su3}}2 \,\, \underset{[N_f=1]}{\overset{\mathfrak{su3}}2} \,\, \overset{\mathfrak{su2}}2 \,\, [SU(1)] : (2,1)\\
(1^3): [SU(4)] \,\, \overset{\mathfrak{su3}}2 \,\,\overset{\mathfrak{su2}}2 \,\, \overset{\mathfrak{su1}}2  : (3) \\
(2,1): [SU(1)] \,\, \overset{\mathfrak{su2}}2 \,\, \underset{[SU(2)]}{\overset{\mathfrak{su3}}2} \,\, \overset{\mathfrak{su2}}2 \,\, [SU(1)] : (2,1)
\end{gather}
where for the purposes of uniformity with higher rank examples we have listed the (trivial) flavor symmetry factor $SU(1)$ which in
F-theory is associated with a component of the discriminant locus with $I_1$ fiber type.

For longer quivers, we could also consider the flows corresponding to partitions $\mu_L = (3), \mu_R =(2,1)$ and $\mu_L=(3) , \mu_R=(3)$.  However, since we only have three hypermultiplets in the case at hand, we are constrained to consider pairs of partitions with no more than four columns, so we need not concern ourselves with such flows.

Similar comments apply for the $BCDEFG$ theories.  We illustrate it with a discussion of $E_6$ nilpotent orbits.  Here, the analog to the ``number of columns of the partition" in the $\mathfrak{su}_N$ case is the distance that the breaking pattern propagates into the interior of the quiver, that is, the number of $E_6$ gauge group factors which are (partially) broken.  For instance, the nilpotent orbits in figure \ref{fig:E6} with B--C labels $0, A_1, 2 A_1, 3 A_1, A_2, A_1+A_1$, and $A_2 + 2 A_1$ do not introduce any breaking into the interior of the quiver.  Even for a theory with a single $\mathfrak{e}_6$ node, it is possible to trigger an RG flow from any of these nilpotent orbits on the left or the right.  Two such examples are
\begin{gather}
A_1 : [SU(6)] \,\, \overset{\mathfrak{su}_3}2 \,\, 1 \,\, \underset{[N_f=1]}{\overset{\mathfrak{e}_6}5} \,\, 1 \,\, [SU(3)] : A_2 + A_1 \\
3 A_1: [SU(2)] \,\, 2 \,\, \underset{[SU(3)]}1 \,\, {\overset{\mathfrak{e}_6}6} \,\, 1 \,\, \overset{\mathfrak{su}_3}3 \,\, 1 \,\,[E_6] : 0
\end{gather}
On the other hand, nilpotent orbits such as the one of B--C label $D_5$ propagate several nodes into the interior of the quiver.  For quivers with a single $\mathfrak{e}_6$ node, we can ignore these nilpotent elements.

\section{Global Symmetries in 6D SCFTs} \label{sec:FLAVOR}

One of the important aspects of the characterization of RG flows in terms of nilpotent orbits is that this is \textit{algebraic} data directly
associated with a conformal fixed point. Assuming the absence of an emergent flavor symmetry in the IR, we can then use the labelling by nilpotent orbits to read off the flavor symmetry for IR fixed points.

Indeed, we have performed a match between a particular class of 6D SCFTs and nilpotent orbits for classical and exceptional algebras.  In many cases, the  global symmetry which is manifest on the tensor branch matches to what is expected from the nilpotent orbit.
An example is the theory
\begin{equation}
[SU(6)] \,\,  \overset{\mathfrak{su_{3}}}2  \,\, 1 \,\, \overset{\mathfrak{e_{6}}}6  \,\, 1 \,\,  \overset{\mathfrak{su_{3}}}3 \,\, 1  \,\, ...[E_6],
\end{equation}
which corresponds to the nilpotent orbit of $E_6$ with B--C label $A_1$.  However, there are other instances in which the global symmetry of a 6D SCFT cannot be easily determined from the theory on the tensor branch.  In particular, as discussed in \cite{Bertolini:2015bwa}, there are instances in which the expected field theoretic global symmetry does not match the global symmetry predicted by F-theory.  An example is the theory with tensor branch,
\begin{equation}
[SO(7)] \,\,  \overset{\mathfrak{su_{2}}}2  \,\, 1 \,\, \overset{\mathfrak{e_{6}}}6  \,\, 1 \,\,  \overset{\mathfrak{su_{3}}}3 \,\, 1  \,\, ...[E_6].
\label{so7example}
\end{equation}
This is the theory associated with nilpotent orbit of $E_6$ with B--C label $2 A_1$.  The ``na\"ive'' field theoretic expectation is that there should be an $SO(8)$ acting on the eight half-hypermultiplets of $SU(2)$, whereas F-theory only permits an $\mathfrak{so}(7)$ flavor curve to meet the $\mathfrak{su}(2)$ gauge algebra.  However, in \cite{Ohmori:2015pia}, it was argued that the na\"ive field theoretic expectation is wrong in this instance, and the correct global symmetry of the field theory matches the prediction from F-theory, with the eight half-hypermultiplets transforming in the spinor of $\mathfrak{so}(7)$.  We note that this also matches the global symmetry predicted from the data of the corresponding nilpotent orbit.

This example dealt with the simple case of an $I_2$ Kodaira fiber type over the leftmost $-2$ curve. But the business of determining global symmetries for 6D SCFTs becomes even more involved once we consider theories with $I_1$, $II$, $III$, and $IV$ fiber types.  The fibers $I_0$, $I_1$, and $II$ all lead to trivial gauge algebras; $I_2$ and $III$ both lead to $\mathfrak{su}(2)$ gauge algebras; and the split $I_3$ and $IV$ fibers both lead to $\mathfrak{su}(3)$ gauge algebras. Nevertheless, the expectation from geometry is that they lead to different global symmetries \cite{Bertolini:2015bwa, Morrison:2016djb}.  This leads to the natural question: do theories with distinct fiber types but identical gauge algebras lead to distinct 6D SCFTs?  If not, what is the correct global symmetry for these theories?  If so, does the F-theory prediction always match the global symmetry seen in field theory?

The analysis of the present paper sheds light on these questions.  We expect that the continuous component of the global symmetry of a 6D SCFT can be read off directly from the commutant of the nilpotent orbit.  Indeed, in all cases in which the global symmetry of the 6D SCFT is well understood, including the subtle case of line (\ref{so7example}), we find this is indeed the case.\footnote{Note that this match holds for $SO(2N+1)$ nilpotent orbits only after we take into account the subtlety of $SO(2N+1) \subset SO(2N+2p)$ for small $p$ discussed in section \ref{ssec:SOodd}.}  Under the assumption that this holds generally, we compare the global symmetries of the 6D SCFTs to the F-theory prediction.  We find that the global symmetry group of a 6D SCFT always contains the global symmetry group predicted by F-theory, and in many cases this containment is proper.  We also find no evidence that theories with identical gauge algebras but distinct fiber types should correspond to distinct 6D SCFTs up to different numbers of free hypermultiplets.

For a first example, consider the theory corresponding to the $E_7$ nilpotent orbit of B--C label $A_3+A_2+A_1$,
$$[SU(2)] \,\,    {\overset{\mathfrak{e_{7}}}5}  \,\, 1 \,\,  ...$$
The global symmetry here is evidently $SU(2)$, rotating the three half-hypermultiplets of $\mathfrak{e}_7$ as a triplet, but as was shown in \cite{Bertolini:2015bwa}, F-theory does not permit any flavor curves to meet a curve carrying gauge algebra $\mathfrak{f}_4, \mathfrak{e}_6, \mathfrak{e}_7$, or $\mathfrak{e_8}$.  Instead, it appears that a flavor symmetry emerges at the origin of the tensor branch (i.e. the SCFT point of the moduli space), matching the field-theoretic expectation (c.f. Table 5.1 of \cite{Bertolini:2015bwa}) rather than the F-theory prediction.

A similar story arises in the case of the $E_7$ theory corresponding to B--C label $2A_2$.  This theory has $G_2 \times SU(2)$ global symmetry.  The gauge algebras of the theory may be realized in several different ways within F-theory, two of which are as follows:
$$
 [I_0^{*,ns}]  \,\, {\overset{IV^{ns}}2}  \,\, {\overset{II}2}\,\,  \overset{I_0}1 \,\, \overset{III^*}8  \,\, ...
$$
\begin{equation}
 [I_3]  \,\, {\overset{I_2}2}  \,\, {\overset{I_1}2}\,\,  \underset{[I_2]}{\overset{I_0}1} \,\, \overset{III^*}8  \,\, ...
\label{2A2}
\end{equation}
Here, the Kodaira fiber types in brackets are supported on non-compact flavor curves.  The first theory has a $G_2$ flavor symmetry living on the non-compact curve with fiber type $I_0^{*,ns}$, but no non-Abelian flavor curve may touch the curve of self-intersection $-1$ with $I_0$ fiber type \cite{miranda1986extremal,miranda1990persson,persson1990configurations}.  In the second case, on the other hand, an $SU(2)$ flavor curve of Kodaira type $I_2$ does touch the $I_0$ curve, but the global symmetry on the left is reduced from $G_2$ to $SU(3)$.  Thus, there is one F-theory configuration in which the $G_2$ flavor symmetry on the left is apparent and one F-theory configuration in which the $SU(2)$ flavor symmetry below is apparent, but there is no F-theory configuration in which the full $G_2 \times SU(2)$ symmetry is realized.  It appears that upon flowing to the IR, the flavor symmetry acting on the hypermultiplets of this theory is the maximal symmetry group acting on those hypermultiplets in any F-theory realization of the model.

Consider next the theory corresponding to B--C label $ A_2 + 2 A_1$:
$$
  [SO(4)]  \,\, {\overset{\mathfrak{su_{2}}}2}  \,\,  \underset{[SU(2)]}{\overset{\mathfrak{su_{2}}}2}\,\, 1 \,\, \overset{\mathfrak{e_{7}}}8  \,\, 1 \,\,  ...
$$
Here, the flavor symmetry expected from F-theory is simply $SU(2) \times SU(2)$, coming from a non-compact $I_2$ flavor curve hitting each of the two $-2$ curves with $\mathfrak{su}_2$ gauge algebras.  However, the symmetry is enhanced from $\mathfrak{su}_2 \times \mathfrak{su}_2$ to $\mathfrak{su}_2 \times \mathfrak{su}_2 \times \mathfrak{su}_2$.

The theory corresponding to the $E_8$ orbit with B--C label $D_4 + A_2$ has an $SU(3)$ global symmetry:
$$
 2\,\, \underset{[SU(3)]}{\overset{\mathfrak{su}_2}2 } \,\,  2  \,\,  1\,\, \overset{\mathfrak{e_{8}}}{11}  \,\, 1 \,\,  ... [E_8]
$$
We should think of this $SU(3)$ as rotating three hypermultiplets charged under the $\mathfrak{su}_2$ gauge symmetry.  An additional half-hypermultiplet of the $\mf{su_2}$ lives at the intersection with each unpaired $-2$ tensor.

Another important point is that theories with identical gauge algebras never show up as distinct nilpotent orbits.  The two F-theory models of (\ref{2A2}) provide one such example.  Another particularly interesting case is the $E_7$ nilpotent orbit with B--C label $A_3 + 2 A_1$:
$$
[SU(2)] \,\, 2 \,\,  \underset{[SU(2)]}1 \,\, \overset{\mathfrak{e_{7}}}7  \,\, 1 \,\,  ...
$$
The gauge algebras shown can be realized in F-theory with either a $I_0$ fiber, an $I_1$ fiber, or a $II$ fiber on the empty $-2$ curve.  The fact that these do not correspond to different nilpotent orbits of $E_7$ is a possible indication that all three of these F-theory realizations give the same 6D SCFT up to different numbers of free hypermultiplets.\footnote{The requirement that Higgs branch flows preserve gravitational anomalies fixes the number of free hypermultiplets, which means that our RG flow analysis will be unable to distinguish between two F-theory models that give 6D SCFTs differing only by a number of free hypermultiplets.}  If we decompactify all base curves besides this $-2$ curve (corresponding to a flow along the tensor branch), we are left with a theory of just a $-2$ curve of fiber type $I_0$, $I_1$, and $II$, respectively.  Assuming that all three of these fiber types do indeed give the same 6D SCFT before this tensor branch flow, we find that the resulting 6D SCFTs after the flow must be identical as well (modulo free hypermultiplets).  Thus, we conjecture that the interacting sector of these three theories are the same and given by the $A_1$ $(2,0)$ 6D SCFT.

Of course, the other possibility is that these distinct F-theory models do give rise to distinct 6D SCFTs, but that only one of them can be realized by an RG flow parameterized by a nilpotent orbit.  This would itself be a rather surprising result.  Determining which solution is the correct one is left as a question for future study.

As a final set of comments, we note that we have also presented evidence for IR fixed points with \textit{abelian} flavor symmetries, a fact which is quite straightforward using the algebraic data of nilpotent orbits. By contrast, identifying such symmetry factors from a geometric perspective can sometimes be subtle. Roughly speaking, we would like to associate such abelian symmetry factors with non-compact components of the discriminant locus supporting a singular $I_1$ fiber. Observe, however, that at least for \textit{gauge} theories (i.e. fibers supported on compact curves), an $I_n$ fiber is expected to realize an $\mathfrak{su}_n$ rather than $\mathfrak{u}_n$ gauge algebra. The distinction boils down to the fact that for a 6D SCFT on its tensor branch, this additional $\mathfrak{u}(1)$ factor is anomalous, and so inevitably decouples anyway via the St\"uckelberg mechanism.
For flavor symmetries, however, there is a priori no such issue. Indeed, in many of the examples encountered earlier, we can clearly see that the presence of an additional $\mathfrak{u}(1)$ correlates tightly with such $I_n$ fibers. We have also seen that in some breaking patterns, there is an overall tracelessness condition, for example with flavor symmetry algebras such as $\mathfrak{s}(\mathfrak{u}(n_1) \oplus ... \oplus \mathfrak{u}(n_l))$. We take this to mean that these $\mathfrak{u}(1)$ flavor symmetries can in general be delocalized in the geometry, that is, they are spread over multiple components of the discriminant locus.

For this reason, we have not assigned the presence of $\mathfrak{u}(1)$'s to specific locations in the diagrams of figures \ref{fig:so8flows}--\ref{fig:f4flows}, as we did for non-abelian symmetries. Their presence can be read off from (\ref{eq:soflavor}) and (\ref{eq:spflavor}) for figures \ref{fig:so8flows}--\ref{fig:so9flows}, and is shown explicitly in the tables of appendix \ref{app:nilp}. In many cases, there is a clear guess as to the origin of the abelian symmetries, coming from the presence of a hypermultiplet localized at the collision of a compact curve with a non-compact curve. In other cases, they are associated to an E-string which has a gauged subgroup of $E_8$ whose commutant has one or more $\mathfrak{u}(1)$. For example, for the $E_6$ nilpotent orbit $2A_1$ in appendix \ref{app:E6} (or figure \ref{fig:E6}) we see an E-string with gauged subalgebra $\mathfrak{su}_2 \oplus \mathfrak{e}_6$; or in theory $(3^2,1^2)$  we see an E-string with
gauged subalgebra $\mathfrak{su}_3 \oplus \mathfrak{so}_8$.

It would be interesting to further explore the extent to which such abelian flavor symmetry factors (both continuous and discrete) can be deduced more directly from the geometric perspective.

\section{Conclusions \label{sec:CONC}}

In this note we have studied renormalization group flows between 6D\ SCFTs
induced by vevs for conformal matter. Focusing on the case of
\textquotedblleft T-brane vacua\textquotedblright\ i.e.~those vacua labeled
by the orbits of nilpotent elements of a flavor symmetry algebra, we have
first of all established a direct correspondence between certain nilpotent
orbits, and a class of F-theory geometries. An important aspect of this
analysis is that the natural notion of partial ordering of elements in the
nilpotent cone of a simple Lie algebra has a direct physical interpretation in
terms of hierarchies of renormalization group flows. Moreover, we have also
used this algebraic data to calculate the unbroken flavor symmetry of the IR fixed point.
To reinforce this point, we have considered explicit examples
of generalized quiver theories with flavor symmetries of type ABCDEFG. We have used these examples to study global symmetries in 6D SCFTs, finding that the global symmetry read off from the nilpotent orbit can be larger than the global symmetry predicted from F-theory.  In the
remainder of this section we discuss some avenues for future investigation.

In the case of $\mathfrak{su}_N$ and $\mathfrak{so}_{\text{even}}$ theories, we remarked that by
taking transposed partitions, our nilpotent hierarchy of RG flows
extends to flows between theories of different maximal gauge group rank such as
\begin{equation}
[SU_{10}]-SU_{10}-...-SU_{10}-[SU_{10}]\rightarrow\lbrack SU_{9}]-SU_{9}-...-SU_{9}%
-[SU_{9}].
\end{equation}
It would be interesting to extend this analysis to exceptional algebras. Establishing this
sort of correspondence in more detail would provide an opportunity to
potentially map out the full class of possible RG\ flows from a UV\ parent
theory. This would bring us significantly closer to the ambitious goal of
classifying \textit{all} RG\ flows between 6D\ SCFTs.

In our analysis, we primarily focused on theories which have a sufficiently
large number of tensor multiplets. Indeed, the parent theories we have started
with all have known holographic duals which take the form $AdS_{7}\times
S^{4}/\Gamma_{ADE}$. The effects of the nilpotent element vevs are primarily
confined to a small region of the quiver theory, which in the holographic dual
will correspond (in units where the radius of the sphere is one) to an order
$1/N$ size effect. It would be quite interesting to confirm this picture
directly in the holographic dual, perhaps by evaluating a protected quantity
such as the conformal anomalies of the 6D\ SCFT.

Finally, it would be interesting to also study how the data of conformal matter vevs as
parameterized by nilpotent orbits shows up in little string theories (see e.g. \cite{Bhardwaj:2015oru}).
We arrive at examples of little string theories by compactifying M5-branes on the background
$S^{1}\times\mathbb{C}^{2}/\Gamma_{ADE}$. When we do so, the independent data
about partitions used to label possible flows are now identified, and always appear with
gauge group factors rather than flavor group factors (there are none for the circular quivers).
This in turn means that the purely local perturbations induced by a choice of partition now
propagate out to the entire generalized quiver, providing a rather novel window
into flows for more general 6D theories.

\section*{Acknowledgements}

We thank J.~Distler, T.~Dumitrescu, S.~Gukov, N.~Mekareeya, D.R.~Morrison and C.~Vafa for helpful discussions. We
also thank the Simons Center for Geometry and Physics 2015 summer workshop for
hospitality during the initial stages of this project. JJH also thanks the theory groups
at Columbia University, the ITS at the CUNY\ graduate center, and the CCPP at
NYU for hospitality during the completion of this work. The work of JJH is supported by NSF CAREER grant PHY-1452037. JJH
also acknowledges support from the Bahnson Fund at UNC Chapel Hill as well as the
R.~J. Reynolds Industries, Inc. Junior Faculty Development Award from the Office
of the Executive Vice Chancellor and Provost at UNC Chapel Hill.
The work of TR is supported by NSF grant PHY-1067976. TR is also supported by the NSF GRF under
DGE-1144152. AT is supported in part by INFN, by the MIUR-FIRB grant RBFR10QS5J \textquotedblleft String
Theory and Fundamental Interactions\textquotedblright, and by the European
Research Council under the European Union's Seventh Framework
Program (FP/2007-2013) - ERC Grant Agreement n. 307286 (XD-STRING).


\appendix

\newpage

\section{Nilpotent Flows for E-type Flavor Symmetries}\label{app:nilp}

In this Appendix we collect the full list of nilpotent orbits for exceptional E-type flavor symmetries, and the corresponding
F-theory model associated with each such flow. We also present the unbroken flavor symmetry for each such model which is predicted by
the choice of a nilpotent element.

\subsection{$E_6$ Nilpotent Orbits}\label{app:E6}

The $E_6$ Nilpotent orbits are as follows. The nilpotent hierarchy is given in figure \ref{fig:E6}.

\begin{longtable}{c|c|c}
B--C Label & Global Symmetry & Theory \\\hline
0 & $E_6$ & $[E_6] \,\, 1 \,\, \overset{\mathfrak{su_{3}}}3  \,\, 1 \,\, \overset{\mathfrak{e_{6}}}6  \,\, 1 \,\,  \overset{\mathfrak{su_{3}}}3 \,\, 1  \,\, ...[E_6]$\\\hline

$A_1$& $SU(6)$& $[SU(6)] \,\,  \overset{\mathfrak{su_{3}}}2  \,\, 1 \,\, \overset{\mathfrak{e_{6}}}6  \,\, 1 \,\,  \overset{\mathfrak{su_{3}}}3 \,\, 1  \,\, ...[E_6]$\\\hline

$2 A_1$ & $ Spin(7)\times U(1)$ & $[SO(7)] \,\,  \overset{\mathfrak{su_{2}}}2  \,\, 1 \,\, \overset{\mathfrak{e_{6}}}6  \,\, 1 \,\,  \overset{\mathfrak{su_{3}}}3 \,\, 1  \,\, ...[E_6]$\\\hline

$3 A_1 $ &$ SU(3) \times SU(2)$ & $[SU(2)] \,\,  2  \,\, \underset{[SU(3)]}1 \,\, \overset{\mathfrak{e_{6}}}6  \,\, 1 \,\,  \overset{\mathfrak{su_{3}}}3 \,\, 1  \,\, ...[E_6]$\\\hline

$A_2$& $SU(3) \times SU(3)$ & $[SU(3)] \,\, 1 \,\, \underset{[SU(3)]}{\underset{1}{\overset{\mathfrak{e_{6}}}6}}  \,\, 1 \,\,  \overset{\mathfrak{su_{3}}}3 \,\, 1  \,\, ...[E_6]$\\\hline

$A_2 + A_1$& $SU(3)\times U(1)$& $[SU(3)] \,\, 1 \,\, {\underset{[N_f=1]}{\overset{\mathfrak{e_{6}}}5}}  \,\, 1 \,\,  \overset{\mathfrak{su_{3}}}3 \,\, 1  \,\, ...[E_6]$\\\hline

$2 A_2$& $G_2$&  $[G_2] \,\, 1 \,\, {\overset{\mathfrak{f_{4}}}5}  \,\, 1 \,\,  \overset{\mathfrak{su_{3}}}3 \,\, 1  \,\,\overset{\mathfrak{e_{6}}}6  \,\, 1 \,\,  ...[E_6]$\\\hline

$A_2 +2 A_1$ & $ SU(2)\times U(1)$& $[SU(2)]\,\,{\overset{\mathfrak{e_{6}}}4}  \,\, 1 \,\,  \overset{\mathfrak{su_{3}}}3 \,\, 1  \,\, \overset{\mathfrak{e_{6}}}6  \,\, 1 \,\,  ...[E_6]$\\\hline

$2 A_2 + A_1 $ & $ SU(2)$ & $ [SU(2)]\,\,{\overset{\mathfrak{f_{4}}}4}  \,\, 1 \,\,  \overset{\mathfrak{su_{3}}}3 \,\, 1  \,\,\overset{\mathfrak{e_{6}}}6  \,\, 1 \,\,  ...[E_6]$\\\hline

$A_3$ &  $Sp(2)\times U(1)$ & $ [Sp(2)] \,\, {\overset{\mathfrak{so_{10}}}4}  \,\, 1 \,\,  \overset{\mathfrak{su_{3}}}3 \,\, 1  \,\, \overset{\mathfrak{e_{6}}}6  \,\, 1 \,\, ...[E_6]$\\\hline

$A_3+A_1$ & $ SU(2)\times U(1)$ & $ [SU(2)] \,\, {\overset{\mathfrak{so_{9}}}4}  \,\, 1 \,\,  \overset{\mathfrak{su_{3}}}3 \,\, 1  \,\, \overset{\mathfrak{e_{6}}}6  \,\, 1 \,\, ...[E_6]$\\\hline

$D_4(a_1)$ & $U(1)^2$ & $ {\overset{\mathfrak{so_{8}}}4}  \,\, 1 \,\,  \overset{\mathfrak{su_{3}}}3 \,\, 1  \,\,\overset{\mathfrak{e_{6}}}6  \,\, 1 \,\,  ...[E_6]$\\\hline

$A_4$ & $SU(2)\times U(1)$&$ [SU(2)] \,\, {\overset{\mathfrak{so_{7}}}3}  \,\,   \overset{\mathfrak{su_{2}}}2 \,\, 1  \,\,\overset{\mathfrak{e_{6}}}6  \,\, 1 \,\,  ...[E_6]$\\\hline

$A_4+A_1 $ & $U(1) $ &  $ {\overset{\mathfrak{g_{2}}}3}  \,\,   {\overset{\mathfrak{su_{2}}}2} \,\, 1  \,\,\overset{\mathfrak{e_{6}}}6  \,\, 1 \,\,  ...[E_6]$\\\hline

$D_4$ & $ SU(3)$ &  $ {\overset{\mathfrak{su_{3}}}3}  \,\,   1  \,\, \underset{[SU(3)]}{\underset{1}{\overset{\mathfrak{e_{6}}}6}}  \,\, 1 \,\,   \overset{\mathfrak{su_{3}}}3 \,\, 1  \,\,\overset{\mathfrak{e_{6}}}6  \,\, 1 \,\,...[E_6]$\\\hline

$A_5$ & $ SU(2)$ &   $ [SU(2)] \,\, {\overset{\mathfrak{g_{2}}}3}  \,\,    1  \,\,\overset{\mathfrak{f_{4}}}5  \,\, 1 \,\,   \overset{\mathfrak{su_{3}}}3 \,\, 1  \,\, \overset{\mathfrak{e_{6}}}6  \,\, 1 \,\,...[E_6]$\\\hline

$D_5 (a_1)$ & $ U(1)$ &  ${\overset{\mathfrak{su_{3}}}3}  \,\,   1  \,\, \underset{[N_f=1]}{\overset{\mathfrak{e_{6}}}5}  \,\, 1 \,\,   \overset{\mathfrak{su_{3}}}3 \,\, 1  \,\,\overset{\mathfrak{e_{6}}}6  \,\, 1 \,\,...[E_6]$\\\hline

$E_6 (a_3)$ & $ 1 $ & $ {\overset{\mathfrak{su_{3}}}3}  \,\,   1  \,\, {\overset{\mathfrak{f_{4}}}5}  \,\, 1 \,\,   \overset{\mathfrak{su_{3}}}3 \,\, 1  \,\,\overset{\mathfrak{e_{6}}}6  \,\, 1 \,\,...[E_6]$\\\hline

$D_5$ & $U(1)$ &  $ {\overset{\mathfrak{su_{2}}}2}  \,\,   {\overset{\mathfrak{so_{7}}}3}  \,\,  \overset{\mathfrak{su_{2}}}2 \,\, 1  \,\,\overset{\mathfrak{e_{6}}}6  \,\, 1 \,\,...[E_6]$ \\\hline

$E_6 (a_1)$ & 1 &  $ {\overset{\mathfrak{su_{2}}}2}  \,\,   {\overset{\mathfrak{g_2}}3}  \,\, 1 \,\, \overset{\mathfrak{f_{4}}}5 \,\, 1 \,\,  {\overset{\mathfrak{su_{3}}}3}  \,\,   1 \,\,\overset{\mathfrak{e_{6}}}6  \,\, 1 \,\,...[E_6]$\\\hline

$E_6$ & 1 &   $2  \,\,   {\overset{\mathfrak{su_{2}}}2}  \,\,   {\overset{\mathfrak{g_2}}3}  \,\, 1 \,\, \overset{\mathfrak{f_{4}}}5 \,\, 1 \,\,  {\overset{\mathfrak{su_{3}}}3}  \,\,   1 \,\,\overset{\mathfrak{e_{6}}}6  \,\, 1 \,\,...[E_6]$ \\\hline
\label{E6list}
\end{longtable}

\subsection{$E_7$ Nilpotent Orbits}

The $E_7$ Nilpotent orbits are as follows. The nilpotent hierarchy can be found for example in \cite[Table 16]{Chacaltana:2012zy}.

\begin{longtable}{c|c|c}
B--C Label & Global Symmetry & Theory \\\hline
0 & $E_7$ &  $[E_7] \,\, 1\,\,  \overset{\mathfrak{su_{2}}}2 \,\, \overset{\mathfrak{so_{7}}}3  \,\,  \overset{\mathfrak{su_{2}}}2\,\, 1 \,\, \overset{\mathfrak{e_{7}}}8  \,\, 1 \,\,  ... [E_7]$ \\\hline
$A_1$ & $SO(12)$ &  $[SO(12)] \,\, \overset{\mathfrak{sp_{1}}}1 \,\, \overset{\mathfrak{so_{7}}}3  \,\,  \overset{\mathfrak{su_{2}}}2\,\, 1 \,\, \overset{\mathfrak{e_{7}}}8  \,\, 1 \,\,  ... [E_7]$ \\\hline
$2A_1$ & $SO(9)\times SU(2)$ &  $[SO(9)] \,\, 1 \,\, \underset{[SU(2)]}{\overset{\mathfrak{so_{7}}}3}  \,\,  \overset{\mathfrak{su_{2}}}2\,\, 1 \,\, \overset{\mathfrak{e_{7}}}8  \,\, 1 \,\,  ... [E_7]$ \\\hline
$(3A_1)'$ & $Sp(3)\times SU(2)$ &  $[Sp(3)]  \,\, \underset{[SU(2)]}{\overset{\mathfrak{so_{7}}}2}  \,\,  \overset{\mathfrak{su_{2}}}2\,\, 1 \,\, \overset{\mathfrak{e_{7}}}8  \,\, 1 \,\,  ... [E_7]$ \\\hline
$(3A_1)''$ & $F_4$ &  $[F_4] \,\, 1 \,\, {\overset{\mathfrak{g_{2}}}3}  \,\,  \overset{\mathfrak{su_{2}}}2\,\, 1 \,\, \overset{\mathfrak{e_{7}}}8  \,\, 1 \,\,  ... [E_7]$ \\\hline
$4A_1$ & $Sp(3)$ &  $[Sp(3)]  \,\, {\overset{\mathfrak{g_{2}}}2}  \,\,  \overset{\mathfrak{su_{2}}}2\,\, 1 \,\, \overset{\mathfrak{e_{7}}}8  \,\, 1 \,\,  ... [E_7]$ \\\hline
$A_2$ & $SU(6)$ &  $[SU(6)]  \,\, {\overset{\mathfrak{su_{4}}}2}  \,\,  \overset{\mathfrak{su_{2}}}2\,\, 1 \,\, \overset{\mathfrak{e_{7}}}8  \,\, 1 \,\,  ... [E_7]$ \\\hline
$A_2+A_1$ & $SU(4)\times U(1)$ &  $[SU(4)]  \,\, {\overset{\mathfrak{su_{3}}}2}  \,\,  \underset{[N_f=1]}{\overset{\mathfrak{su_{2}}}2}\,\, 1 \,\, \overset{\mathfrak{e_{7}}}8  \,\, 1 \,\,  ... [E_7]$ \\\hline
$A_2+2 A_1$ & $SU(2) \times SU(2) \times SU(2)$ &  $[SO(4)]  \,\, {\overset{\mathfrak{su_{2}}}2}  \,\,  \underset{[SU(2)]}{\overset{\mathfrak{su_{2}}}2}\,\, 1 \,\, \overset{\mathfrak{e_{7}}}8  \,\, 1 \,\,  ... [E_7]$ \\\hline
$2A_2$ & $G_2 \times SU(2)$ &  $[G_2]  \,\, {\overset{\mathfrak{su_{2}}}2}  \,\, {\overset{\mathfrak{su_{1}}}2}\,\,  \underset{[SU(2)]}1 \,\, \overset{\mathfrak{e_{7}}}8  \,\, 1 \,\,  ... [E_7]$ \\\hline
$A_2+ 3A_1$ & $G_2$ &  $ {\overset{\mathfrak{su_{1}}}2}  \,\, \underset{[G_2]}{\overset{\mathfrak{su_{2}}}2}\,\,  1 \,\, \overset{\mathfrak{e_{7}}}8  \,\, 1 \,\,  ... [E_7]$ \\\hline
$2A_2+ A_1$ & $SU(2) \times SU(2)$ &  $[SU(2)] \,\, {2}  \,\, 2 \,\,  \underset{[SU(2)]}1 \,\, \overset{\mathfrak{e_{7}}}8  \,\, 1 \,\,  ... [E_7]$ \\\hline
$A_3$ & $SO(7) \times SU(2)$ &  $[SO(7)] \,\, \overset{\mathfrak{so}_7}2 \,\, 1 \,\, \underset{[SU(2)]}{\underset{1}{\overset{\mathfrak{e_{7}}}8}}  \,\, 1 \,\,  ... [E_7]$ \\\hline
$(A_3 +A_1)'$ & $SU(2) \times SU(2) \times SU(2)$ &  $[SU(2)] \,\, 2 \,\,  \underset{[SU(2)]}1 \,\, \underset{[SU(2)]}{\underset{1}{\overset{\mathfrak{e_{7}}}8}}  \,\, 1 \,\,  ... [E_7]$ \\\hline
$(A_3 +A_1)''$ & $SO(7) $ &  $[SO(7)] \,\, \overset{\mathfrak{su_{2}}}2 \,\,  1 \,\, \overset{\mathfrak{e_{7}}}7  \,\, 1 \,\,  ... [E_7]$ \\\hline
$A_3 +2A_1$ & $SU(2) \times SU(2) $ &  $[SU(2)] \,\, 2 \,\,  \underset{[SU(2)]}1 \,\, \overset{\mathfrak{e_{7}}}7  \,\, 1 \,\,  ... [E_7]$ \\\hline
$D_4(a_1)$ & $SU(2) \times SU(2) \times SU(2)$ &  $[SU(2)] \,\,1 \,\, \overset{[SU(2)]}{\overset{1}{\underset{[SU(2)]}{\underset{1}{\overset{\mathfrak{e_{7}}}8}}}}  \,\, 1 \,\,  ... [E_7]$ \\\hline
$D_4(a_1)+A_1$ & $SU(2) \times SU(2)$ &  $[SU(2)] \,\,1 \,\, \overset{[SU(2)]}{\overset{1}{{\overset{\mathfrak{e_{7}}}7}}}  \,\, 1 \,\,  ... [E_7]$ \\\hline
$A_3+A_2$ & $SU(2) \times U(1)$ &  $[SU(2)] \,\,  1  \,\, \underset{[N_f=1]}{\overset{\mathfrak{e_{7}}}6}  \,\, 1 \,\,  ... [E_7]$
 \\\hline
$D_4$ & $Sp(3)$ &  $[Sp(3)]  \,\, {\overset{\mathfrak{so_{12}}}4}  \,\,{\overset{\mathfrak{sp_{1}}}1}  \,\,{\overset{\mathfrak{so_{7}}}3}  \,\, {\overset{\mathfrak{su_{1}}}2}\,\,  1 \,\, \overset{\mathfrak{e_{7}}}8  \,\, 1 \,\,  ... [E_7]$ \\\hline
$A_3+A_2+A_1$ & $SU(2) $ &  $[SU(2)] \,\,    {\overset{\mathfrak{e_{7}}}5}  \,\, 1 \,\,  ... [E_7]$
 \\\hline
$A_4$ & $SU(3) \times U(1)$ &  $[SU(3)] \,\,  1  \,\, {\overset{\mathfrak{e_{6}}}6}  \,\, 1 \,\,  \overset{\mathfrak{su_{2}}}2 \,\, \overset{\mathfrak{so_{7}}}3 \,\,   ... [E_7]$
 \\\hline
$A_4+A_1$ & $U(1)^2 $ &  $ \underset{[N_f=1]}{\overset{\mathfrak{e_{6}}}5}  \,\, 1  \,\,  \overset{\mathfrak{su_{2}}}2 \,\, \overset{\mathfrak{so_{7}}}3 \,\,  ... [E_7]$
 \\\hline
$D_4+A_1$ & $Sp(2)$ &  $[Sp(2)]  \,\, {\overset{\mathfrak{so_{11}}}4}  \,\, \underset{[N_f=\frac{1}{2}]}{\overset{\mathfrak{sp_{1}}}1}  \,\,{\overset{\mathfrak{so_{7}}}3}  \,\, {\overset{\mathfrak{su_{2}}}2}\,\,  1 \,\, \overset{\mathfrak{e_{7}}}8  \,\, 1 \,\,  ... [E_7]$ \\\hline
$D_5(a_1)$ & $SU(2)\times U(1)$ &  $[Sp(1)]  \,\, {\overset{\mathfrak{so_{10}}}4}  \,\, \underset{[N_f=1]}{\overset{\mathfrak{sp_{1}}}1}  \,\,{\overset{\mathfrak{so_{7}}}3}  \,\, {\overset{\mathfrak{su_{2}}}2}\,\,  1 \,\, \overset{\mathfrak{e_{7}}}8  \,\, 1 \,\,  ... [E_7]$ \\\hline
$A_4+A_2$ & $1 $ &  $  {\overset{\mathfrak{f_{4}}}5}  \,\, \underset{[SU(2)]}1   \,\,  \overset{\mathfrak{su_{2}}}2 \,\, \overset{\mathfrak{so_{7}}}3 \,\, ... [E_7]$ \\\hline
$A_5''$ & $G_2$ &  $ [G_2]\,\, 1 \,\, {\overset{\mathfrak{f_{4}}}5}  \,\, 1    \,\, \overset{\mathfrak{g_{2}}}3 \,\, ... [E_7]$ \\\hline
$A_5+A_1$ & $SU(2)$ &  $ [Sp(1)]\,\,  {\overset{\mathfrak{f_{4}}}4}  \,\, 1    \,\, \overset{\mathfrak{g_{2}}}3 \,\, ... [E_7]$ \\\hline
$D_5(a_1) + A_1$ & $SU(2)$ &  $ {\overset{\mathfrak{so_{9}}}4}  \,\, \underset{[SO(3)]}{\overset{\mathfrak{sp_{1}}}1}  \,\,{\overset{\mathfrak{so_{7}}}3}  \,\, {\overset{\mathfrak{su_{2}}}2}\,\,  1 \,\, \overset{\mathfrak{e_{7}}}8  \,\, 1 \,\,  ... [E_7]$ \\\hline
$A_5'$ & $SU(2) \times SU(2)$ &  $[SU(2)]\,\, {\overset{\mathfrak{so_{9}}}4}  \,\, 1  \,\, \underset{[SU(2)]}{\overset{\mathfrak{so_{7}}}3}  \,\, {\overset{\mathfrak{su_{2}}}2}\,\,  1 \,\, \overset{\mathfrak{e_{7}}}8  \,\, 1 \,\,  ... [E_7]$ \\\hline
$D_6(a_2) $ & $SU(2)$ &  $ [SU(2)] \,\, {\overset{\mathfrak{so_{9}}}4}  \,\, 1  \,\, {\overset{\mathfrak{g_{2}}}3}  \,\, {\overset{\mathfrak{su_{2}}}2} \,\,  1 \,\, \overset{\mathfrak{e_{7}}}8  \,\, 1 \,\,  ... [E_7]$ \\\hline
$E_6(a_3) $ & $SU(2)$ &  $  {\overset{\mathfrak{so_{8}}}4}  \,\, 1  \,\, \underset{[SU(2)]}{\overset{\mathfrak{so_{7}}}3}  \,\, {\overset{\mathfrak{su_{2}}}2} \,\,  1 \,\, \overset{\mathfrak{e_{7}}}8  \,\, 1 \,\,  ... [E_7]$ \\\hline
$E_7(a_5) $ & $1$ &  $ {\overset{\mathfrak{so_{8}}}4}  \,\, 1  \,\, {\overset{\mathfrak{g_{2}}}3}  \,\, {\overset{\mathfrak{su_{2}}}2} \,\,  1 \,\, \overset{\mathfrak{e_{7}}}8  \,\, 1 \,\,  ... [E_7]$ \\\hline
$D_5 $ & $SU(2) \times SU(2)$ &  $ [SU(2)]\,\, {\overset{\mathfrak{so_{7}}}3}  \,\, {\overset{\mathfrak{su_{2}}}2} \,\,  1 \,\, \underset{[SU(2)]}{\underset{1}{\overset{\mathfrak{e_{7}}}8}}  \,\, 1 \,\,  ... [E_7]$ \\\hline
$A_6 $ & $SU(2) $ &  $ [SU(2)]\,\, {\overset{\mathfrak{so_{7}}}3}  \,\, {\overset{\mathfrak{su_{2}}}2} \,\,  1 \,\, \overset{\mathfrak{e_{7}}}7  \,\, 1 \,\,  ... [E_7]$ \\\hline
$D_6(a_1) $ & $SU(2) $ &  $ {\overset{\mathfrak{g_{2}}}3}  \,\,  {\overset{\mathfrak{su_{2}}}2} \,\,{2}\,\,  \underset{[SU(2)]}1 \,\, {\overset{\mathfrak{e_{7}}}8}  \,\, 1 \,\,  ... [E_7]$ \\\hline
$D_5 +A_1$ & $SU(2) $ &  $  {\overset{\mathfrak{g_{2}}}3}  \,\, {\overset{\mathfrak{su_{2}}}2} \,\,  1 \,\, \underset{[SU(2)]}{\underset{1}{\overset{\mathfrak{e_{7}}}8}}  \,\, 1 \,\,  ... [E_7]$ \\\hline
$E_7(a_4)$ & $1$ &  $  {\overset{\mathfrak{g_{2}}}3}  \,\, {\overset{\mathfrak{su_{2}}}2} \,\,  1 \,\, \overset{\mathfrak{e_{7}}}7  \,\, 1 \,\,  ... [E_7]$ \\\hline
$D_6$ & $SU(2) $ &  $[SU(2)] \,\, {\overset{\mathfrak{g_{2}}}3}    \,\, 1 \,\, {\overset{\mathfrak{f_{4}}}5}  \,\, 1\,\, {\overset{\mathfrak{g_{2}}}3}  \,\, {\overset{\mathfrak{su_{2}}}2}  \,\, 1 \,\, \overset{\mathfrak{e_{7}}}8 \,\,  ... [E_7]$ \\\hline
$E_6(a_1)$ & $U(1)$ &  $ {\overset{\mathfrak{su_{3}}}3}    \,\, 1 \,\, {\overset{\mathfrak{e_{6}}}6}  \,\, 1\,\, {\overset{\mathfrak{su_{2}}}2}  \,\, {\overset{\mathfrak{so_{7}}}3}  \,\, {\overset{\mathfrak{su_{2}}}2}  \,\, 1 \,\, \overset{\mathfrak{e_{7}}}8 \,\,  ... [E_7]$ \\\hline
$E_6$ & $SU(2) $ &  $ {\overset{\mathfrak{su_{2}}}2} \,\, {\overset{\mathfrak{so_{7}}}3} \,\, {\overset{\mathfrak{su_{2}}}2}   \,\, 1 \,\, \underset{[SU(2)]}{\underset{1}{\overset{\mathfrak{e_{7}}}8}}  \,\, 1\,\, {\overset{\mathfrak{su_{2}}}2} \,\, {\overset{\mathfrak{so_{7}}}3}  \,\, {\overset{\mathfrak{su_{2}}}2}  \,\,  ... [E_7]$ \\\hline
$E_7(a_3)$ & $1 $ &  $  {\overset{\mathfrak{su_{3}}}3}    \,\, 1 \,\, {\overset{\mathfrak{f_{4}}}5}  \,\, 1\,\,  {\overset{\mathfrak{g_{2}}}3}  \,\, {\overset{\mathfrak{su_{2}}}2}  \,\, 1 \,\, \overset{\mathfrak{e_{7}}}8 \,\, ... [E_7]$ \\\hline
$E_7(a_2)$ & $1 $ &  $ {\overset{\mathfrak{su_{2}}}2} \,\, {\overset{\mathfrak{so_{7}}}3} \,\, {\overset{\mathfrak{su_{2}}}2}   \,\, 1 \,\, \overset{\mathfrak{e_{7}}}7  \,\, 1\,\, {\overset{\mathfrak{su_{2}}}2} \,\, {\overset{\mathfrak{so_{7}}}3}  \,\, {\overset{\mathfrak{su_{2}}}2}  \,\,  ... [E_7]$ \\\hline
$E_7(a_1)$ & $1 $ &  $  {\overset{\mathfrak{su_{2}}}2}\,\,  {\overset{\mathfrak{g_{2}}}3}    \,\, 1 \,\, {\overset{\mathfrak{f_{4}}}5}  \,\, 1\,\,  {\overset{\mathfrak{g_{2}}}3}  \,\, {\overset{\mathfrak{su_{2}}}2}  \,\, 1 \,\, \overset{\mathfrak{e_{7}}}8 \,\, ... [E_7]$ \\\hline
$E_7$ & $1 $ &  $  2 \,\, {\overset{\mathfrak{su_{2}}}2}\,\,  {\overset{\mathfrak{g_{2}}}3}    \,\, 1 \,\, {\overset{\mathfrak{f_{4}}}5}  \,\, 1\,\,  {\overset{\mathfrak{g_{2}}}3}  \,\, {\overset{\mathfrak{su_{2}}}2}  \,\, 1 \,\, \overset{\mathfrak{e_{7}}}8 \,\, ... [E_7]$ \\\hline
\caption{6D SCFTs associated with $E_7$ nilpotent orbits.}
\label{E7list}
\end{longtable}

\subsection{$E_8$ Nilpotent Orbits}\label{app:E8}

The $E_8$ Nilpotent orbits are as follows. The nilpotent hierarchy can be found for example in \cite[Table 19]{Chacaltana:2012zy}.

\begin{longtable}{c|c|c}
B--C Label & Global Symmetry & Theory \\\hline
0 & $E_8$ &  $[E_8] \,\, 1\,\, 2 \,\, \overset{\mathfrak{su_{2}}}2 \,\, \overset{\mathfrak{g_{2}}}3  \,\, 1 \,\, \overset{\mathfrak{f_{4}}}5 \,\,  1 \,\, \overset{\mathfrak{g_{2}}}3 \,\, \overset{\mathfrak{su_{2}}}2 \,\, 2 \,\, 1\,\, \overset{\mathfrak{e_{8}}}{12}  \,\, 1 \,\,  ... [E_8]$ \\\hline
$A_1$ & $E_7$ &  $[E_7] \,\, 1\,\,  \overset{\mathfrak{su_{2}}}2 \,\, \overset{\mathfrak{g_{2}}}3  \,\, 1 \,\, \overset{\mathfrak{f_{4}}}5 \,\,  1 \,\, \overset{\mathfrak{g_{2}}}3 \,\, \overset{\mathfrak{su_{2}}}2 \,\, 2 \,\, 1\,\, \overset{\mathfrak{e_{8}}}{12}  \,\, 1 \,\,  ... [E_8]$ \\\hline
$2 A_1$ & $SO(13)$ &  $[SO(13)] \,\, \overset{\mathfrak{sp_{1}}}1  \,\, \overset{\mathfrak{g_{2}}}3  \,\, 1 \,\, \overset{\mathfrak{f_{4}}}5 \,\,  1 \,\, \overset{\mathfrak{g_{2}}}3 \,\, \overset{\mathfrak{su_{2}}}2 \,\, 2 \,\, 1\,\, \overset{\mathfrak{e_{8}}}{12}  \,\, 1 \,\,  ... [E_8]$ \\\hline
$3 A_1$ & $F_4 \times SU(2)$ &  $[F_4] \,\, 1  \,\, \underset{[Sp(1)]}{\overset{\mathfrak{g_{2}}}3}  \,\, 1 \,\, \overset{\mathfrak{f_{4}}}5 \,\,  1 \,\, \overset{\mathfrak{g_{2}}}3 \,\, \overset{\mathfrak{su_{2}}}2 \,\, 2 \,\, 1\,\, \overset{\mathfrak{e_{8}}}{12}  \,\, 1 \,\,  ... [E_8]$ \\\hline
$A_2$ & $E_6$ &  $[E_6] \,\, 1  \,\, \overset{\mathfrak{su_{3}}}3  \,\, 1 \,\, \overset{\mathfrak{f_{4}}}5 \,\,  1 \,\, \overset{\mathfrak{g_{2}}}3 \,\, \overset{\mathfrak{su_{2}}}2 \,\, 2 \,\, 1\,\, \overset{\mathfrak{e_{8}}}{12}  \,\, 1 \,\,  ... [E_8]$ \\\hline
$4 A_1$ & $Sp(4)$ &  $[Sp(4)] \,\, \overset{\mathfrak{g_{2}}}2  \,\, 1 \,\, \overset{\mathfrak{f_{4}}}5 \,\,  1 \,\, \overset{\mathfrak{g_{2}}}3 \,\, \overset{\mathfrak{su_{2}}}2 \,\, 2 \,\, 1\,\, \overset{\mathfrak{e_{8}}}{12}  \,\, 1 \,\,  ... [E_8]$ \\\hline
$A_2+A_1$ & $SU(6)$ &  $[SU(6)] \,\, \overset{\mathfrak{su_{3}}}2  \,\, 1 \,\, \overset{\mathfrak{f_{4}}}5 \,\,  1 \,\, \overset{\mathfrak{g_{2}}}3 \,\, \overset{\mathfrak{su_{2}}}2 \,\, 2 \,\, 1\,\, \overset{\mathfrak{e_{8}}}{12}  \,\, 1 \,\,  ... [E_8]$ \\\hline
$A_2+2A_1$ & $SO(7) \times SU(2)$ &  $[SO(7)] \,\, \overset{\mathfrak{su_{2}}}2  \,\, \underset{[SU(2)]}1 \,\, \overset{\mathfrak{f_{4}}}5 \,\,  1 \,\, \overset{\mathfrak{g_{2}}}3 \,\, \overset{\mathfrak{su_{2}}}2 \,\, 2 \,\, 1\,\, \overset{\mathfrak{e_{8}}}{12}  \,\, 1 \,\,  ... [E_8]$ \\\hline
$A_3$ & $SO(11)$ &  $[SO(11)] \,\, \overset{\mathfrak{sp_{1}}}1  \,\,  \overset{\mathfrak{so_{9}}}4 \,\,  1 \,\, \overset{\mathfrak{g_{2}}}3 \,\, \overset{\mathfrak{su_{2}}}2 \,\, 2 \,\, 1\,\, \overset{\mathfrak{e_{8}}}{12}  \,\, 1 \,\,  ... [E_8]$ \\\hline
$A_2+3 A_1$ & $G_2 \times SU(2)$ &  $[SU(2)] \,\, 2  \,\, \underset{[G_2]}1 \,\, \overset{\mathfrak{f_{4}}}5 \,\,  1 \,\, \overset{\mathfrak{g_{2}}}3 \,\, \overset{\mathfrak{su_{2}}}2 \,\, 2 \,\, 1\,\, \overset{\mathfrak{e_{8}}}{12}  \,\, 1 \,\,  ... [E_8]$ \\\hline
$2 A_2$ & $G_2 \times G_2$ &  $[G_2] \,\, 1 \,\, \underset{[G_2]}{\underset{1}{\overset{\mathfrak{f_{4}}}5}} \,\,  1 \,\, \overset{\mathfrak{g_{2}}}3 \,\, \overset{\mathfrak{su_{2}}}2 \,\, 2 \,\, 1\,\, \overset{\mathfrak{e_{8}}}{12}  \,\, 1 \,\,  ... [E_8]$ \\\hline
$2 A_2+A_1$ & $G_2 \times SU(2)$ &  $[G_2] \,\, 1 \,\, \underset{[Sp(1)]}{\overset{\mathfrak{f_{4}}}4} \,\,  1 \,\, \overset{\mathfrak{g_{2}}}3 \,\, \overset{\mathfrak{su_{2}}}2 \,\, 2 \,\, 1\,\, \overset{\mathfrak{e_{8}}}{12}  \,\, 1 \,\,  ... [E_8]$ \\\hline
$ A_3+A_1$ & $SO(7) \times SU(2)$ & $[SO(7)]   \,\, 1  \,\,
\underset{[SU(2)] }{\overset{\mathfrak{so}_9}4} \,\,  1 \,\, \overset{\mathfrak{g_{2}}}3 \,\, \overset{\mathfrak{su_{2}}}2 \,\, 2 \,\, 1\,\, \overset{\mathfrak{e_{8}}}{12}  \,\, 1 \,\,  ... [E_8]$ \\\hline
$ 2A_2+2 A_1$ & $Sp(2)$ & $[Sp(2)]   \,\,  {\overset{\mathfrak{f}_4}3} \,\,  1 \,\, \overset{\mathfrak{g_{2}}}3 \,\, \overset{\mathfrak{su_{2}}}2 \,\, 2 \,\, 1\,\, \overset{\mathfrak{e_{8}}}{12}  \,\, 1 \,\,  ... [E_8]$ \\\hline
$ D_4(a_1)$ & $SO(8)$ & $[SO(8)]  \,\, 1  \,\,  {\overset{\mathfrak{so}_8}4} \,\,  1 \,\, \overset{\mathfrak{g_{2}}}3 \,\, \overset{\mathfrak{su_{2}}}2 \,\, 2 \,\, 1\,\, \overset{\mathfrak{e_{8}}}{12}  \,\, 1 \,\,  ... [E_8]$ \\\hline
$ A_3 + 2 A_1$ & $Sp(2) \times SU(2)$ & $[Sp(2)]  \,\, \underset{[SU(2)]}{\overset{\mathfrak{so}_9}3} \,\,  1 \,\, \overset{\mathfrak{g_{2}}}3 \,\, \overset{\mathfrak{su_{2}}}2 \,\, 2 \,\, 1\,\, \overset{\mathfrak{e_{8}}}{12}  \,\, 1 \,\,  ... [E_8]$ \\\hline
$ D_4(a_1)+A_1$ & $SU(2) \times SU(2) \times SU(2)$ & $[SU(2) \times SU(2) \times SU(2)] \,\,  {\overset{\mathfrak{so}_8}3} \,\,  1 \,\, \overset{\mathfrak{g_{2}}}3 \,\, \overset{\mathfrak{su_{2}}}2 \,\, 2 \,\, 1\,\, \overset{\mathfrak{e_{8}}}{12}  \,\, 1 \,\,  ... [E_8]$ \\\hline
$ D_4$ & $F_4$ & $[F_4]  \,\, 1 \,\, \overset{\mathfrak{g_{2}}}3 \,\, \overset{\mathfrak{su_{2}}}2 \,\, 2 \,\, 1\,\, \overset{\mathfrak{e_{8}}}{11}  \,\, 1 \,\,  ... [E_8]$ \\\hline
$ A_3+ A_2$ & $Sp(2)\times U(1)$ & $[Sp(2)]  \,\, \overset{\mathfrak{so}_7}3 \,\, 1 \,\,  \overset{\mathfrak{g}_2}3 \,\, \overset{\mathfrak{su_{2}}}2 \,\, 2 \,\, 1\,\, \overset{\mathfrak{e_{8}}}{12}  \,\, 1 \,\,  ... [E_8]$ \\\hline
$ A_4$ & $SU(5)$ & $[SU(5)]  \,\, \overset{\mathfrak{su}_4}2 \,\, \overset{\mathfrak{su}_3}2 \,\,  \overset{\mathfrak{su}_2}2 \,\, \overset{\mathfrak{su_{1}}}2 \,\,  1\,\, \overset{\mathfrak{e_{8}}}{12}  \,\, 1 \,\,  ... [E_8]$ \\\hline
$ A_3+A_2+A_1$ & $SU(2) \times SU(2)$ & $[Sp(1)] \,\, \overset{\mathfrak{g}_2}3 \,\, \underset{[SU(2)]}1 \,\,  \overset{\mathfrak{g}_2}3 \,\, \overset{\mathfrak{su_{2}}}2 \,\, 2 \,\, 1\,\, \overset{\mathfrak{e_{8}}}{12}  \,\, 1 \,\,  ... [E_8]$ \\\hline
$ D_4+A_1$ & $Sp(3)$ & $[Sp(3)]  \,\, \overset{\mathfrak{g}_2}2 \,\,  \overset{\mathfrak{su}_2}2 \,\, 2 \,\,  1\,\, \overset{\mathfrak{e_{8}}}{11}  \,\, 1 \,\,  ... [E_8]$ \\\hline
$ D_4(a_1)+A_2$ & $SU(3) $ & $ \overset{\mathfrak{su}_3}3 \,\, \underset{[SU(3)]}1 \,\,  \overset{\mathfrak{g}_2}3 \,\, \overset{\mathfrak{su_{2}}}2 \,\, 2 \,\, 1\,\, \overset{\mathfrak{e_{8}}}{12}  \,\, 1 \,\,  ... [E_8]$ \\\hline
$ A_4+A_1$ & $SU(3)\times U(1)$ & $[SU(3)]  \,\, \overset{\mathfrak{su}_3}2 \,\, \underset{[N_f=1]}{\overset{\mathfrak{su}_3}2} \,\,  \overset{\mathfrak{su}_2}2 \,\, \overset{\mathfrak{su_{1}}}2 \,\,  1\,\, \overset{\mathfrak{e_{8}}}{12}  \,\, 1 \,\,  ... [E_8]$ \\\hline
$ 2 A_3$ & $Sp(2)$ & $\overset{\mathfrak{su}_2}2 \,\, \underset{[Sp(2)]}{\overset{\mathfrak{g}_2}2} \,\,  \overset{\mathfrak{su}_2}2 \,\, 2 \,\,  1\,\, \overset{\mathfrak{e_{8}}}{12}  \,\, 1 \,\,  ... [E_8]$ \\\hline
$ D_5(a_1)$ & $SU(4)$ & $[SU(4)] \,\, \overset{\mathfrak{su}_3}2 \,\,  \overset{\mathfrak{su}_2}2 \,\, \overset{\mathfrak{su}_1}2 \,\,  1\,\, \overset{\mathfrak{e_{8}}}{11}  \,\, 1 \,\,  ... [E_8]$ \\\hline
$ A_4 + 2 A_1$ & $SU(2)\times U(1)$ & $\underset{[N_f=1]}{\overset{\mathfrak{su}_2}2}\,\,  \underset{[SU(2)]}{\overset{\mathfrak{su}_3}2 } \,\,  \overset{\mathfrak{su}_2}2 \,\, \overset{\mathfrak{su}_1}2 \,\,  1\,\, \overset{\mathfrak{e_{8}}}{12}  \,\, 1 \,\,  ... [E_8]$ \\\hline
$ A_4 + A_2$ & $SU(2) \times SU(2)$ & $[SO(4)] \,\, \overset{\mathfrak{su}_2}2\,\, {\overset{\mathfrak{su}_2}2 } \,\,  \underset{[N_f=1]}{\overset{\mathfrak{su}_2}2} \,\, \overset{\mathfrak{su}_1}2 \,\,  1\,\, \overset{\mathfrak{e_{8}}}{12}  \,\, 1 \,\,  ... [E_8]$ \\\hline
$ D_5(a_1) + A_1$ & $SU(2) \times SU(2)$ & $[SO(4)] \,\,  {\overset{\mathfrak{su}_2}2 } \,\,  \underset{[N_f=1]}{\overset{\mathfrak{su}_2}2} \,\, \overset{\mathfrak{su}_1}2 \,\,  1\,\, \overset{\mathfrak{e_{8}}}{11}  \,\, 1 \,\,  ... [E_8]$ \\\hline
$A_4+A_2+ A_1$ & $ SU(2)$ & $  {\overset{\mathfrak{su}_1}2 } \,\,   \underset{[N_f=1]}{\overset{\mathfrak{su}_2}2} \,\,  \underset{[N_f=1]}{\overset{\mathfrak{su}_2}2} \,\, \overset{\mathfrak{su}_1}2 \,\,  1\,\, \overset{\mathfrak{e_{8}}}{12}  \,\, 1 \,\,  ... [E_8]$ \\\hline
$A_5$ & $ G_2 \times SU(2)$ & $ [G_2] \,\, {\overset{\mathfrak{su}_2}2} \,\, 2 \,\,  1\,\, \underset{[SU(2)]}{\underset{2}{\underset{1}{\overset{\mathfrak{e_{8}}}{12}}}}  \,\, 1 \,\,  ... [E_8]$ \\\hline
$A_4+A_3$ & $  SU(2)$ &  $[SU(2)] \,\, 2\,\, {2 } \,\,  2 \,\, 2 \,\,  1\,\, \overset{\mathfrak{e_{8}}}{12}  \,\, 1 \,\,  ... [E_8]$ \\\hline
$D_4+A_2$ & $  SU(3)$ &  $ 2\,\, \underset{[SU(3)]}{\overset{\mathfrak{su}_2}2 } \,\,  2  \,\,  1\,\, \overset{\mathfrak{e_{8}}}{11}  \,\, 1 \,\,  ... [E_8]$ \\\hline
$E_6(a_3)$ & $  G_2$ &  $ [G_2] \,\, {\overset{\mathfrak{su}_2}2 } \,\,  2  \,\,  1\,\, \overset{\mathfrak{e_{8}}}{10}  \,\, 1 \,\,  ... [E_8]$ \\\hline
$A_5+A_1$ & $  SU(2) \times SU(2)$ & $ [SU(2)] \,\, {2} \,\, 2 \,\,  1\,\, \underset{[SU(2)]}{\underset{2}{\underset{1}{\overset{\mathfrak{e_{8}}}{12}}}}  \,\, 1 \,\,  ... [E_8]$ \\\hline
$D_5(a_1) + A_2$ & $  SU(2)$ & $ [SU(2)] \,\, 2\,\, {2} \,\, 2 \,\,  1\,\, \overset{\mathfrak{e_{8}}}{11}   \,\, 1 \,\,  ... [E_8]$ \\\hline
$E_6(a_3) + A_1$ & $  SU(2)$ & $ [SU(2)] \,\,  {2} \,\, 2 \,\,  1\,\, \overset{\mathfrak{e_{8}}}{10}   \,\, 1 \,\,  ... [E_8]$ \\\hline
$D_6(a_2) $ & $  SU(2) \times SU(2)$ & $ [SU(2)] \,\,   2 \,\,  1\,\, \underset{[SU(2)]}{\underset{2}{\underset{1}{\overset{\mathfrak{e_{8}}}{11}}}}   \,\, 1 \,\,  ... [E_8]$ \\\hline
$D_5 $ & $  SO(7)$ & $ [SO(7)] \,\,   \overset{\mathfrak{su_{2}}}2 \,\,  1\,\, \overset{\mathfrak{e_{7}}}8  \,\, 1\,\,  \overset{\mathfrak{su_{2}}}2 \,\, \overset{\mathfrak{g_{2}}}3  \,\, 1 \,\, \overset{\mathfrak{f_{4}}}5 \,\,  1 \,\,  ... [E_8]$ \\\hline
$E_7(a_5) $ & $  SU(2)$ & $ [SU(2)] \,\,   2 \,\,  1\,\, \overset{\mathfrak{e_{8}}}{9}   \,\, 1 \,\,  ... [E_8]$ \\\hline
$D_5 +A_1$ & $  SU(2) \times SU(2)$ & $ [SU(2)] \,\,   2 \,\,  \underset{[SU(2)]}1 \,\, \overset{\mathfrak{e_{7}}}8  \,\, 1\,\,  \overset{\mathfrak{su_{2}}}2 \,\, \overset{\mathfrak{g_{2}}}3  \,\, 1 \,\, \overset{\mathfrak{f_{4}}}5 \,\,  1 \,\,  ... [E_8]$ \\\hline
$E_8(a_7)$ & $ 1$ &$    \overset{\mathfrak{e_{8}}}{7}   \,\, 1 \,\, 2 \,\, \overset{\mathfrak{su_{2}}}2 \,\, \overset{\mathfrak{g_{2}}}3  \,\, 1 \,\, \overset{\mathfrak{f_{4}}}5 \,\,  1 \,\,  ... [E_8]$ \\\hline
$D_6(a_1)$ & $  SU(2) \times SU(2)$ & $ [SU(2)] \,\,   1 \,\, \underset{[SU(2)]}{\underset{1}{\overset{\mathfrak{e_{7}}}8}}  \,\, 1\,\,  \overset{\mathfrak{su_{2}}}2 \,\, \overset{\mathfrak{g_{2}}}3  \,\, 1 \,\, \overset{\mathfrak{f_{4}}}5 \,\,  1 \,\,  ... [E_8]$ \\\hline
$A_6$ & $  SU(2) \times SU(2)$ & $ [SU(2)] \,\,   1 \,\, {\overset{\mathfrak{e_{7}}}8}  \,\, \underset{[SU(2)]}1\,\, 2\,\, \overset{\mathfrak{su_{2}}}2 \,\, \overset{\mathfrak{g_{2}}}3  \,\, 1 \,\, \overset{\mathfrak{f_{4}}}5 \,\,  1 \,\,  ... [E_8]$ \\\hline
$E_7(a_4)$ & $  SU(2)$ & $ [SU(2)]  \,\, 1 \,\, \overset{\mathfrak{e_{7}}}7 \,\, 1\,\,  \overset{\mathfrak{su_{2}}}2 \,\, \overset{\mathfrak{g_{2}}}3  \,\, 1 \,\, \overset{\mathfrak{f_{4}}}5 \,\,  1 \,\,  ... [E_8]$ \\\hline
$A_6+A_1$ & $  SU(2) $ & $ \overset{\mathfrak{e_{7}}}7  \,\, \underset{[SU(2)]}1\,\, 2\,\, \overset{\mathfrak{su_{2}}}2 \,\, \overset{\mathfrak{g_{2}}}3  \,\, 1 \,\, \overset{\mathfrak{f_{4}}}5 \,\,  1 \,\,  ... [E_8]$ \\\hline
$E_6(a_1)$ & $  SU(3) $ & $ [SU(3)] \,\, 1 \,\, {\overset{\mathfrak{e_{6}}}6}  \,\, 1\,\, \overset{\mathfrak{su_{3}}}3  \,\, 1 \,\, \overset{\mathfrak{f_{4}}}5 \,\,  1 \,\,  ... [E_8]$ \\\hline
$D_5+A_2$ & $  U(1) $ & $\underset{[N_f=1]}{\overset{\mathfrak{e_{7}}}6}  \,\, 1  \,\, \overset{\mathfrak{su_{2}}}2 \,\,\overset{\mathfrak{g_{2}}}3  \,\, 1 \,\, \overset{\mathfrak{f_{4}}}5 \,\,  1 \,\,  ... [E_8]$ \\\hline
$D_7(a_2)$ & $  U(1) $ & ${\overset{\mathfrak{e_{6}}}6}  \,\, 1  \,\, \overset{\mathfrak{su_{2}}}2 \,\,\overset{\mathfrak{g_{2}}}3  \,\, 1 \,\, \overset{\mathfrak{f_{4}}}5 \,\,  1 \,\,  ... [E_8]$ \\\hline
$ E_6$ & $G_2$ & $[G_2] \,\, 1 \,\, {\overset{\mathfrak{f}_4}5} \,\,  1 \,\, \overset{\mathfrak{g_{2}}}3 \,\, \overset{\mathfrak{su_{2}}}2 \,\, 2 \,\, 1\,\, \overset{\mathfrak{e_{8}}}{11}  \,\, 1 \,\,  ... [E_8]$ \\\hline
$ A_7$ & $SU(2)$ & ${\overset{\mathfrak{f_{4}}}5}  \,\, 1   \,\,  \underset{[Sp(1)]}{\overset{\mathfrak{g_{2}}}3}  \,\, 1 \,\, \overset{\mathfrak{f_{4}}}5 \,\,  1  \,\, \overset{\mathfrak{g_{2}}}3 \,\, \overset{\mathfrak{su_{2}}}2 \,\, 2 \,\, 1\,\, \overset{\mathfrak{e_{8}}}{12}  \,\, 1 \,\,  ... [E_8]$  \\\hline
$ E_6(a_1)+A_1$ & $U(1)$ & $\underset{[N_f=1]}{\overset{\mathfrak{e_{6}}}5}  \,\, 1   \,\,  {\overset{\mathfrak{su_{3}}}3}  \,\, 1 \,\, \overset{\mathfrak{f_{4}}}5 \,\,  1  \,\, \overset{\mathfrak{g_{2}}}3 \,\, \overset{\mathfrak{su_{2}}}2 \,\, 2 \,\, 1\,\, \overset{\mathfrak{e_{8}}}{12}  \,\, 1 \,\,  ... [E_8]$  \\\hline
$ E_8(b_6)$ & $1$ & ${\overset{\mathfrak{f_{4}}}5}  \,\, 1   \,\,  {\overset{\mathfrak{su_{3}}}3}  \,\, 1 \,\, \overset{\mathfrak{f_{4}}}5 \,\,  1  \,\, \overset{\mathfrak{g_{2}}}3 \,\, \overset{\mathfrak{su_{2}}}2 \,\, 2 \,\, 1\,\, \overset{\mathfrak{e_{8}}}{12}  \,\, 1 \,\,  ... [E_8]$  \\\hline
$ D_6$ & $Sp(2)$ & $ [Sp(2)] \,\, \overset{\mathfrak{so_{11}}}4  \,\, \overset{\mathfrak{sp}_1}1 \,\, \overset{\mathfrak{so_{9}}}4 \,\,  1  \,\, \overset{\mathfrak{g_{2}}}3 \,\, \overset{\mathfrak{su_{2}}}2 \,\, 2 \,\, 1\,\, \overset{\mathfrak{e_{8}}}{12}  \,\, 1 \,\,  ... [E_8]$  \\\hline
$ E_7(a_3)$ & $SU(2)$ & $ [Sp(1)] \,\, \overset{\mathfrak{so_{10}}}4  \,\, \underset{[N_f=\frac{1}{2}]}{\overset{\mathfrak{sp}_1}1} \,\, \overset{\mathfrak{so_{9}}}4 \,\,  1  \,\, \overset{\mathfrak{g_{2}}}3 \,\, \overset{\mathfrak{su_{2}}}2 \,\, 2 \,\, 1\,\, \overset{\mathfrak{e_{8}}}{12}  \,\, 1 \,\,  ... [E_8]$  \\\hline
$ D_7(a_1)$ & $U(1)$ & $ \overset{\mathfrak{so_{9}}}4  \,\, \underset{[N_f=1]}{\overset{\mathfrak{sp}_1}1} \,\, \overset{\mathfrak{so_{9}}}4 \,\,  1  \,\, \overset{\mathfrak{g_{2}}}3 \,\, \overset{\mathfrak{su_{2}}}2 \,\, 2 \,\, 1\,\, \overset{\mathfrak{e_{8}}}{12}  \,\, 1 \,\,  ... [E_8]$  \\\hline
$ E_6 +A_1$ & $SU(2)$ & $ [Sp(1)] \,\, \overset{\mathfrak{f_{4}}}4 \,\,  1  \,\, \overset{\mathfrak{g_{2}}}3 \,\, \overset{\mathfrak{su_{2}}}2 \,\, 2 \,\, 1\,\, \overset{\mathfrak{e_{8}}}{11}  \,\, 1 \,\,  ... [E_8]$  \\\hline
$ E_7(a_2)$ & $SU(2)$ & $ [Sp(1)] \,\, \overset{\mathfrak{so_{9}}}4 \,\,  1  \,\, \overset{\mathfrak{g_{2}}}3 \,\, \overset{\mathfrak{su_{2}}}2 \,\, 2 \,\, 1\,\, \overset{\mathfrak{e_{8}}}{11}  \,\, 1 \,\,  ... [E_8]$  \\\hline
$ E_8(a_6)$ & $1$ & $ \overset{\mathfrak{so_{8}}}4  \,\, {1} \,\, \overset{\mathfrak{so_{8}}}4 \,\,  1  \,\, \overset{\mathfrak{g_{2}}}3 \,\, \overset{\mathfrak{su_{2}}}2 \,\, 2 \,\, 1\,\, \overset{\mathfrak{e_{8}}}{12}  \,\, 1 \,\,  ... [E_8]$  \\\hline
$ E_8(b_5)$ & $1$ & $ \overset{\mathfrak{so_{8}}}4 \,\,  1  \,\, \overset{\mathfrak{g_{2}}}3 \,\, \overset{\mathfrak{su_{2}}}2 \,\, 2 \,\, 1\,\, \overset{\mathfrak{e_{8}}}{11}  \,\, 1 \,\,  ... [E_8]$  \\\hline
$E_7(a_1)$ & $  SU(2)$ & $ [SU(2)] \,\,    \overset{\mathfrak{so}_7}3 \,\, \overset{\mathfrak{su_{2}}}2 \,\, 1 \,\, {\overset{\mathfrak{e_{7}}}8}  \,\, 1 \,\,  \overset{\mathfrak{su_{2}}}2 \,\, \overset{\mathfrak{g_{2}}}3  \,\, 1 \,\, \overset{\mathfrak{f_{4}}}5 \,\,  1 \,\,  ... [E_8]$ \\\hline
$D_7$ & $  SU(2) $ & $  \,\, \overset{\mathfrak{g_{2}}}3 \,\, \overset{\mathfrak{su_{2}}}2 \,\, 2 \,\,  1\,\, \underset{[SU(2)]}{\underset{2}{\underset{1}{\overset{\mathfrak{e_{8}}}{12}}}}  \,\, 1 \,\,  ... [E_8]$ \\\hline
$E_8(a_5)$ & $  1 $ & $  \,\, \overset{\mathfrak{g_{2}}}3 \,\, \overset{\mathfrak{su_{2}}}2 \,\, 2 \,\,  1\,\, \overset{\mathfrak{e_{8}}}{10}  \,\, 1 \,\,  ... [E_8]$ \\\hline
$E_8(b_4)$ & $ 1$ & $   \overset{\mathfrak{g}_2}3 \,\, \overset{\mathfrak{su_{2}}}2 \,\, 1 \,\, {\overset{\mathfrak{e_{7}}}8}  \,\, 1 \,\,  \overset{\mathfrak{su_{2}}}2 \,\, \overset{\mathfrak{g_{2}}}3  \,\, 1 \,\, \overset{\mathfrak{f_{4}}}5 \,\,  1 \,\,  ... [E_8]$ \\\hline
$E_7$ & $SU(2)$ &  $[SU(2)] \,\, \overset{\mathfrak{g_{2}}}3  \,\, 1 \,\, \overset{\mathfrak{f_{4}}}5 \,\,  1 \,\, \overset{\mathfrak{g_{2}}}3 \,\, \overset{\mathfrak{su_{2}}}2 \,\, 2 \,\, 1\,\, \overset{\mathfrak{e_{8}}}{11}  \,\, 1 \,\,  ... [E_8]$ \\\hline
$E_8(a_4)$ & $ 1$ & $   \overset{\mathfrak{su}_3}3 \,\, 1 \,\, {\overset{\mathfrak{e_{6}}}6}  \,\, 1 \,\,   \overset{\mathfrak{su_{3}}}3  \,\, 1 \,\, \overset{\mathfrak{f_{4}}}5 \,\,  1 \,\,  ... [E_8]$ \\\hline
$E_8(a_3)$ & $1$ &  $ \overset{\mathfrak{su_{3}}}3  \,\, 1 \,\, \overset{\mathfrak{f_{4}}}5 \,\,  1 \,\, \overset{\mathfrak{g_{2}}}3 \,\, \overset{\mathfrak{su_{2}}}2 \,\, 2 \,\, 1\,\, \overset{\mathfrak{e_{8}}}{11}  \,\, 1 \,\,  ... [E_8]$ \\\hline
$E_8(a_2)$ & $ 1$ & $\overset{\mathfrak{su_{2}}}2 \,\,   \overset{\mathfrak{so}_7}3 \,\, \overset{\mathfrak{su_{2}}}2 \,\, 1 \,\, {\overset{\mathfrak{e_{7}}}8}  \,\, 1 \,\,  \overset{\mathfrak{su_{2}}}2 \,\, \overset{\mathfrak{g_{2}}}3  \,\, 1 \,\, \overset{\mathfrak{f_{4}}}5 \,\,  1 \,\,  ... [E_8]$ \\\hline
$E_8(a_1)$ & $1$ &  $ {\overset{\mathfrak{su_{2}}}2} \,\, \overset{\mathfrak{g_{2}}}3  \,\, 1 \,\, \overset{\mathfrak{f_{4}}}5 \,\,  1 \,\, \overset{\mathfrak{g_{2}}}3 \,\, \overset{\mathfrak{su_{2}}}2 \,\, 2 \,\, 1\,\, \overset{\mathfrak{e_{8}}}{11}  \,\, 1 \,\,  ... [E_8]$ \\\hline
$E_8$ & $1$ &  $2 \,\, \overset{\mathfrak{su_{2}}}2 \,\, \overset{\mathfrak{g_{2}}}3  \,\, 1 \,\, \overset{\mathfrak{f_{4}}}5 \,\,  1 \,\, \overset{\mathfrak{g_{2}}}3 \,\, \overset{\mathfrak{su_{2}}}2 \,\, 2 \,\, 1\,\, \overset{\mathfrak{e_{8}}}{11}  \,\, 1 \,\,  ... [E_8]$ \\\hline

\caption{6D SCFTs associated with $E_8$ nilpotent orbits.}
\label{E8list}
\end{longtable}

\newpage

\bibliographystyle{utphys}
\bibliography{6Dtabflow}

\end{document}